\documentclass[fleqn,usenatbib]{mnras}
\usepackage{newtxtext,newtxmath}
\usepackage[T1]{fontenc}
\usepackage{orcidlink}
\usepackage{multirow}
\usepackage{makecell} 
\usepackage{subcaption}
\usepackage{ulem}
\usepackage{soul}
\usepackage{graphicx}	%
\usepackage{amsmath}	
\usepackage{subcaption} 
\usepackage{appendix}

\title[AGN feedback in a multiphase ICM]{AGN feedback in the multiphase inner intracluster medium of a cosmological super-resolved galaxy cluster} 

\author[M. Prunier et al.]{Marine Prunier\orcidlink{0009-0003-0932-2487}$^{1,2,3}$\thanks{E-mail: prunier@mpia.de},
Rahul Ramesh\orcidlink{0000-0002-7259-7454}$^{4,5}$, Dylan Nelson$^{6,7}$, Annalisa Pillepich\orcidlink{0000-0003-1065-927}$^{3}$, \newauthor Julie Hlavacek-Larrondo$^{1,2}$, Katrin Lehle$^{6}$, and Joey Braspenning$^{3}$
\\
$^{1}$Département de Physique, Université de Montréal, Succ. Centre-Ville, Montréal, Québec, H3C 3J7, Canada\\
$^{2}$Centre de recherche en astrophysique du Québec (CRAQ)\\
$^{3}$Max-Planck-Institut für Astronomie, Königstuhl 17, D-69117 Heidelberg, Germany\\
$^{4}$Kavli Institute for the Physics and Mathematics of the Universe (WPI), UTIAS, The University of Tokyo, Kashiwa, Chiba 277-8583, Japan\\
$^{5}$Center for Data-Driven Discovery, Kavli IPMU (WPI), UTIAS, The University of Tokyo, Kashiwa, Chiba 277-8583, Japan\\
$^{6}$Universität Heidelberg, Zentrum für Astronomie (ZAH), ITA, Albert-Ueberle-Str. 2, 69120 Heidelberg, Germany\\
$^{7}$Universität Heidelberg, Interdisziplinäres Zentrum für Wissenschaftliches Rechnen (IWR), INF 205, 69120 Heidelberg, Germany}
\date{}
\pubyear{\the\year{}}

\begin{document}
\label{firstpage}
\maketitle

\begin{abstract} 
{The intracluster medium (ICM) of galaxy cluster cores is a multiphase environment regulated by active galactic nuclei (AGN) feedback. Extended cool gas nebulae with complex morphologies and kinematics are commonly observed in cool-core clusters, but their origin and interaction with AGN feedback remain poorly understood. 
We present a cosmological magnetohydrodynamical zoom-in simulation of a low-mass cool-core galaxy cluster (\(M_{\mathrm{200c}}=2.3\times10^{14}\,\mathrm{M}_\odot\)), using initial conditions from TNG-Cluster and the IllustrisTNG galaxy formation model. Through super-Lagrangian refinement, we achieve a baryonic mass resolution of \(2.3\times10^{4}\,\mathrm{M}_\odot\) and a spatial resolution of \(\sim40\text{--}90\) pc in the cool (\(10^{4}\,\mathrm{K}<T<10^{5.5}\,\mathrm{K}\)) gas phase within the central \(370\) ckpc, while retaining the full cosmological context.
We investigate the {properties and evolution} of cool clouds and filaments and their interaction with AGN feedback at $z\leq0.35$. The cluster core hosts a rich population of dense, low-entropy filaments, including inflowing and outflowing structures and occasional rotationally supported disks. Comparing simulations with AGN kinetic feedback turned off and on, we find that AGN activity {fragments and redistributes} the cool gas and increases its covering fraction without destroying the cool gas reservoir. Individual outbursts drive transient increases in the velocities and velocity dispersions of both cool and hot gas, on short, $\lesssim 50$ Myr, timescales. We also identify low-entropy gas uplifted by AGN-driven outflows. While feedback globally heats the ICM and offsets cooling, {it also creates conditions that may favor local thermal instabilities -- through shock compressions, mixing, turbulence, and uplift.}}
\end{abstract}

\section{Introduction}

The intracluster medium (ICM) in galaxy clusters -- hereafter clusters -- is a hot, diffuse, and highly ionized plasma with temperatures reaching $10^{7}$--$10^{8} \mathrm{K}$ \citep[e.g.,][]{mohr_properties_1999}. At these temperatures, the gas emits primarily via thermal bremsstrahlung and line emission in the X-ray band, making clusters luminous X-ray sources. In the cores of relaxed, cool-core (CC) clusters, the radiative cooling time of the ICM is much shorter than the Hubble time. In the absence of a heating mechanism, this would lead to substantial gas condensation and inflows onto the central brightest cluster galaxy (BCG), producing high star formation rates (SFR). However, the observed SFRs are far below those predicted by classical cooling-flow models \citep[e.g.,][]{1999Crawford}, implying that heating sources offset radiative losses and prevent catastrophic cooling.

This heating is thought to be orchestrated by feedback from active supermassive black holes (AGN, SMBH) hosted in the BCGs. AGN outbursts are believed to prevent excessive cooling in cluster core atmospheres through a combination of processes. These include heating of the gas via the inflation and rise of X-ray cavities \citep[e.g.,][]{2004Birzan}, turbulence \citep[e.g.,][]{2014Zhuravleva,2022Zhang}, cosmic-ray injection and diffusion \citep[e.g.,][]{2007Pfrommer,Sijacki2008,2018Ehlert}, as well as AGN-driven shocks \citep[e.g.,][]{Fabian2006_Perseus_Shock} and sound waves \citep[e.g.,][]{2008Sanders_soundwaves}. All of these feedback channels likely interact in complex ways and operate across a range of spatial and temporal scales to globally balance radiative cooling in cool-core clusters.

Although rare, the presence of massive cold gas reservoirs, in the centers of CC clusters in particular, indicates that substantial cooling still occurs \citep[][]{2019Olivares_mol}. Additionally, recent evidence has identified so-called `hidden' cooling flows in ellipticals, groups, and clusters \citep{Fabian_hiddend1}. Hot gas ($ \geq 10^{5.5} \mathrm{K}$), warm ionized gas ($\sim 10^{5} \mathrm{K}$), cold atomic gas ($\sim 10^{4} \mathrm{K}$), and molecular gas ($\lesssim 10^{3} \mathrm{K}$) can clearly coexist within cluster cores, and cooler phases can be observationally detected through several tracers, including as H$\alpha$ nebulae \citep[e.g.,][]{2024Vigneron, 2025Olivares_Nature}, H I or Mg II quasar absorbers \citep[e.g.,][]{2024Fresco}, molecular CO \citep[e.g.,][]{2019Olivares_mol, 2024Omoruyi}, and cooler components in soft X-ray emission \citep{2016Shin}.

Among these gas phases, warm ionized filamentary structures are common in BCGs and particularly striking. They can extend over tens of kiloparsecs (kpc), with thicknesses as small as $\sim 50$ pc, and display a wide range of morphologies—from coherent, disk-like structures to clumpy, extended filaments. Well-known examples include the 90-kpc H$\alpha$ filaments in Perseus \citep[e.g.,][]{2025Rhea} or the rotating disk in Hydra A \citep[e.g.,][]{2014Hamer}. Some filaments appear relaxed, while others reveal clear signs of dynamical disturbance, hinting at the ongoing interplay between cooling, feedback, and cluster dynamics \citep[e.g.,][]{2022Olivares_MUSE}. These filamentary nebular emissions are often co-spatial with cold molecular clouds \citep[e.g.,][]{2025Li_VSF} and X-ray structures, frequently wrapped around or in the wake of X-ray cavities \citep[e.g.,][]{2008Salome,2017Russel}. This spatial correlation suggests that warm filaments may be actively shaped by AGN feedback processes. 

Despite extensive observations, the origin of these colder phases remains debated. The observed spatial and quantitative correlation between H$\alpha$ emission and soft X-ray surface brightness in filaments of CC clusters \citep[][]{2025Olivares_Nature,2026Poitras}, suggests a close coupling between the warm and hot gas phases. Such coupling is consistent with a top-down multiphase condensation cascade, in which both the warm ionized and cold molecular components condense from the hot phase and "rain" toward the cluster center \citep{2012Sharma}. Hot cluster atmospheres can become locally thermally unstable when the ratio of the cooling to free-fall timescales, $t_{\rm cool}/t_{\rm ff}$, falls below $\sim 10$–$20$ \citep[e.g.,][]{2011Gaspari_Melioli,2016McNamara,2018Prasad, 2022DonahueVoit}. In this regime, cool cloud formation can arise from several \textit{in-situ} channels.

Cool filaments may condense directly from the hot gas in regions of thermal instabilities (TI), around overdensities or small cold gas clouds acting like seeds. Gas-rich mergers or satellite ram-pressure stripping can deliver multiphase material to the cluster core, providing seeds for further cold gas growth \citep[e.g.,][]{2006Mayer,2014Sparks,2023Ramesh_tng50}. Low-entropy gas lifted by AGN-inflated X-ray cavities or shocks can reach altitudes where it becomes thermally unstable, leading to precipitation that rains back toward the center \citep[e.g.,][]{2005Pizzolato,2012McCourt,2015LiBryan,2017Voit,2018Tremblay,2024Duan_uplift}. Turbulence, induced either by AGN feedback or by structure formation, can source overdensities and enhance cooling in the ICM \citep{2013Gaspari,2014LiBryan}. 

Simulations provide insights into the origin and coupling of multiphase gas in cluster cores. A multiphase ICM naturally forms in both idealized and cosmological numerical simulations of galaxy clusters \citep{2014Guo,2013Gaspari, 2019Beckmann, 2021Chadayammuri,2025Staffehl}. In idealized setups, \citealt{2008Revaz} and \citealt{2015Brighenti} show how cold gas can condense in the wake and at the edges of rising cavities, falling back towards the center of the gravitational potential. {\citet{2012Sharma} showed that local thermal instability can produce a multiphase medium, with cold filaments condensing out of the hot intracluster medium when the ratio of the cooling time to the free-fall time becomes sufficiently low.}
More recently, \citet{2026Sotira} show that AGN jet injection of fast hot material compresses the surrounding ICM, locally reducing the cooling time and triggering the formation of cold gas. Magnetic fields have been shown to influence the physical properties and evolution of cool gas clouds \citep[in TNG50;][]{2020Nelson_tng50}, in part by magnetic draping effects \citep[in GIBLE;][]{2024Ramesh_draping}. Magnetic fields can also help form elongated morphologies while suppressing massive cold discs that are otherwise common in purely hydrodynamic simulations \citep[e.g.,][]{2019Qiu,2024Fournier}. Simulations also suggest that the kinematics of cold gas may retain a "memory" of the hot phase from which it condensed \citep[e.g.,][]{2018Gaspari_Shaken,2021Wang}, though this remains debated \citep[][]{2025Sotira,2025Fournier}. 

While idealized high-resolution simulations of isolated clusters have been instrumental in studying multiphase gas and its interaction with AGN feedback, their idealized setup and lack of a cosmological environment prevent them from capturing the full complexity of small-scale structure formation. Cosmological simulations provide an alternative approach, evolving halo gas over cosmic time and scales, capturing the interplay between galactic outflows, large-scale gas accretion, and satellite interactions, offering a self-consistent representation of the ICM. However, these simulations are typically run at coarser numerical resolution compared to idealized setups, which can limit their ability to resolve small-scale structure.

To bridge the resolution gap separating typical cosmological simulations from their idealized counterparts, a promising approach is to enhance the numerical resolution in the ICM using halo-targeting refinement schemes \citep{2019Suresh,2019Hummels}. These methods deliver high resolution where it is most needed while maintaining relatively coarser resolution elsewhere, significantly reducing computational cost compared to uniformly high-resolution simulations. Such setups have recently been employed to investigate the formation and long-term evolution of clouds in the CGM \citep{2025Ramesh_origin,2025Augustin}, to resolve the interaction of AGN jet lobes with the ICM \citep{2021Bourne_weak_sim,2026Sotira}, and to study halo gas in high-redshift protoclusters \citep{2026Bennett}.

In this work, we combine the benefits of high-resolution refinement with the full cosmological and galaxy formation context, and simulate a cool-core, low-mass galaxy cluster ($M_{500\mathrm{c}} = 1.4 \times 10^{14}\ \mathrm{M}_\odot$ at $z=0$). The initial conditions (ICs) are selected from the TNG-Cluster\footnote{\url{www.tng-project.org/cluster/}} magnetohydrodynamical cosmological simulation \citep{2024Nelson}. We employ a super-Lagrangian refinement scheme \citep{2019Suresh,2024Ramesh_GibleI} within the central $370$ comoving kpc\footnote{Throughout this paper, ckpc denotes comoving kiloparsec, and kpc denotes physical kpc.} of the BCG, achieving a target gas mass resolution of $2.3 \times 10^{4}\ \mathrm{M}_\odot$, 512 times higher than in the fiducial TNG-Cluster run, resolving gas cells down to a few tens of parsecs. We present the simulation and characterize the properties, dynamics, and evolution of the multiphase gas in cluster cores at low redshift. {This work also provides the foundation for a follow-up study that will incorporate greater observational realism and investigate the coupling between gas phases through their surface-brightness distributions and kinematics.}

In Section~\ref{sec:meth}, we describe the methodology and simulation setup. Section~\ref{sec:results} presents our main simulation results. In Section~\ref{sec:discussion}, we examine the resolution, the origin of the cold clouds, and compare our findings with observations, before concluding in Section~\ref{sec:conclusion}.


\section{Methods}\label{sec:meth}

We simulate a cosmological MHD zoom-in of a cool-core galaxy cluster with $M_{500\mathrm{c}} = 1.4 \times 10^{14}\  \mathrm{M}_\odot$ ({ $M_{200\mathrm{c}} = 2.3 \times 10^{14} \ \mathrm{M}_\odot$)} at $z=0$, using initial conditions drawn from TNG-Cluster \citep{2024Nelson}. To better resolve the multiphase structure of the central ICM, we apply a super-Lagrangian refinement scheme in AREPO within the central $370$ ckpc around the BCG, achieving a target gas mass resolution of $2.3 \times 10^{4} \ \mathrm{M}_\odot$. In the following section, we describe the galaxy formation model, cluster selection, refinement strategy, and simulation suite presented in this work.

\subsection{TNG Galaxy Formation Model and Multiphase ICM}\label{subsec:meth_model}

Our zoom-in cosmological cluster is evolved with the unchanged IllustrisTNG galaxy formation model \citep[hereafter TNG;][]{2017Weinberger,2018Pillepich}, with ICs selected from the TNG-Cluster \citep[][]{2024Nelson}. The TNG model includes gas cooling and heating, star formation, metal enrichment, stellar feedback, SMBH seeding and growth, and feedback. It includes magnetic fields, but does not model cosmic rays or thermal conduction. The gas is prevented from cooling below an effective temperature floor of $\sim 10^{4} \mathrm{K}$, {and} molecular phases and dust are not modeled. Star formation follows the prescription of \citet{2003SpringelHernquist}, where gas above the density threshold of $n_{\mathrm{H}} > 0.1 \mathrm{cm}^{-3}$ forms stars stochastically according to the Kennicutt-Schmidt relation, and assuming a Chabrier initial mass function.

AGN feedback in TNG is implemented as a dual-mode subgrid model. At high accretion rates, i.e., when the instantaneous Bondi-Hoyle accretion rate exceeds a given fraction of the Eddington limit, SMBHs inject continuous thermal, isotropic energy (quasar mode), while at low rates, feedback shifts to kinetic, directional outbursts (kinetic mode). The feedback energy rate is $\Delta \dot{E}_{\text{kin}} = \epsilon \dot{M}_{\text{SMBH}} c^2$, with a maximum coupling efficiency of $\epsilon = 0.2$ \citep{2017Weinberger}, resulting in a typical power of feedback event in clusters of $10^{41} - 10^{45} \, \text{erg s}^{-1}$ at low redshift \citep{Prunier2025a}. TNG kinetic feedback naturally produces a wealth of AGN-driven X-ray morphological structures that emerge naturally in the simulations. These include eROSITA-like bubbles in Milky Way-mass galaxies \citep{2021PillepichErosita} and X-ray cavities in the BCGs, shocks, and pressure waves, \citep{2024Truong,Prunier2025a,Prunier2025b,2025cPrunier}. This demonstrates that the model can reproduce X-ray morphological features similar to many of those observed, despite not explicitly modeling relativistic AGN jets.

Previous studies have shown that the gas in TNG galaxy clusters is multiphase, with massive halos hosting various amounts of cool gas over cosmic time \cite[e.g.,][]{2020Zinger,2026Chatzigiannakis}. However, the central ICM has either been excluded from these analyses \citep[][who consider only FoF gas within the aperture $0.15$-$1.0,R_{200\mathrm{c}}$ and exclude satellite gas from their analysis]{2025Rohr_cool}, or investigated primarily on larger cluster scales and in connection with satellite populations \citep{2025Staffehl}. As shown in Figure~\ref{fig:meth_bcgs_tngcluster}, the central cold (r $<50\ \mathrm{kpc}$, T $< 10^{4.5}\mathrm{K}$) gas mass as a function of halo mass at $z \sim 0$ demonstrates that many halos retain a substantial cold gas reservoir in the TNG-Cluster simulation. Moreover, strong cool-core clusters, which undergo more frequent AGN feedback \citep{2024Lehle,Prunier2025a}, tend to host even larger reservoirs.

The combination of a realistic galaxy formation model, ICM physics, and the presence of multiphase gas in the core of clusters provides an ideal set-up to study in detail how these components interact. However, the TNG-Cluster simulation has a relatively coarse gas mass resolution of $10^{7} \mathrm{M}_\odot$, which limits our ability to study small-scale structure. In this paper, we therefore select one cluster from TNG-Cluster and re-simulate it from its ICs at significantly higher resolution using a super-Lagrangian refinement scheme.

\subsection{Initial Conditions}\label{subsec:meth_ics}

\begin{figure}
\centering\includegraphics[width=0.52\textwidth]{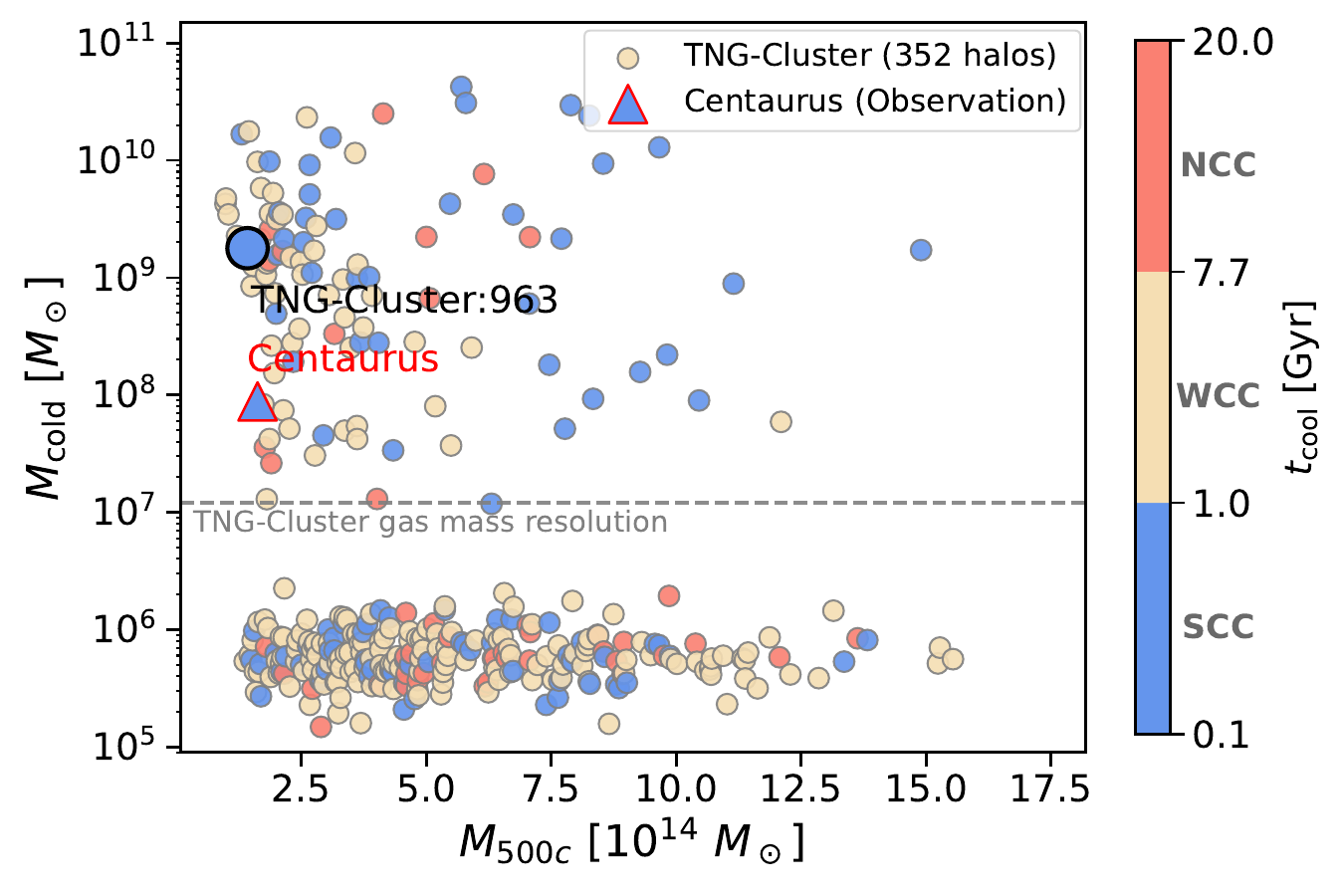}

    \caption{Cluster selection.{We show the mean cold gas mass ($T < 10^{4.5}\,\mathrm{K}$) within the central 50 kpc of TNG-Cluster halos over $z=0$-$0.1$, as a function of halo mass at $z=0$. Halos are color-coded according to \citet{2024Lehle} as strong cool-core (SCC), weak cool-core (WCC), and non cool-core (NCC), based on their central cooling time at $z=0$.} The red-outlined triangle corresponds to the Centaurus cluster \citep[][]{2019Olivares_mol}, and the black-circled marker shows the TNG-Cluster halo, whose ICs were selected for our super-resolved simulation (halo original ID 963). For visualization, halos with no cold gas are assigned a random value $<10^{7} \mathrm{M}_\odot$. Many halos contain a $>10^{7} \mathrm{M}_\odot$ reservoir of cold gas, all of them being either SCC or WCC, with a weak correlation between cold gas mass and halo mass \citep[see also][]{2025Rohr_cool}.}
    \label{fig:meth_bcgs_tngcluster}
\end{figure}

Not all TNG-Cluster BCGs\footnote{By BCG we mean the most massive galaxy of the friends-of-friends (FoF) halo corresponding to each cluster.} are quenched at $z=0$. In fact, most clusters undergo multiple phases of quiescence and rejuvenation over their lifetimes, often accompanied by transitions between cool-core and non cool-core states and the presence or absence of central cold gas reservoirs \citep{2025Lehle}, particularly at high redshift \citep{2026Gottschewski}. At redshift zero, 8\% of the BCGs are classified as star-forming, defined by a specific star formation rate above the quenching threshold of $\mathrm{sSFR} = 10^{-11} \mathrm{yr}^{-1}$ \citep[][]{2024Nelson}, while 15\% have a cold gas reservoir in the central 50 kpc at low redshifts (Figure \ref{fig:meth_bcgs_tngcluster}). These clusters exhibit a diverse range of central cool-gas morphologies, even at the native TNG-Cluster resolution, from relatively relaxed, disk-like structures to highly disturbed filamentary configurations (\textcolor{blue}{Lehle et al. in prep.}).

\begin{figure*}
    \centering
    \includegraphics[width=0.48\textwidth]{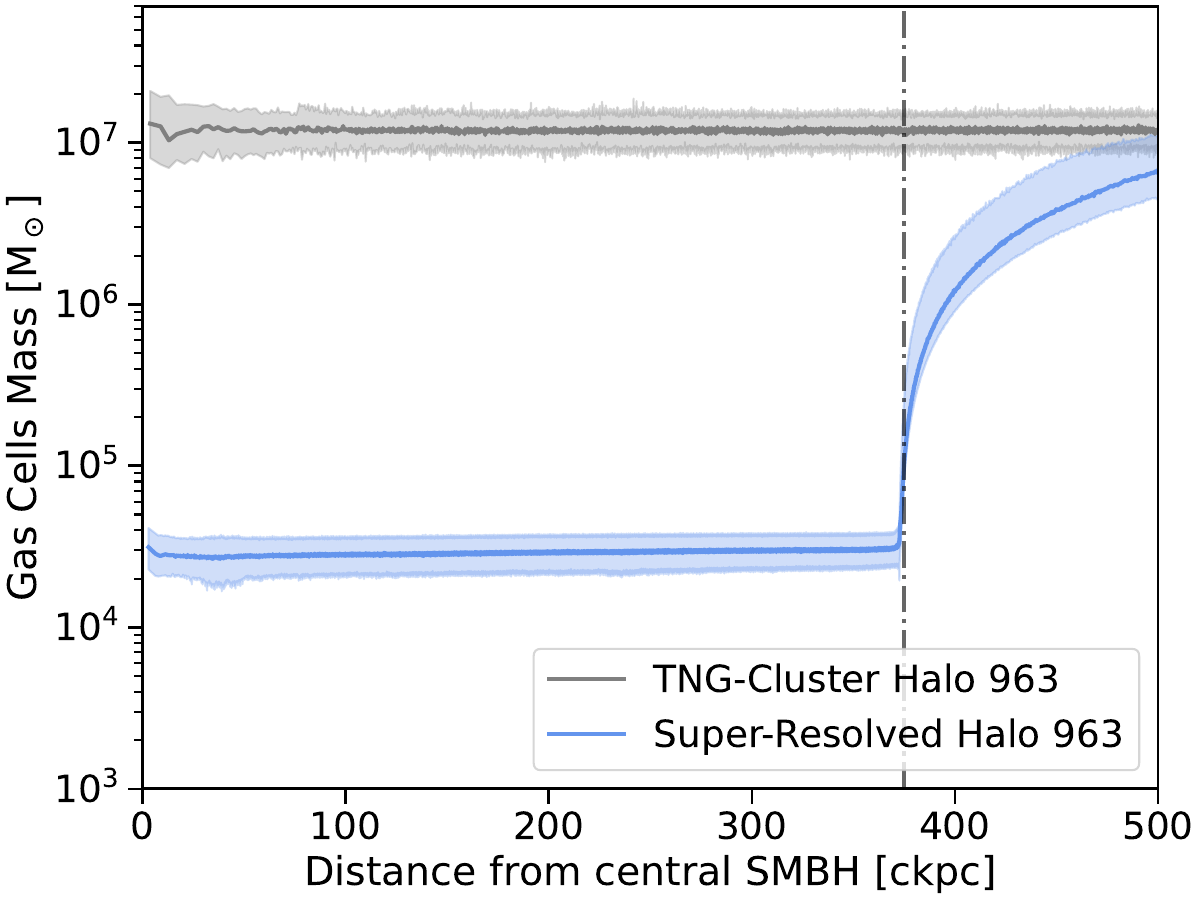}
    \includegraphics[width=0.48\textwidth]{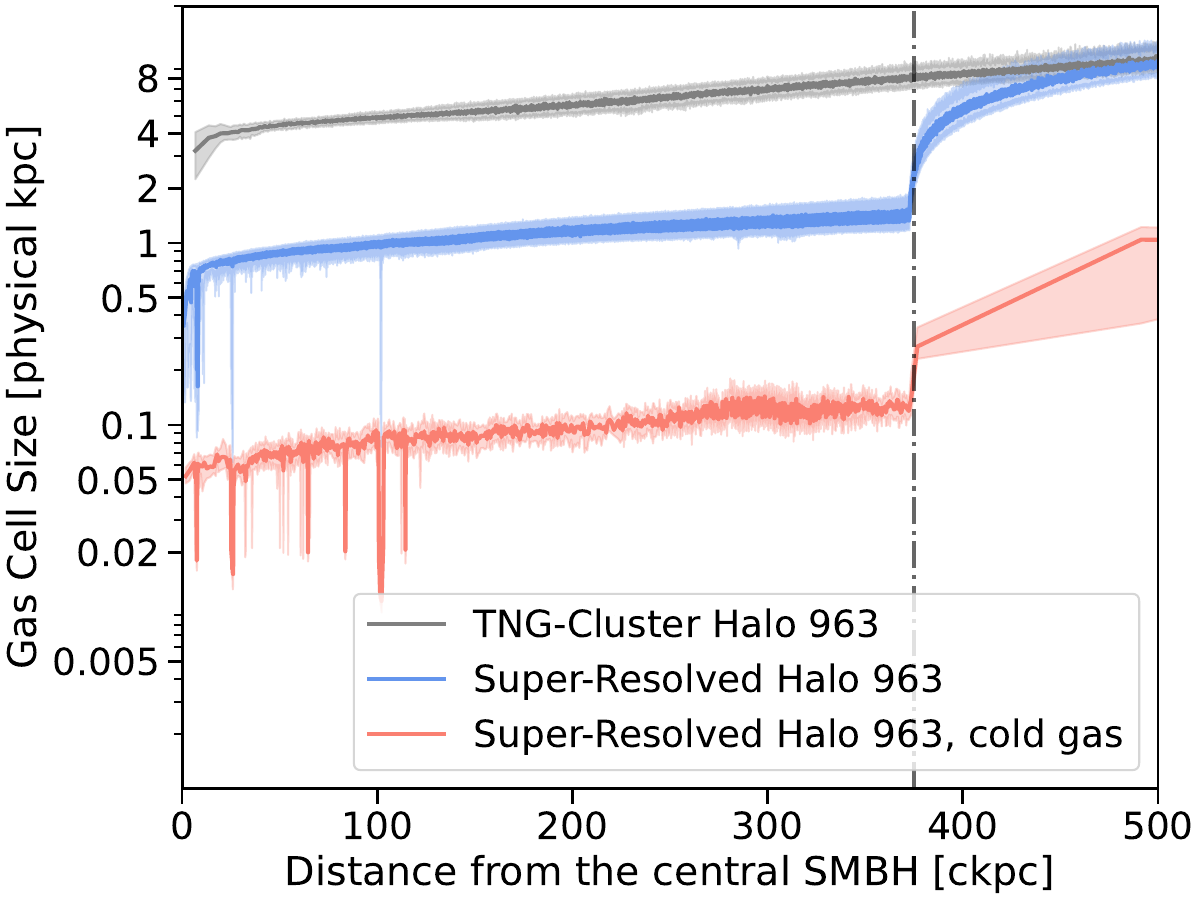}
    \caption{Numerical resolution. Radial profiles of gas mass resolution (left) and spatial resolution (right) within $R_\text{500c}$ of h963 from our super-resolved run (blue, all gas cells; light red, cold gas cells only) compared to TNG-Cluster (grey, all gas cells). Shaded regions indicate the 16th–84th percentile range. The dashed grey vertical line marks the boundary of the refinement region. The distance-dependent target mass of our ICM refinement scheme boosts both mass and spatial resolution in the BCG core. The smallest gas cells are dominated by the cold phase. The {dips} in the blue/grey curve (right panel) correspond to satellites or streams of dense, cold gas inflowing or outflowing from the galaxy.}  
    \label{fig:meth_refinement}
\end{figure*}

We select from the TNG-Cluster zoom-ins ICs, a low-mass, Centaurus-like cluster with $M_{500c} = 1.4 \times 10^{14}\ \mathrm{M}_\odot$ (original halo ID 963, hereafter h963). h963 is characterized at $z = 0$ by a cooling time below 1 Gyr ($t_{\rm cool} = 0.75$ Gyr) and an X-ray luminosity of $L_{\rm 500,X} \sim 10^{44} \mathrm{erg\ s}^{-1}$. {The system hosts a cold gas reservoir of $1.8 \times 10^{9}\,\mathrm{M}_\odot$ (averaged over $z=0$--$0.1$, within the central 50 kpc),} and shows X-ray cavities and weak AGN-driven shocks, making it well-suited for investigating the formation of multiphase gas in interaction with AGN feedback. According to \citet{2024Lehle}, it is classified as either a strong or weak cool-core system over most of its lifetime. At low redshift, h963 exhibits low but non-zero star formation, with a specific star formation rate in the central 50 kpc region of $\mathrm{sSFR} = 3 \times 10^{-11}\,\mathrm{yr}^{-1}$.

The selection of this cluster is motivated by our requirement for a system that retains cold gas at redshift $z = 0$. We aim to avoid the further expense associated with extreme star formation rates and the presence of massive, dense gas disks. 

\subsection{Super-Lagrangian Refinement and Simulation Setup}\label{subsec:meth_slr}

We use the refinement scheme developed for the GIBLE simulations \citep{2024Ramesh_GibleI} and described in detail therein. We modify the refinement region, setting it as a spherical volume centered on the central SMBH of the BCG and extending to a radius of $r_{\rm max} = 370$ ckpc. The scheme defines distance from a seed black hole, flagged once its host halo crosses a mass threshold of $\sim 7 \times 10^{10} \mathrm{M}_\odot$. When the flagged black hole merges, the flag is inherited by the descendant, ensuring its continuous tracking throughout the simulation.

\textsc{Arepo} gas fluid dynamics are solved on an adaptive Voronoi-based moving mesh, which is locally density-dependent: higher-density regions are sampled by smaller cells, while (de)refinement maintains a nearly fixed gas cell mass. The refinement scheme builds upon the existing (de)refinement operations of \textsc{Arepo}, which normally maintain gas cells within a factor of two of a constant target mass. Here, we require a distance-dependent target mass, $m_{\rm dist,target}(r)$, of $2.3 \times 10^{4} \mathrm{M}_\odot$ within $r < r_{\rm max}$. Outside the refinement region, this smoothly transitions to the fiducial TNG-Cluster gas mass resolution of $1.2 \times 10^{7} \mathrm{M}_\odot$. 

Figure~\ref{fig:meth_refinement} shows how mass and spatial resolution vary with distance with the super-Lagrangian refinement active. The curves converge with TNG-Cluster beyond $r_{\rm max}$, where the median gas cell sizes are 4-8~kpc, while in the super-resolved region they are 0.5-1~kpc in extent. This increased resolution substantially boosts our ability to study gas cooling and small gas structures in the inner ICM of this cluster. 

\begin{figure*}
    \centering
    \includegraphics[width=0.4985\linewidth, trim={0.0 0.0cm 0.0cm 0.0cm},clip]{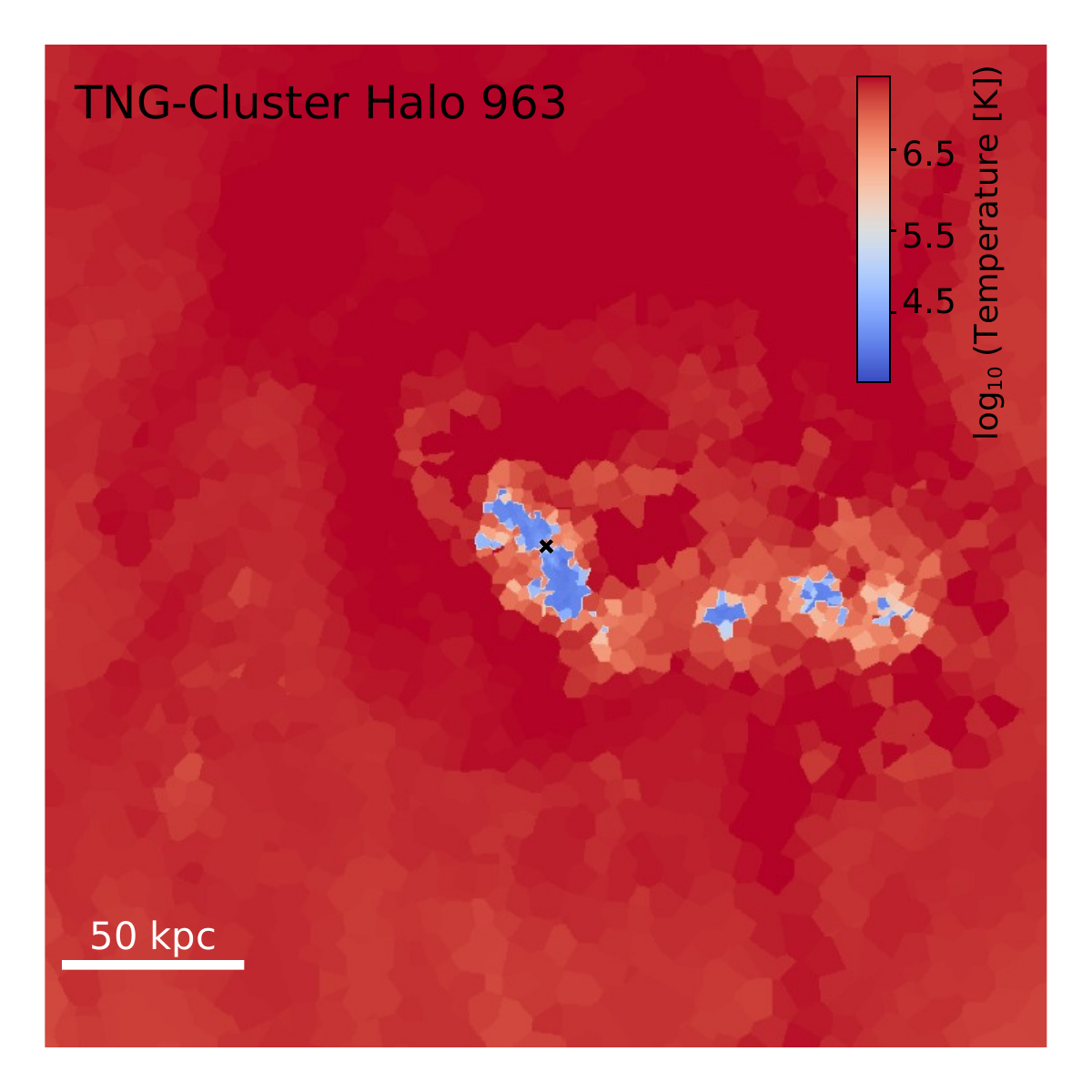}
    \includegraphics[width=0.495\linewidth, trim={0.0cm 0.0cm 0.2cm 0.0cm},clip]{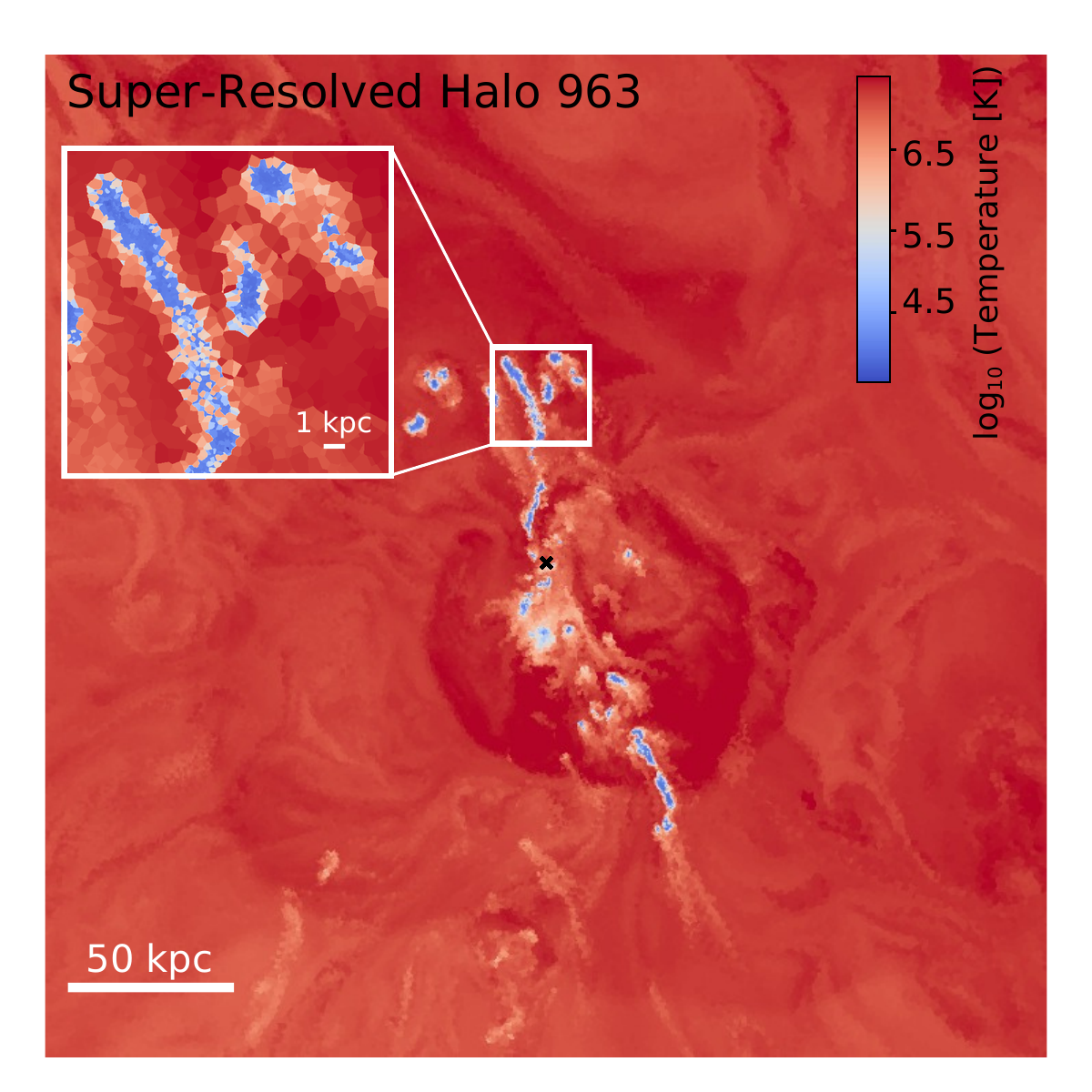}
    \vspace{-0.5cm}
    \caption{Illustration of the super-Lagrangian scheme. The left panel shows the halo h963 in TNG-Cluster, while the right panel presents our super-resolved simulation of h963. We show a cross-section of the Voronoi tessellation of the gas, taken through a plane passing through the cluster center, color-coded by temperature. The black crosses denote the position of the central SMBH.  Owing to the quasi-Lagrangian nature of \textsc{Arepo} and the fixed target gas cell mass ($\sim 10^{7} \mathrm{M}_\odot$ in the TNG-Cluster halo versus $2.3 \times 10^{4} \mathrm{M}_\odot$ in the super-resolved halo), higher-density regions are naturally resolved by smaller cells. As a result, in the right plot, finer structures are resolved in the core: a higher number of small, cold structures emerges surrounded by layers of warmer gas, AGN-driven hot fronts, and gas turbulent motions with swirls and eddies are more sharply defined, all embedded within a more finely resolved hot halo.}
    \label{fig:meth_refinement_img}
\end{figure*}

Figure~\ref{fig:meth_refinement_img} visualizes the underlying \textsc{Arepo} Voronoi tessellation for our simulation (h963 super-resolved, right) and its counterpart in TNG-Cluster (left). {Dense, small-scale cold clouds are more highly resolved, with cold gas cell sizes of $35$ pc at the 5th percentile of the distribution across redshifts in the refinement volume.} We also resolve intermediate-density, warm-phase interface regions. This preferential increase of resolution in the colder gas and its surroundings is a consequence of the GIBLE `uniform mass factor' type criteria \citep{2024Ramesh_GibleI}, a feature not shared by other CGM refinement schemes that instead use `uniform/maximum volume' type criteria \citep[e.g. SURGE and FOGGIE;][]{2025Rintoul,2025Augustin}.

We activate the refinement scheme for $z \leq 0.6$, maintaining a constant comoving volume within $r_{\rm max}$ at high resolution down to $z=0$. Our focus is on the formation and dynamics of cold gas in the core and at low redshift ($z \lesssim 0.35$), where clusters begin to be observationally accessible in spatially resolved detail and where most cool-core filamentary nebulae are detected. This redshift is in the range of the highest-redshift systems for which both X-ray cavities and cold gas have been robustly characterized, such as Abell~1835 at $z = 0.25$, or the Phoenix cluster at $z=0.59$ \citep[][]{2025Li_VSF,2025Reefe}. Previous studies of galaxy clusters motivate our choice of $r_{\rm max}$. Observationally, cold gas filaments are typically found within a few tens of kpc, with some of the most extended ionised warm nebula structures reaching $\sim 100-150$kpc \citep[][]{2019Olivares_mol}. Similarly, X-ray cavities are generally observed within $<$200 kpc from the cluster center \citep[e.g.,][]{2014Panagoulia_vol}. 

To assess the response of the cluster after refinement is introduced at $z \leq 0.6$, we perform additional tests in which the refinement is instead activated at $z=0.5$ and $0.4$. We find that key quantities, such as the SFR, gas density, total cool gas mass, cloud mass, and cloud size distributions, converge within $\Delta z \sim 0.15$. {This corresponds to a lookback time significantly longer than the dynamical timescale of the halo.} We therefore consider the simulation results to be reliable from $z \lesssim 0.45$ onward. 

From $z=0.2$, we output snapshots every 100 Myr. We additionally output three high–temporal-cadence segments, with snapshots spaced by 5 Myr and each lasting 50–80 Myr, in order to study the gas dynamics during phases of AGN feedback with enhanced temporal resolution. Finally, we perform a variation run in which AGN kinetic feedback is switched off from $z=0.6$ onward (see Table\ref{tab:tab_sim_run}). This allows us to completely isolate and quantify the impact of feedback on the cluster core and on the properties and dynamics of the multiphase gas. The total computational cost of the main simulation run is $4.3 \times 10^{5}$ CPU hours.

{We do not perform a systematic resolution convergence study, given the computational cost of the super-Lagrangian refinement. However, we can offer some insights. Our baryonic mass resolution is $\sim4$ times finer than that of TNG50 \citep{2019Nelson, 2019Pillepich}, placing us in a regime where the TNG model has been extensively explored \citep[e.g.,][]{2018Pillepich, 2019Pillepich}, albeit mostly in lower-mass halos, whereas here it is reached in a cluster-mass halo. Nevertheless, in TNG, SMBH growth and feedback histories have been shown to be sensitive to resolution and the same injected energy may couple differently to better-resolved gas. We have checked (albeit do not show) that within the TNG model, these effects are less pronounced for progressively more massive halos, where SMBH mergers dominate growth and kinetic feedback reduces central gas availability \citep[e.g.,][]{2019Lakhchaura,2026Gottschewski}. To further assess the impact of refinement, we compare the SMBH mass, accretion rate, and cumulative kinetic energy injection in the super-resolved h963 with those of its TNG-Cluster analog. At $z<0.4$, these quantities agree within $\sim0.3$--$0.8$ dex, indicating broadly similar SMBH growth and cumulative feedback energetics. The BCG star formation histories, and the star-forming gas masses within the full refinement region, are also similar with median values of $7\times10^{9}\,\mathrm{M}_\odot$ and $9\times10^{9}\,\mathrm{M}_\odot$ in the super-resolved halo and its TNG-Cluster analog, respectively. This suggests that refinement in h963 has a limited impact on SMBH growth, cumulative feedback energetics, and the star-forming gas reservoir.}

{\renewcommand{\arraystretch}{1.4}
\begin{table}
\centering
\caption{{Summary of simulation runs. In all runs, super-Lagrangian refinement is activated at $z \leq 0.6$. The  `Analysis Time' column corresponds to the simulation redshift range considered in this work. Ranges in spatial resolution give 16--84th percentiles.}}
\begin{tabular}{l l l} 
\hline
\textbf{Mass resolution} &
\multicolumn{2}{c}{\textbf{Cell resolution for $T<10^{5.5}$ K gas}} \\
\hline

\multirow{2}{*}{$m_{\rm gas}=2.3\times10^{4}\,{\rm M}_\odot$}
& \multicolumn{2}{l}{$39$--$85$ pc (in the refinement region $<370$ ckpc)} \\
& \multicolumn{2}{l}{$30$--$55$ pc (in the core $50$ ckpc)} \\

\hline
\vspace{0.05cm}
\textbf{Run} & \textbf{Description} & \textbf{Analysis time} \\
\hline

Main & AGN kinetic feedback on & $z=0.35 \rightarrow 0.00$ \\

\multirow{2}{*}{Variation} 
& AGN kinetic feedback off &
$z=0.35 \rightarrow 0.17$ \\
& (from $z=0.60$ onward) & \\[0.3ex]
\hline
\end{tabular}
 \label{tab:tab_sim_run}
\end{table}
}

\subsection{Physical Definitions and Cloud Identification}\label{subsec:meth_gas_clouds}

We divide the gas into three temperature phases \citep[similar to][]{Rohr2024_cool}: cold and star forming gas with $T \leq 10^{4.5} \mathrm{K}$, warm gas with $10^{4.5} < T < 10^{5.5}\mathrm{K}$, and hot gas with $T \geq 10^{5.5} \mathrm{K}$. In the following, we also sometimes consider the cold and warm phase together, i.e. gas with $T < 10^{5.5}\mathrm{K}$. We adopt this fixed temperature-based classification throughout the analysis, independent of redshift.

Following  \citet{2020Nelson_tng50,Ramesh_Nelson_23_TNG50_II}, we identify clouds as spatially contiguous sets of cold ($T \leq 10^{4.5}\mathrm{K}$) gas cells in the unstructured \textsc{Arepo} Voronoi tessellation. Each cloud must contain at least 10 resolution elements (neighboring cells). Structures composed of fewer than 10 cells are considered unresolved. Gas cells gravitationally bound to satellite galaxies, as identified by the \textsc{Subfind} substructure identification algorithm, are excluded.

\begin{figure*}
    \centering
    \includegraphics[width=0.49\textwidth]{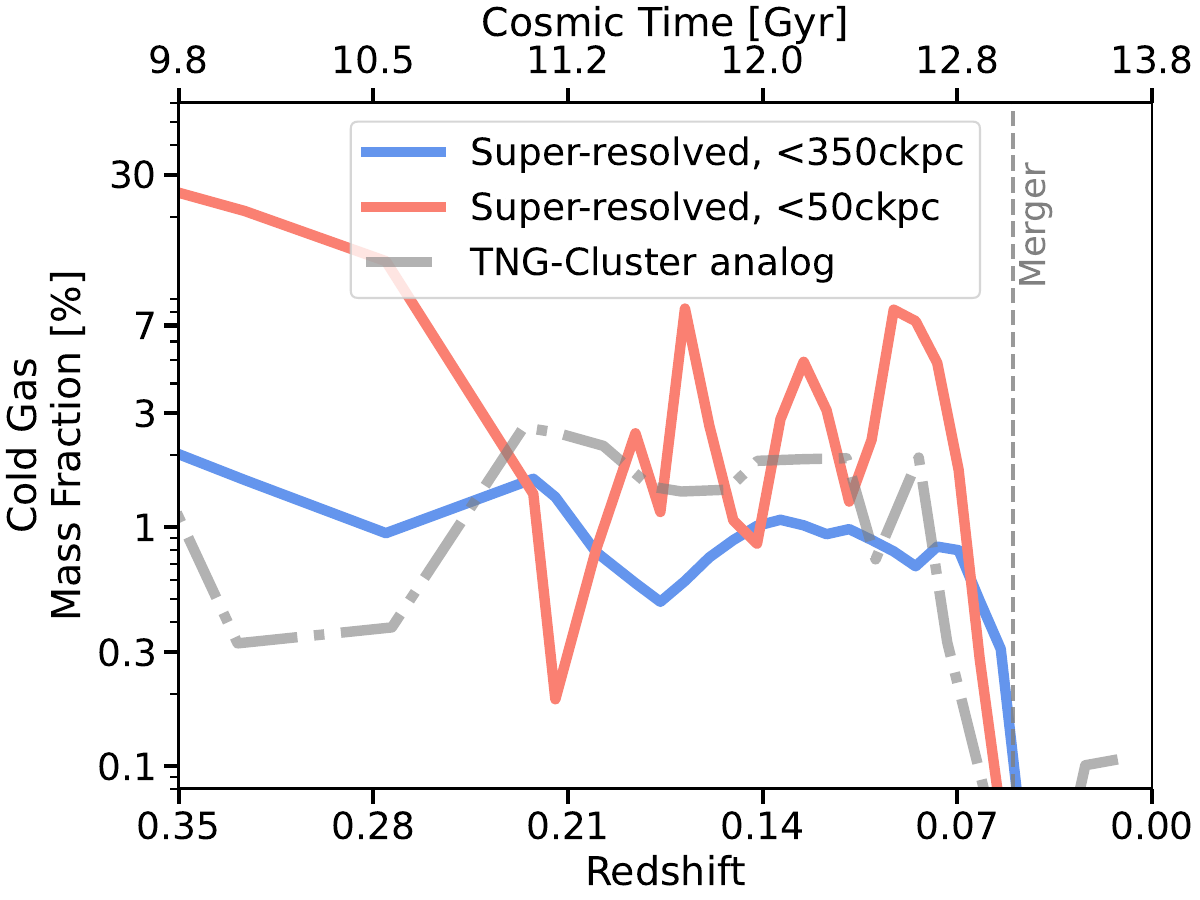}
    \includegraphics[width=0.49\textwidth]{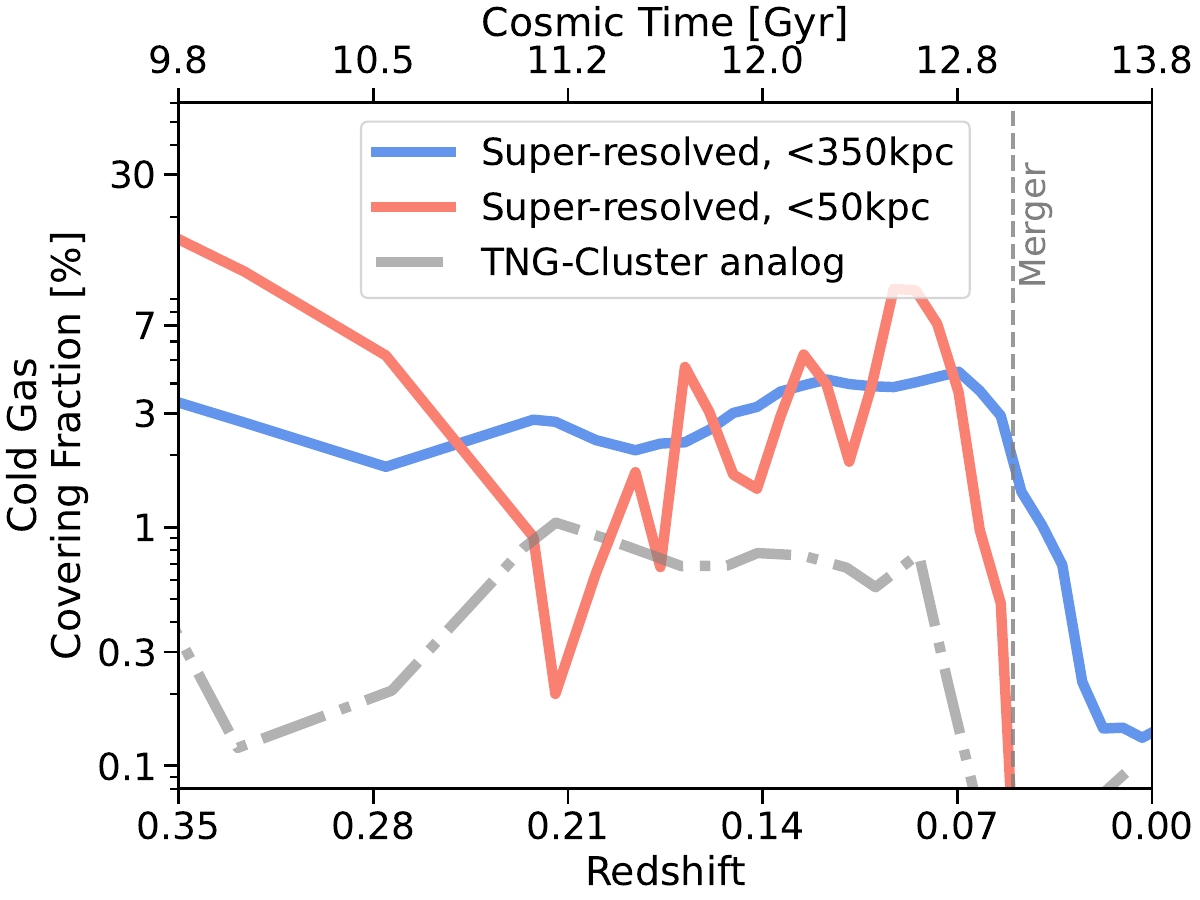}
    \caption{Cold gas content. We show the cold gas mass fraction (left) and covering fraction (right, i.e., the surface fraction covered by cold gas) as a function of redshift. Solid curves show the quantities within the full super-resolved region (blue) or the core, $<50$~kpc, region (light red), while the grey dotted-dashed lines correspond to the fractions for the TNG-Cluster analog, computed within the same aperture as the full super-resolved region. The cold gas mass fractions are similar for both simulations and remain low ($<2\%$, blue and grey curves)  until $z \sim 0.05$, when a merger sweeps out most of the remaining cold gas. However, the cold gas covering fraction is a factor 3-5 higher in the super-resolved halo, reflecting the larger spread of cold gas cloudlets. The fractions are higher in the core (up to 10 \%), with the fluctuations corresponding to transient, more or less extended filamentary and/or disk-like structures in the central region.}
    \label{fig:res_covering_frac}
\end{figure*}

\section{Results}\label{sec:results}

\subsection{Overview of the multiphase gas properties}

\paragraph*{Cold gas reservoir.}

Cold and warm gas at the center of the BCG are present throughout nearly the entire simulation, although their relative abundances and morphologies evolve with time. At most epochs, a persistent cold gas reservoir of $\sim10^{8}$–$10^{10}\, \mathrm{\mathrm{M}_\odot}$ is maintained within the core central 50 ckpc. 

Figure~\ref{fig:res_covering_frac} (left panel) shows the cold gas mass fraction in the super-resolved region\footnote{{We restrict here the analysis to a 350 ckpc aperture to minimize potential effects associated with the boundary of the super-resolved region.}} (blue) and the core (light red). The fraction of cold gas in the super-resolved region stays low and relatively constant around 1\%, similarly to its TNG-Cluster equivalent, despite the resolution difference. 
To explain this small difference, we also consider the cumulative cloud mass distribution, and find that the most massive clouds ($> 10^{6.8}\ \mathrm{M}_\odot$) already contain the bulk of the cold gas mass ($\sim 90\%$), indicating that ever more numerous low-mass clouds and increasing numerical resolution contribute negligibly to the total cold gas mass. The invariance of the total cold gas mass to resolution may also reflect a balance of numerical effects at the poorly resolved interface layers of cool gas clouds \citep{2026Lachlan}.

This cold reservoir survives down to $z \sim 0.05$, when a satellite galaxy sweeps out most of the cold gas around the central SMBH. The cold gas fraction within the central 50 ckpc of the BCG is generally higher, indicating that cold gas is more concentrated in the core, and shows variations with redshift between 1-10 \%, which correspond to central inflows and outflows of cold gas and, occasionally, the presence of a disky structure in the center. {For the TNG-Cluster equivalent, the cold gas fraction within the central 50 ckpc follows similar overall trends (not shown in Fig.~\ref{fig:res_covering_frac} for clarity); however, the sharp dips seen in the high-resolution run (light red curve) are mirrored by periods of $<$0.1\% cold gas in the TNG-Cluster case.}

To characterize the spatial distribution of the cold gas, Figure~\ref{fig:res_covering_frac} shows its covering fraction (right panel), quantifying the projected area occupied by the cold phase. {The covering fraction is computed from projected maps and defined as the fraction of pixels containing cold gas below the adopted temperature threshold. To reduce projection effects, we compute this quantity along three orthogonal projections and report the average value.} The covering fraction of the cold gas is systematically higher in our super-resolved cluster than in its TNG-Cluster counterpart, by $3-10$\%, likely due to resolution. Indeed, at lower resolution, small-scale structures become visibly smoothed out, as shown qualitatively in Figure~\ref{fig:meth_refinement_img}, and fine filamentary features extending several tens of kpc are largely absent.

\paragraph*{Morphologies.}

\begin{figure*}
    \includegraphics[width=\linewidth, trim={0cm 4.5cm 0cm 2.5cm},clip]{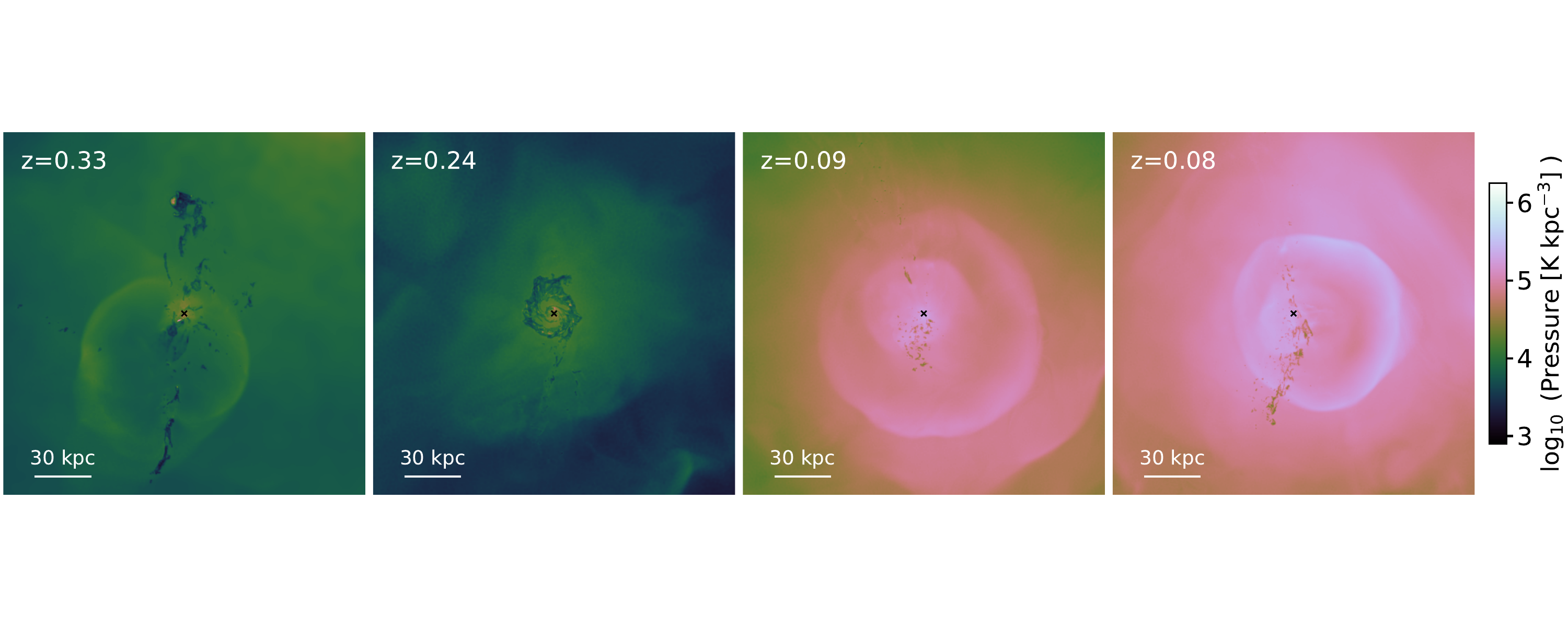}    
    \includegraphics[width=\linewidth,trim={0cm 4.5cm 0cm 4cm},clip]{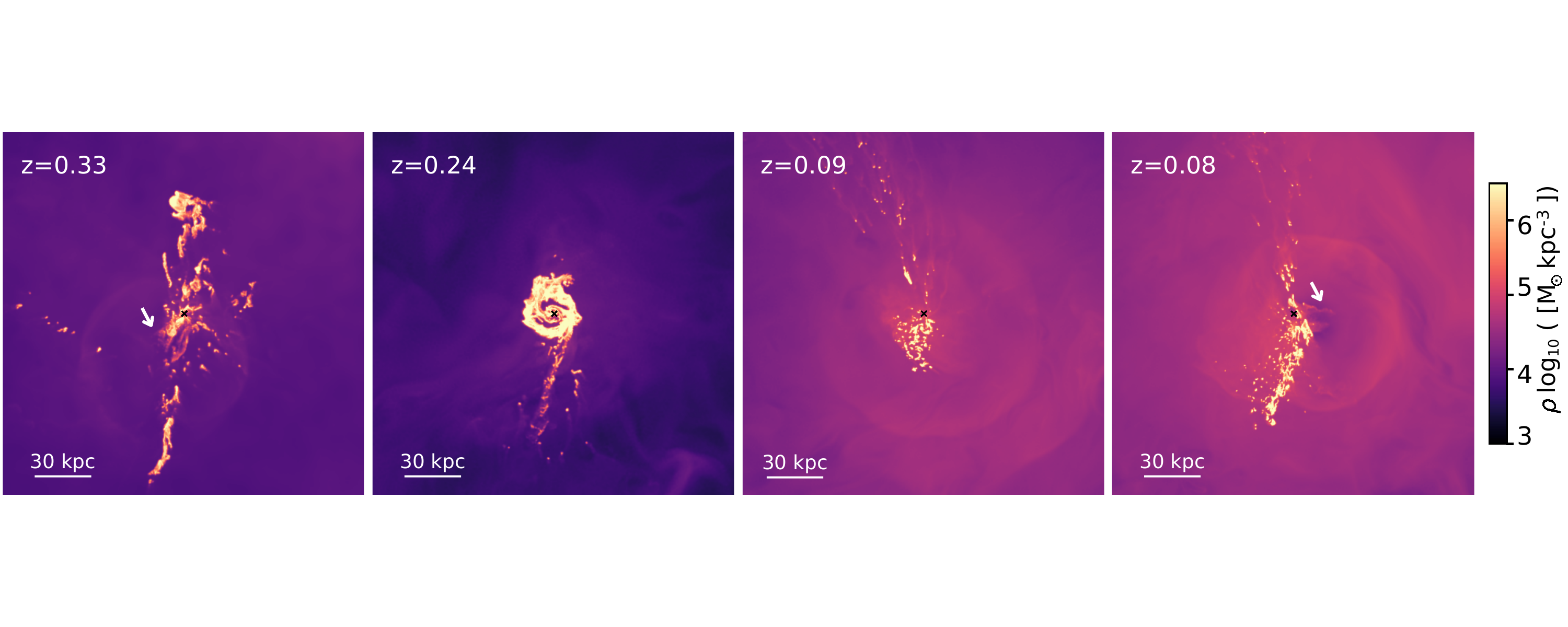} 
    \includegraphics[width=1.015\linewidth,trim={0.2cm 0cm 0cm 0.0cm},clip]{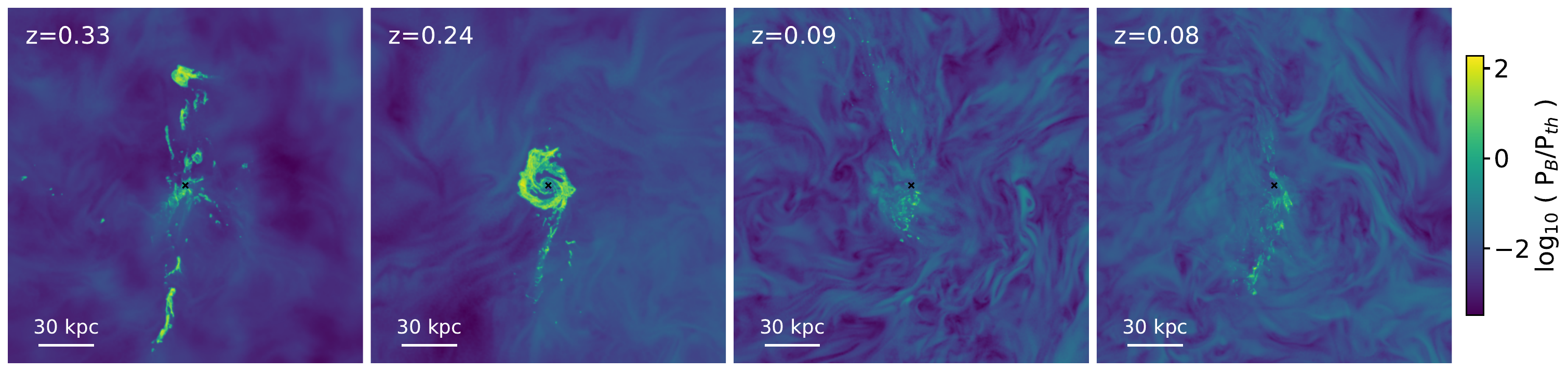} 
    \includegraphics[width=1.013\linewidth,trim={0.05cm 4.0cm 0cm 4.5cm},clip]{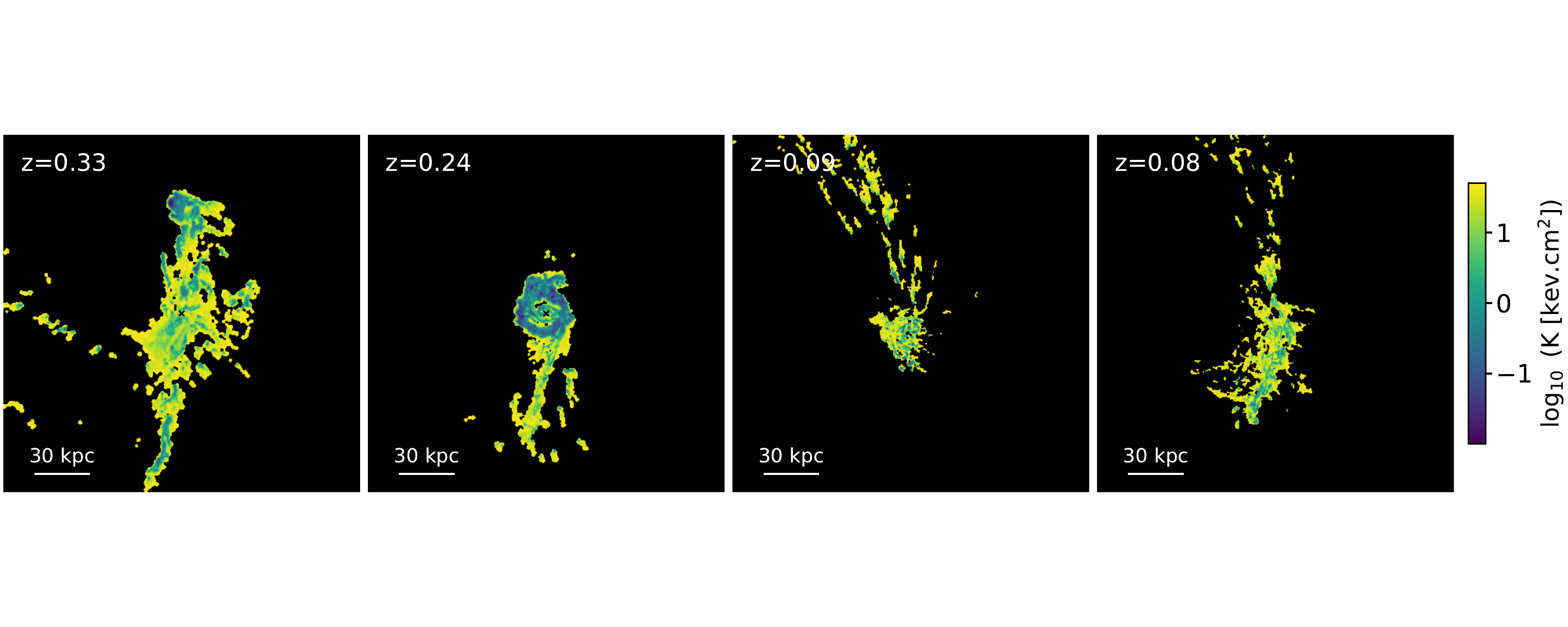} 
    \includegraphics[width=1.031\linewidth,trim={0.17cm 4.5cm 0cm 4.6cm},clip] {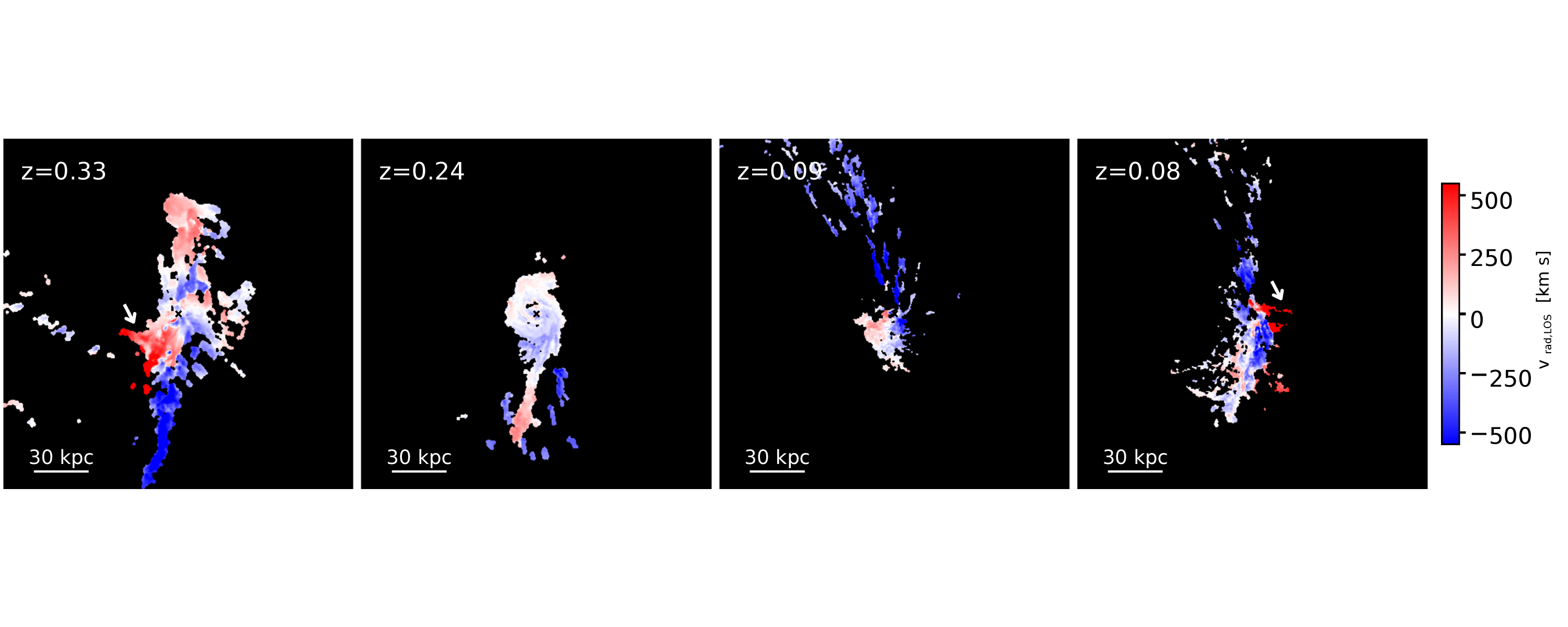}
    \caption{The complex configurations and kinematics of the multiphase gas in h963. We show 200 by 200 kpc maps (+/- 20 kpc depth) at four different redshifts, with rows displaying gas pressure, density, {inverse plasma beta (magnetic pressure over thermal gas pressure)}, entropy of the low-entropy gas \citep[$K < 50~\mathrm{keV\ cm^2}$;][]{2008Cavagnolo,2009Cavagnolo}, and mass-weighted gas radial velocity of the low-entropy gas (with respect to the central SMBH, velocity color scale is capped to better highlight the inflowing and outflowing gas). The black crosses denote the position of the central most massive SMBH. {The white arrows at $z=0.33$ and $z=0.08$ point to high-velocity outflowing gas in the path of AGN feedback outbursts.} The cool, low-entropy, and higher-density gas can take a variety of configurations, from galactic disks, extended filaments, and clumpy structures to small cloudlets. These interact with AGN feedback-driven outflows, cavities, and shock fronts, visible in the density and pressure maps. Filaments are mainly inflowing, with radial velocities spanning $ -1000\ \text{to}\ 600~\mathrm{km\ s^{-1}}$.}
    \label{fig:res_panel}
\end{figure*}

While the cold gas mass fraction changes little across redshift, its morphology and spatial distribution in the cluster core vary significantly. Below, we qualitatively describe the main gaseous structures that emerge as a function of time.

In Figure~\ref{fig:res_panel} we show four representative simulation snapshots at redshifts $z=0.33, 0.24, 0.09,$ and $0.08$ (columns, from left to right). From top to bottom, the rows show: gas thermal pressure ($=n_e kT$), the gas column density, the plasma beta, the entropy of the low-entropy gas ($K = k_{\mathrm{B}} T n_e^{-2/3} < 50~\mathrm{keV cm^2}$), and the mass-weighted line-of-sight radial velocity. {Specifically, gas coordinates and velocities are rotated into the galaxy angular-momentum frame and measured relative to the position and velocity of the BCG. The radial velocity of each gas cell is $ v_{\rm rad}=\vec{v}\cdot\hat{r}$, where $\vec{v}$ is relative to the BCG motion and $\hat{r}$ is the radial unit vector with origin at the central SMBH of the BCG. The maps show the mass-weighted mean within a slice of thickness $\pm 20$ kpc around the projection plane.}

These four epochs sample different portions of the AGN feedback cycle. Columns 1, 3, and 4 correspond to phases shortly, or a few tens of Myr, after a powerful AGN kinetic feedback outburst, whereas the second column shows a comparatively quiescent phase, occurring more than a hundred Myr after the last major outburst.

\begin{figure*}
    \includegraphics[width=0.266\paperwidth]{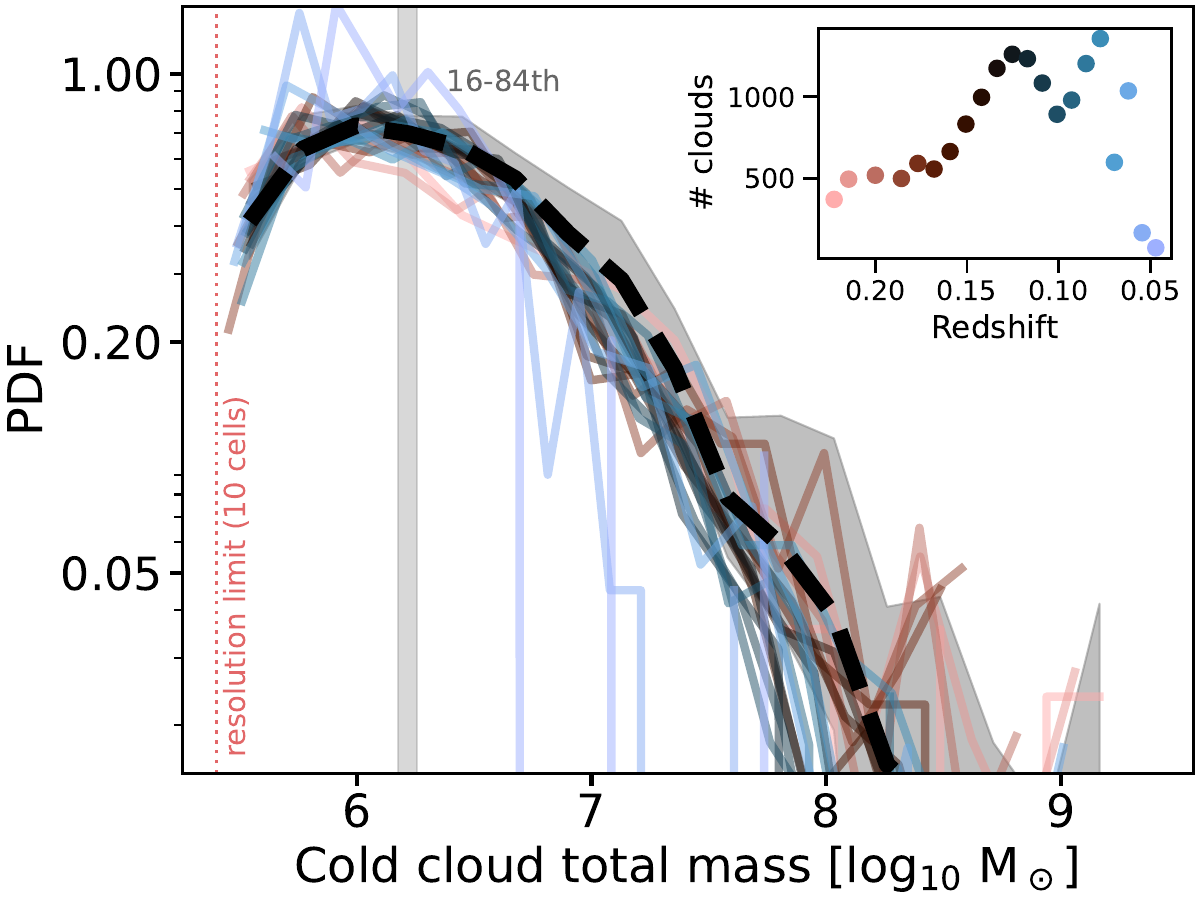}
    \includegraphics[width=0.266\paperwidth]{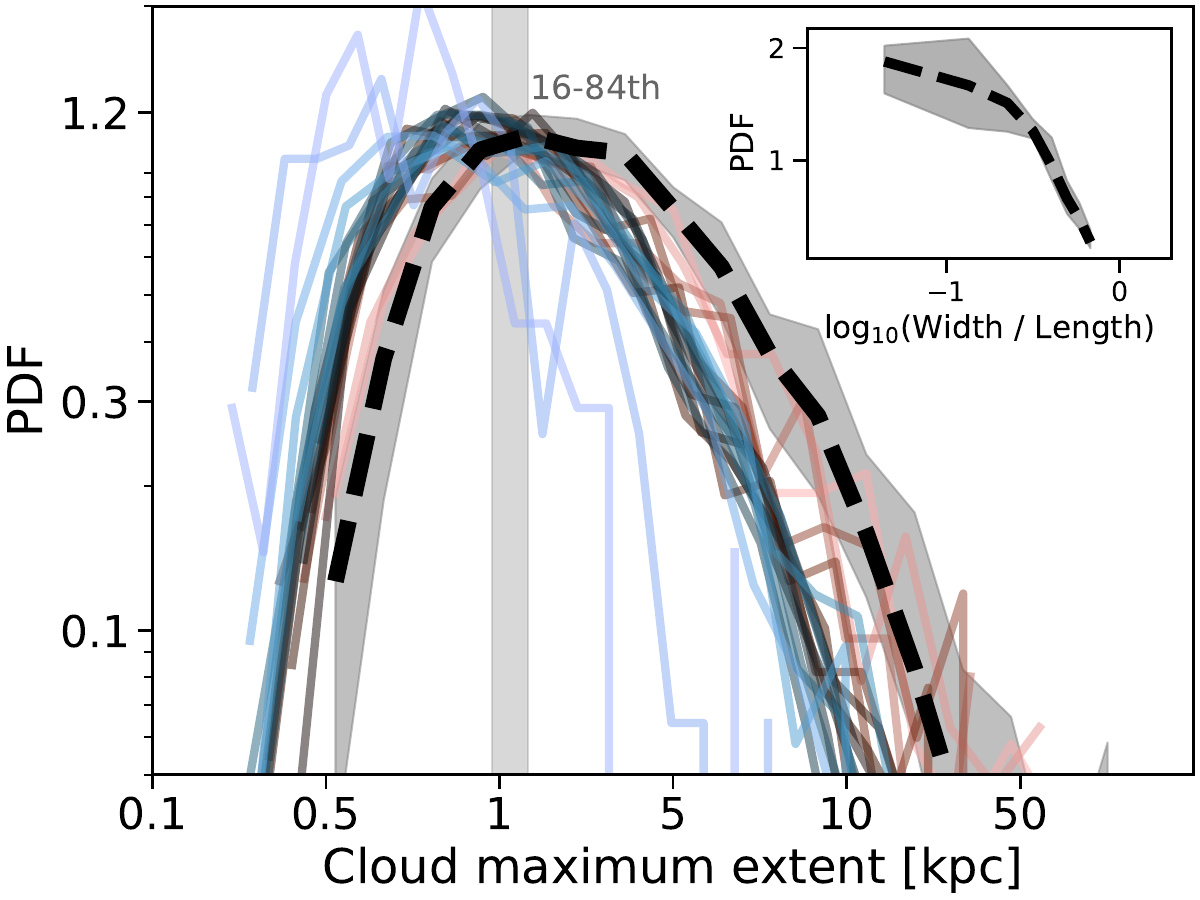}
    \includegraphics[width=0.312\paperwidth]{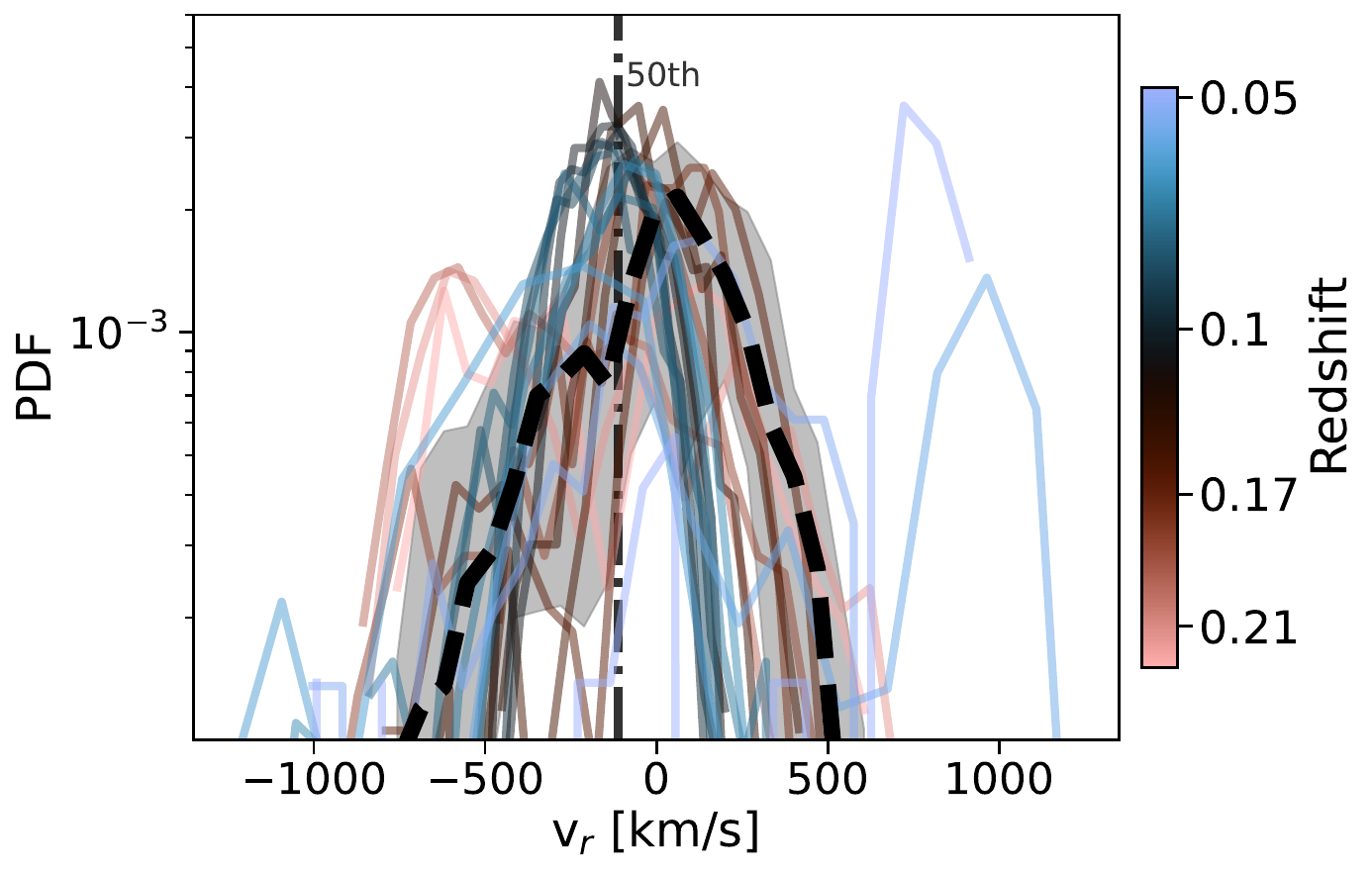}
    \caption{Properties of cold gas clouds. Probability distribution functions (PDFs) of clouds within a 250 kpc aperture around the central SMBH as a function of (left) mass, (middle) characteristic size, and (right) mass-weighted radial velocity, color-coded by redshift. {The dashed black curve shows the median PDF across redshifts, with the shaded grey region indicating the 16th--84th percentile range. Insets show the evolution of the number of clouds with redshift (left) and the cloud width-to-length ratio distribution (middle). }The overall PDF shapes remain broadly stable over redshifts, although after $z = 0.05$, a massive satellite disrupts the cold gas, and no clouds are present until $z=0$. {At this resolution and given our cloud definition, their }masses span $10^{5.3-9}\ M_{\odot}$, with sizes ranging $\sim 0.2$--$40\ \mathrm{kpc}$. Filamentary structures extend up to a few tens of kpc. Clouds are typically inflowing but also exist in outflows, with mass-weighted radial velocities sometimes $>$ 500 km/s.}
    \label{fig:results_cloud_stats}
\end{figure*}

The pressure and density maps reveal AGN-inflated cavities and over-pressurized shells associated with weak shocks \citep[see][for a characterization of X-ray cavities and shocks in TNG-Cluster]{Prunier2025a,Prunier2025b,2025cPrunier}. These structures shape the distribution and kinematics of multiphase gas, contributing to its diverse morphologies. In the vicinity of, or trailing behind, these cavities and shock fronts, the gas density maps reveal overdense small clouds and extended filamentary structures. 

{These dense filamentary structures are magnetically dominated (third row; $\beta^{-1} = P_{\rm B}/P_{\rm th}$, where $P_{\rm B} = B^2/(8\pi)$ and $P
_\mathrm{th} = n k T$), with magnetic-to-thermal pressure ratios exceeding unity. The surrounding halo gas, with typical field strengths $B \sim 0.3$--$0.6\,\mu$G as expected in clusters \citep[e.g.,][]{2008DuboisTessier, 2026Lehle}, remains thermally dominated ($\beta^{-1} \ll 1$). It also exhibits significant structure, including regions of tangential or radially oriented magnetic field configurations. The clouds reach $B \sim 2$--$30\,\mu$G (5-95th percentile), i.e., up to two orders of magnitude larger than the background, suggesting that a significant fraction of their support arises from magnetic field compression and amplification during formation \citep{2020Nelson_tng50}. In the simple case of isotropic compression of a flux-frozen magnetic field, $B\propto\rho^{2/3}$ and $\beta$ therefore scales as $\beta\propto T \rho^{-1/3}$. This provides a simple expectation for magnetic-field amplification as gas cools and becomes denser. In our simulations, the transition from the diffuse hot phase ($T\sim10^7\ \mathrm{K}$, $\rho\sim10^4\  M_\odot \ \mathrm{kpc}^{-3}$) to dense cold gas ($T\sim10^4\ \mathrm{K}$, $\rho\sim10^6\ M_\odot\ \mathrm{kpc}^{-3}$) is accompanied by a two- to three-order-of-magnitude increase in $\beta^{-1}$.}

Dense filamentary structures are also low-entropy  $<50~\mathrm{keV cm^2}$ (Fig.~\ref{fig:res_panel}, fourth row), and display significant morphological diversity, extending from the very core to cluster-centric distances of several hundred kpc. In several instances, we see compact nuclear, rotationally supported disks confined to the central few kpc (e.g., at $z=0.24$ in Fig.~\ref{fig:res_panel}). Beyond disk-like configurations, most times exhibit clumpy or filamentary structures with little or no clear rotational symmetry. Entropy gradients are present: gas located in boundary layers, where filaments or disks interact with the surrounding hot atmosphere, tends to exhibit higher entropy than the gas in the interior of filaments, which remains more effectively shielded from mixing and heating. Figure~\ref{fig:meth_refinement_img} shows how cool clouds have temperature gradients, with the innermost regions the coldest (blue colors), while the outer layers are relatively warmer (light red colors). The clouds are embedded in a warm interface layer that lies between the cloud and the hot background \citep[e.g.][]{2022Dutta}.

These low-entropy gaseous structures also display strong internal kinematic variations (bottom row), with large diversity in radial velocity directions and magnitudes, between filaments and also within a single filament (e.g., $z = 0.33$). Mostly infalling, some show outflowing wings of low-entropy gas in the central regions with high positive radial velocities in the direction of the AGN-induced pressure wave (two examples are shown at $z = 0.33$ and $z = 0.08$ {(columns 1 and 4, pointed with white arrows)}.

\paragraph*{Cloud characteristics.}

To quantitatively characterize the small-scale cold structures, we identify and analyze discrete cold gas clouds. Figure~\ref{fig:results_cloud_stats} shows the cloud probability distributions as a function of mass, size, and radial velocity, as well as their evolution with redshift. 

{Given the target gas mass resolution and our cloud definition ($>$10 cold contiguous Voronoi cells), the minimum cloud mass is intrinsically $2.3 \times 10^{5}\,\rm{M}_\odot$.} Typical cloud masses lie in the range $10^{5.3}$–$10^{9}\ \mathrm{\mathrm{M}_\odot}$ with most clouds around $10^{6.3}\, \mathrm{\mathrm{M}_\odot}$. The low-mass end of the distribution should be interpreted with caution: the combination of the target gas mass and the adopted cloud definition imposes a lower limit of $\sim 10^{5} \,\mathrm{\mathrm{M}_\odot}$, and the lowest-mass clouds are not well resolved. The turn-over visible at $\sim 10^6 \,\rm{M}_\odot$ represents the resolution limit of the simulation \citep[see][]{2026Biba}. 

We define cloud sizes using a principal component analysis (PCA) of the spatial distribution of cells belonging to each cloud. The PCA computes the covariance matrix of the cloud cell positions and extracts the eigenvectors and eigenvalues, with the first principal axis corresponding to the direction of maximum spatial extent. As shown in the middle panel of Fig.~\ref{fig:results_cloud_stats}, {clouds span a broad range of spatial extents, from a minimum scale of $\sim$ 0.2 kpc, set by our cloud definition and the simulation resolution, up to $\sim$ 30–40 kpc}, reflecting the morphology of these structures, from cloudlets up to disky structures or long filaments of cold gas. {In the inset, we show the cumulative distribution of the width-to-length ratio of cold clouds across redshift, providing a measure of their filamentarity. The distribution is by construction below 1 and rises steeply toward low ratios ($< 10^{-1}$). This corresponds to typical length-to-width ratios of order $\sim 10$, implying that most clouds are significantly filamentary rather than compact, isotropic clumps.}

The overall shape of the cloud mass and size distributions does not vary significantly with redshift, although the total number of cold clouds (inset) fluctuates between redshifts, ranging from $\sim320$ to $\sim1900$. In addition, the SMBH-centric distances of the clouds span the full extent of the super-resolved region, with the bulk of the cloud population found between 120 and 150 kpc (16-84th percentile; not shown). We find no strong correlation ($\mid$pearson index$\mid$ < 0.3) between cloud properties and their distance from the central AGN, with only weak positive (size) and negative (mass) trends. Clouds spanning a wide range of masses and sizes are found across a broad range of radial distances within the cluster core, and under diverse background density and temperature conditions.

Fig.~\ref{fig:results_cloud_stats} also shows the radial velocities of clouds (right panel). The shape of this velocity distribution varies with time. In some cases, the radial velocity distribution is double-peaked (with distinct inflowing and outflowing components), while at most times the distributions are single-peaked and centered at negative v$_\text{rad}$, i.e., predominantly inflowing, or around zero. Across most redshifts, a high-velocity tail extends to positive values of up to $500~\mathrm{km\ s^{-1}}$, while the bulk of the cloud population remains infalling. 

At $z = 0.058$, a merger disrupts the halo, and the number of clouds begins to decrease. At the same time, their radial velocity distribution (in the rightmost panel of Fig.~\ref{fig:results_cloud_stats}, light blue curve) shows a double-peaked shape with $\geq 800 $km/s outflowing gas tail, reflecting the large-scale motions induced by the merger. Subsequently, almost no cold clouds are present until $z = 0$.

\subsection{AGN feedback in interaction with the multiphase gas}

\subsubsection{{Cumulative effect of AGN feedback:} Variation run with AGN kinetic feedback turned off.} \label{sec:results:parag:var_run}

\begin{figure}
\centering\includegraphics[width=0.49\textwidth]{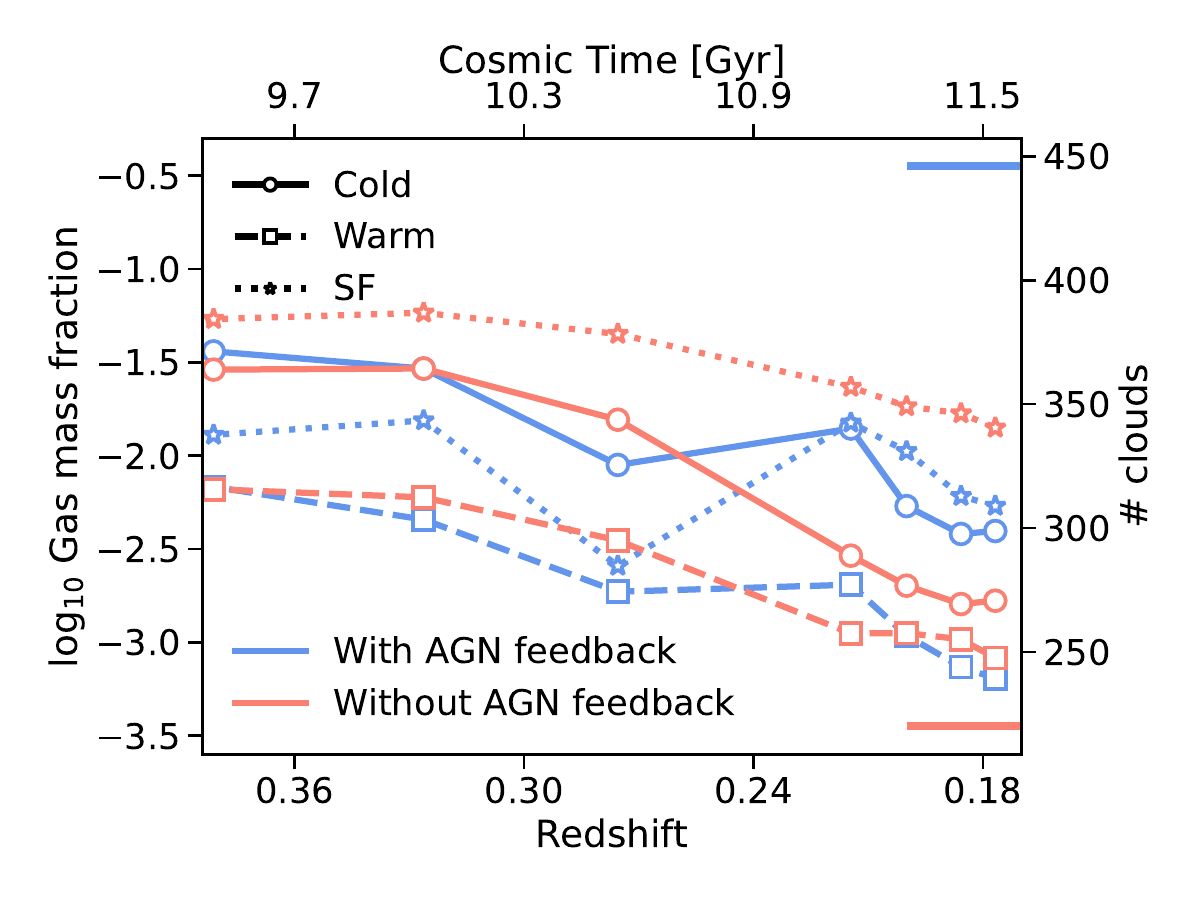}
\vspace{-0.8cm}
\caption{{Gas phases with AGN kinetic feedback on or off. We show the gas mass fractions within 200~ckpc of the BCG center for cold, warm, and star-forming gas as a function of time, for the runs with AGN kinetic feedback on (blue) and off (light red).  We focus on this central region to highlight the impact of AGN feedback. The left $y$-axis shows the median number of cold clouds over the redshift range considered. With AGN feedback active, the number of cold clouds increases from $\sim200$ to $\sim450$ despite similar cold- and warm-gas mass fractions, while the feedback-off run develops a larger reservoir of star-forming gas, indicating more efficient conversion of cold gas into stars.}} 
\label{fig:glob_quant_with_without}
\end{figure}

To isolate the global impact of AGN feedback, we perform a variation run in which the AGN kinetic feedback is switched off after $z = 0.6$. We let the simulation evolve until $z = 0.17$ (running for $\sim 3.7$~Gyr after $z = 0.6$, see Table \ref{tab:tab_sim_run}), allowing sufficient time to assess the consequences \footnote{{The dynamical timescale in the refinement region at $z=0.6$ is $\sim0.6$~Gyr, so the variation run extends to more than six dynamical times.}}.

\paragraph*{Gas mass fractions}
{In Figure~\ref{fig:glob_quant_with_without}, we compare the cold, warm, and star-forming gas mass fractions between the two runs. Between the two runs, the total cold gas reservoir at $z=0.17$ is almost unchanged (M$_{\rm cold} \sim 9\times$ 10$^{10}$ M$_\odot$), only $0.1$ dex higher when feedback is disabled. Over the interval from $z=0.6$ to $0.17$, the total cold-gas mass differs by less than $0.2$ dex between the two runs, while the warm-gas mass fraction remains nearly unchanged with time (Fig.~\ref{fig:glob_quant_with_without}, circle and square markers). Conversely, the feedback-off run develops a substantially larger reservoir of star-forming gas, by $\sim0.5$--$1$ dex over time, indicating more efficient conversion of cold gas into stars in the absence of AGN feedback (Fig.~\ref{fig:glob_quant_with_without}, star markers).}

\begin{figure*}
    \hspace{1.3cm}
    \includegraphics[width=0.8\linewidth, trim={0cm 4.5cm 0cm 2.5cm},clip]{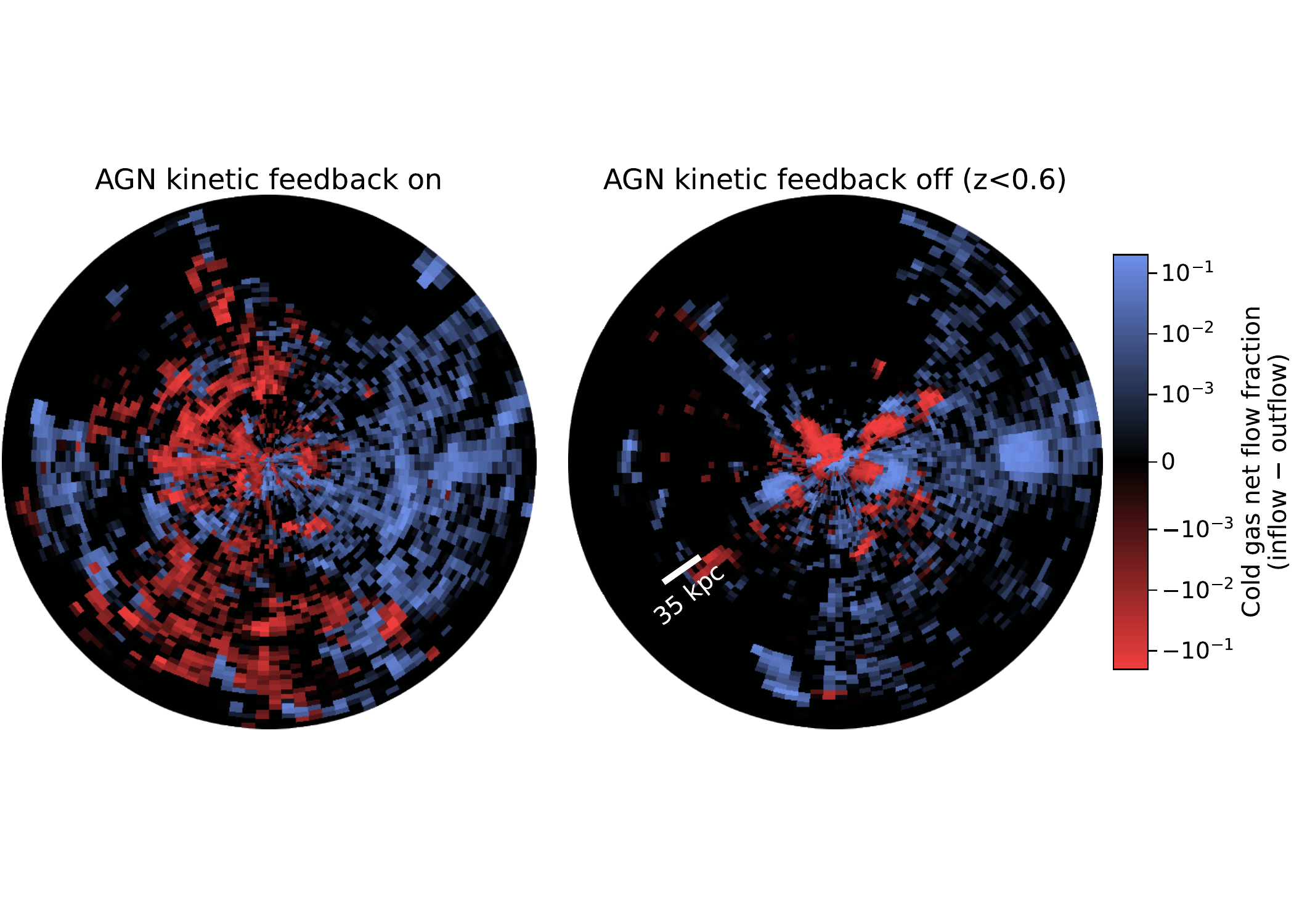} \caption{Spatial distribution of inflowing versus outflowing cold gas. Polar distribution of the cold gas within 200 ckpc of the central SMBH in the fiducial run (left) and after the kinetic feedback mode has been disabled (right) (for the gas bound to the BCG, i.e., satellite excised). {Each pixel shows the cumulative cold gas mass fraction carried by inflowing gas minus that carried by outflowing gas, over 1.8 Gyr; positive (blue) values indicate net inflow, negative (red) values indicate net outflow.} The color scale is symmetric-log (linear near zero, logarithmic $> 10^{-3}$ and capped at $\pm 0.12$). With kinetic feedback turned off, the cold gas is predominantly inflowing and forms a central disk, whereas the feedback-on run shows a redistribution of the cold gas and net outward radial transport.}
    \label{fig:res_nobh}
\end{figure*}

\begin{figure*}
    \centering
    \includegraphics[width=0.81\linewidth]{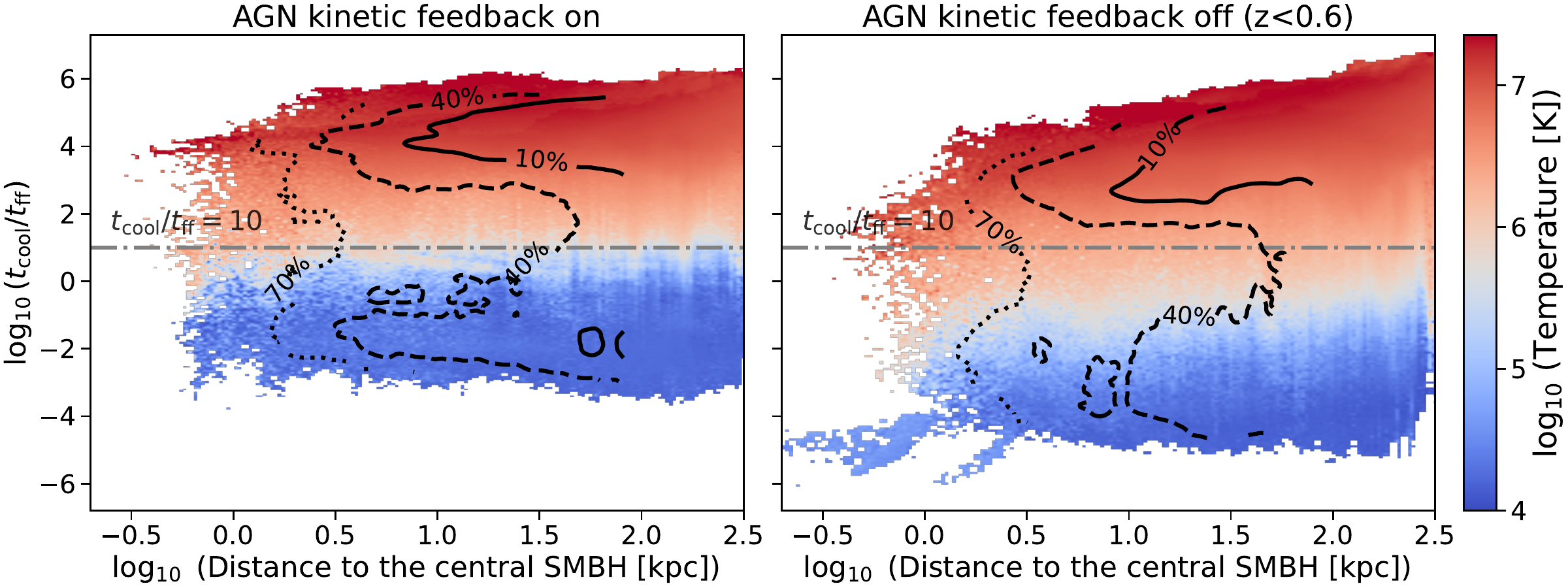}
    \includegraphics[width=0.82\linewidth]{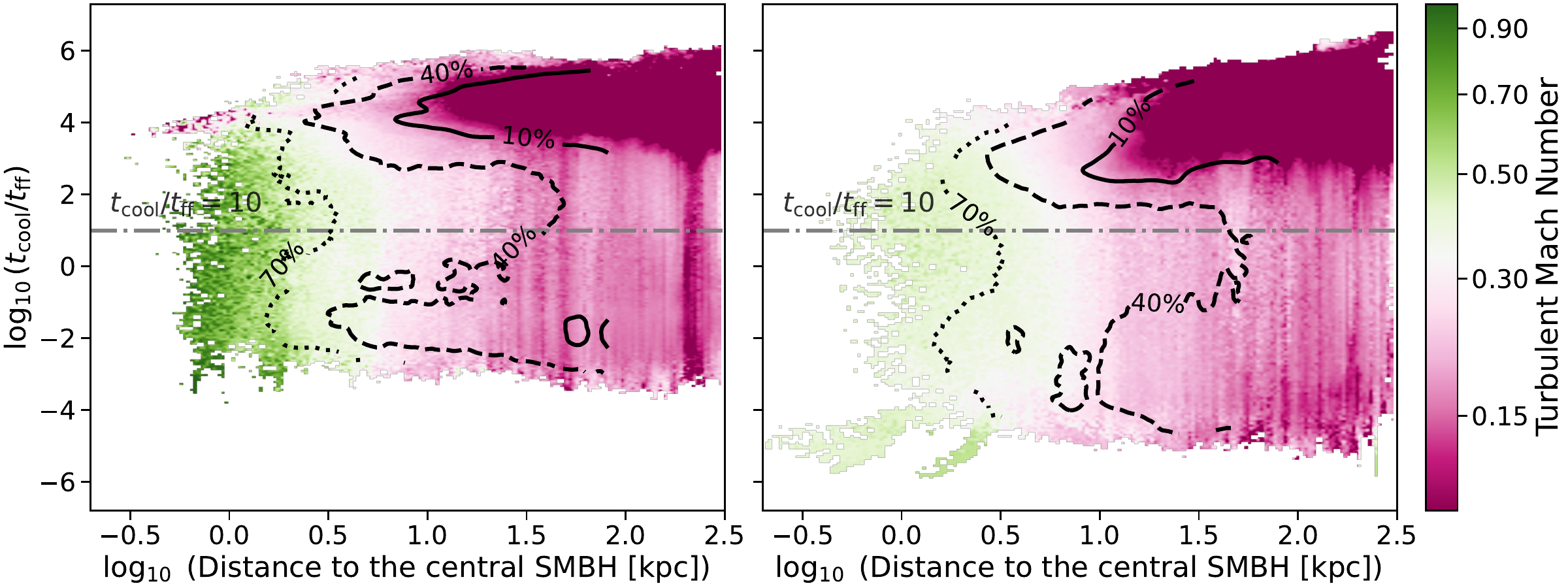}    
    \caption{{Cumulative effect of AGN feedback on the cooling properties of the gas.} We show the distribution of gas in the plane of $t_{\rm cool}/t_{\rm ff}$ as a function of distance from the SMBH, integrated over several Gyr, color-coded with mean temperature \textbf{(top)} and \textbf{(bottom)} {turbulent Mach number (i.e., defined as the gas cell velocity dispersion over the sound speed of the hot gas within a fixed sphere of 5 ckpc, colorbar limited to 5th-95th percentile).} The dashed grey line marks the canonical instability threshold at $t_{\rm cool}/t_{\rm ff}=10$. Contours indicate the regions containing fractions of total gas mass relative to the maximum (between 0.3 and 100 kpc): solid = 10\%, dashed = 40\%, dotted = 70\%. With feedback active, the cooling time in the central region is systematically increased, shifting a large fraction of the warm gas phase above the instability threshold. {In addition, the inner 10 kpc region reaches turbulent Mach numbers $>0.5$, reflecting the cumulative kinematic impact of recurrent feedback outbursts.}}
    \label{fig:res_phase_space}
\end{figure*}

While the gas fractions stay similar, a significant difference lies in the spatial distribution of cold gas within the ICM. This is immediately visible in its spatial distribution in the halo, as illustrated in the polar plots of Fig.~\ref{fig:res_nobh}. {We show the cumulative net radial transport of cold gas over several snapshots spanning 1.8~Gyr, with the color indicating the direction of transport: blue denotes net inflow of cold gas, while red denotes net outflow. When feedback is turned off (right panel Fig.~\ref{fig:res_nobh}), the cold gas remains centrally concentrated and dominated by inflow; in contrast, with feedback active (left), the cold gas becomes more uniformly distributed and outflowing over a broad range of radii, leading to larger covering and volume-filling fractions.} Indeed, in the absence of AGN feedback, after $1$~Gyr, a dense, disk-like structure forms at the center of the BCG. This disc gradually becomes more extended over time, although it is intermittently disrupted by streams of dense, cool gas inflowing towards the center. In contrast, in the run with AGN feedback active over the same time interval, no such disc forms, and the cold gas is more spread out.

\paragraph*{Clouds properties}
{Despite similar cold- and warm-gas mass fractions, the number of cold clouds increases from $220$ to $446$ when AGN feedback is active (right y-axis, Fig.~\ref{fig:glob_quant_with_without}, averaged over $z=0.19$-$0.17$)}, possibly due to cloud shattering from successive feedback outbursts. The cloud morphologies are also modestly affected. In the feedback-on run, clouds are on average less filamentary, with width-to-length ratios closer to unity, and slightly less dense. However, the median cloud masses and sizes differ by less than $0.1$ dex between the two runs. These differences may be driven by the cloud environment, which shifts toward higher entropy and temperature when feedback is active, or by cloud deformation and interactions induced by the outburst. Because the total cold gas reservoir varies only weakly between the two runs, the reduced number of clouds in the feedback-off case suggests that more of the cold gas resides in {the central formed disky structure}. Indeed, without feedback, the most massive cloud alone holds ${\sim}36\%$ of the cold gas mass and the ten most massive clouds ${\sim}87\%$, compared with ${\sim}9\%$ and ${\sim}48\%$ in the feedback-on run (averaged over  $z=0.19-0.17$)

\paragraph*{ICM susceptibility to thermal instability}
{The hot gaseous halo is strongly affected by AGN feedback, which regulates its radiative cooling and, consequently, its susceptibility to TI and in-situ condensation.} In the feedback-off run, the cooling-time profile within the inner 50~kpc remains below $1$~Gyr, while the entropy remains below $10~\mathrm{keV\ cm^2}$, consistent with a strong cool-core state. In contrast, when feedback is active, both profiles fluctuate substantially in the core, with the cooling time frequently exceeding $\sim3$~Gyr and the entropy reaching above $\sim50~\mathrm{keV\ cm^2}$.

Figure~\ref{fig:res_phase_space} further shows the global impact of feedback on key gas thermodynamic properties. We show the ratio $t_{\rm cool}/t_{\rm ff}$ as a function of distance from the SMBH, integrated over several snapshots. The free-fall time of a gas cell located at a distance $r$ is defined as $t_{\mathrm{ff}} = \left( \frac{2 r^{3}}{G M_\text{halo}\text{(r)}} \right)^{1/2}$, and $t_{\rm cool} = \frac{3}{2}\frac{(n_e+n_i)k_{\rm B}T}{n_e n_i \Lambda}$ where $n_i$/$n_e$ the ion/electron number density, $T$ the gas temperature, $\Lambda$ the cooling function, and $k_{\rm B}$ the Boltzmann constant. The dashed line at $t_{\rm cool}/t_{\rm ff} = 10$ marks the typical theoretical threshold below which gas is expected to become thermally unstable and condense \citep[][]{2012Sharma}. Cells are color-coded by average temperature (top) and turbulent Mach number (bottom, defined as the velocity dispersion within a 5 kpc sphere relative to the local sound speed of the background ICM, $c_s = \sqrt{\gamma k_B T/(\mu m_{\rm H})}$, with $\gamma=5/3$.). In order to give a sense of the gas mass distribution, we overlay contours indicating regions enclosing fixed fractions of the total gas mass within 0.3–100 kpc from the SMBH. 

In the ICM of both the feedback-on and feedback-off runs, gas exhibits a clear multiphase structure spanning a broad range of temperatures. In both cases, an extended distribution is present: the diffuse hot medium ($T > 10^{6}\,\mathrm{K}$,  $n_{\rm H} < 10^{-2.5}\ \mathrm{cm}^{-3}$) predominantly occupies regions with $t_{\rm cool}/t_{\rm ff} \gtrsim 100$, while cooler gas is concentrated at lower values, down to $t_{\rm cool}/t_{\rm ff} \lesssim 1$ \citep{2020Nelson_tng50}. While the hot phase represents the bulk of the gas mass (traced by top 10\% contours), the 40-70\% contours highlight the presence of denser, cooler phases at all radii. However, the gas with temperatures $T \sim 10^{6}\ \mathrm{K}$ constitutes only a small fraction of the total gas mass at all radii. This suggests that gas does not remain long in this state, but instead rapidly transitions through it, likely condensing out to a cooler phase \citep[e.g.,][]{2017Voit}.

With AGN feedback active, gas at intermediate temperatures ($T \sim 10^{5\text{--}6}\,\mathrm{K}$) is preferentially located near the canonical instability threshold $t_{\rm cool}/t_{\rm ff} \sim 10$ {(see also Discussion \ref{sec:discussion:cooling_heating})}. In contrast, when feedback is off, a significant fraction of both warm and hot gas falls below this threshold, suggesting that this gas should be more prone to rapid cooling towards $T \sim 10^{4}\ \mathrm{K}$. 

Globally, the presence of feedback increases the cooling time of the central gas, shifting a larger fraction of material above the instability threshold. The gas also shows a higher fraction of high-temperature and low-density gas ($T > 10^{7.5}\ \mathrm{K}$, $n_H < 10^{-3.5}\ \mathrm{cm}^{-3}$), reflecting the cumulative heating and ejective effects of recurrent outbursts.

Although star-forming gas cells are not included in Figure~\ref{fig:res_phase_space}, distinct ``wings'' of rapidly cooling gas, close to the SMBH, approaching the star formation threshold in TNG (n$_{\rm H} > 10^{-1}\ \mathrm{cm}^{-3}$), are visible in the feedback-off run, but are absent when feedback is active. This is consistent with the globally more elevated SFR found when feedback is off (Fig.~\ref{fig:glob_quant_with_without}). 

{We also show the turbulent Mach number $\mathcal{M}_{\rm turb}$ of the gas in the bottom panels of Figure ~\ref{fig:res_phase_space}. In both the feedback-on and feedback-off runs, most of the gas exhibits subsonic turbulence, with a low median $\mathcal{M}_{\rm turb}\sim0.1$ in the hot and $t_{\rm cool}/t_{\rm ff}>10^3$ phase. In the feedback-on run (left panel), the central $<10$ kpc region systematically reaches higher turbulent Mach numbers, with $\mathcal{M}_{\rm turb}>0.5$ across all gas phases, from the shortest cooling times to the hot phase. By contrast, the feedback-off run remains at $\mathcal{M}_{\rm turb}<0.4$ throughout this region. This enhancement likely reflects the cumulative impact of repeated kinetic feedback outbursts, which increase the bulk velocities and velocity dispersion of the different gas phases, particularly in the vicinity of the SMBH \citep[][]{2026Saha}. Consistent with this picture,  \citet{2025cPrunier} shows using mock X-ray observations that kinetic AGN feedback drives expanding, overpressurized shells of hot gas that generate weak shocks, with typical Mach numbers of $\sim1$--$1.5$, extending up to $\sim200$ kpc from the AGN.}

The comparison with the variation run highlights that AGN kinetic feedback in the TNG galaxy formation model, in addition to its well-established role in regulating and quenching star formation \citep[e.g.,][]{2018Pillepich,2017Weinberger,2020Zinger}, also plays an important role in shaping the spatial extent and distribution of cold gas. At the same time, AGN feedback modifies the thermodynamic and kinematic state of the ICM, influencing the conditions under which the cold phase can condense without preventing its continued presence.

\begin{figure*}
    \centering
    \includegraphics[width=0.99\linewidth]{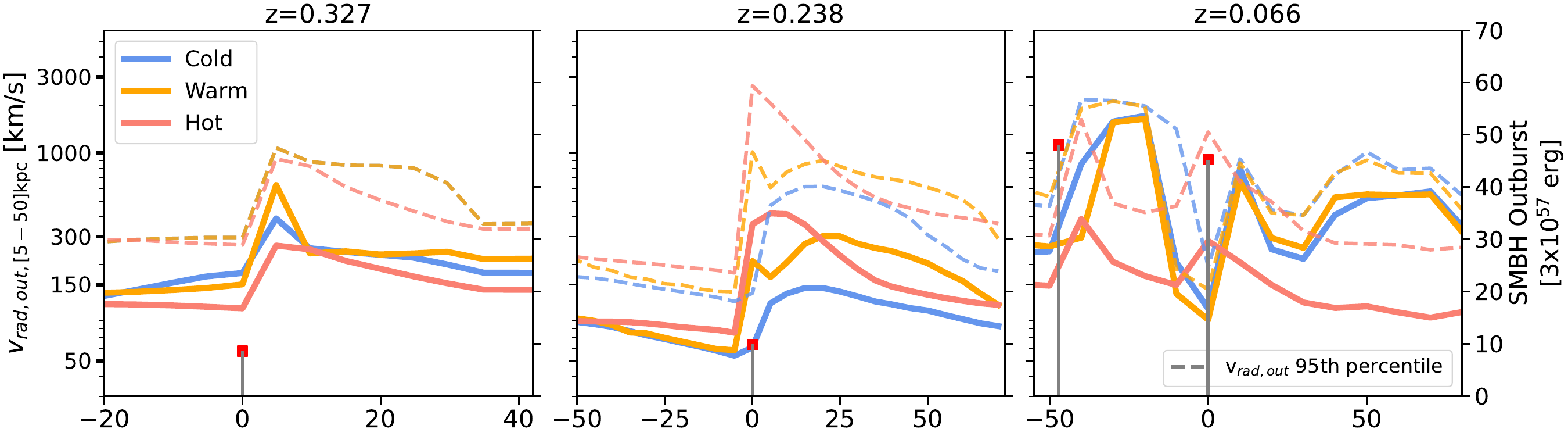}
    \includegraphics[width=0.99\linewidth, trim=0cm 0cm 0cm 0cm, clip ]{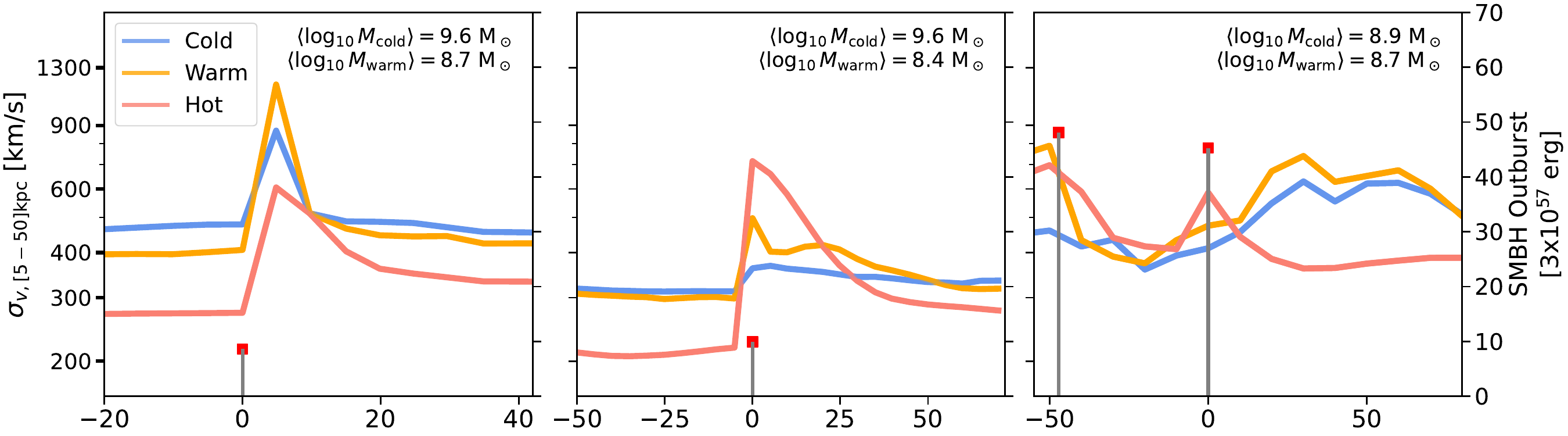}
    \includegraphics[width=0.99\linewidth, trim=0 0cm 0 0cm,
    clip ]{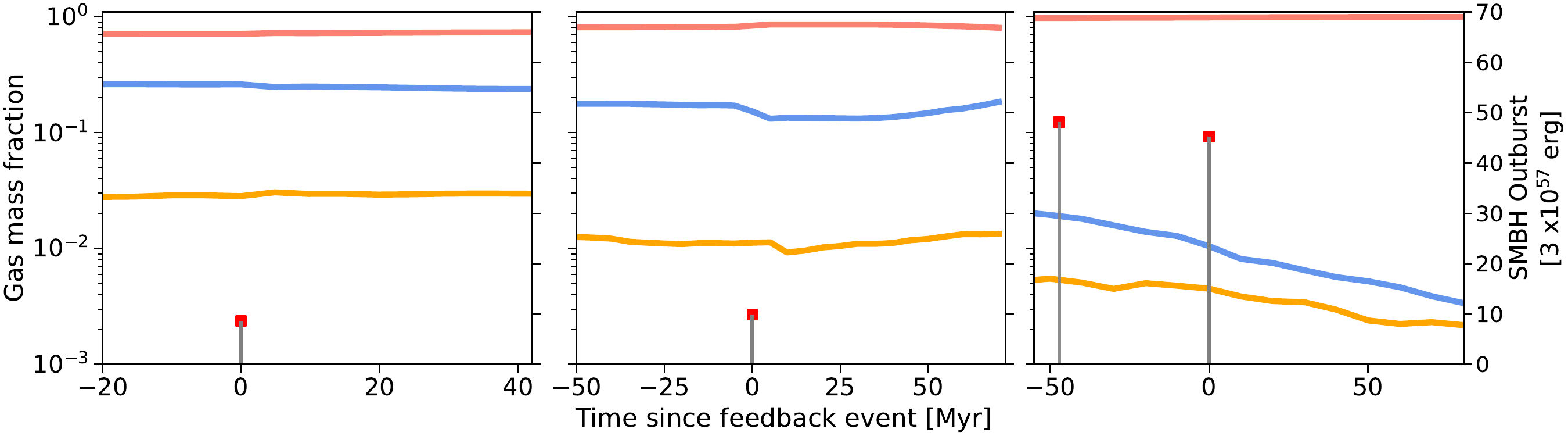}
    \caption{Instantaneous effects of AGN feedback on the gas. Time evolution of gas kinematics {and gas mass fractions}, { within 5–50 kpc shells} centered on the SMBH, around four representative AGN kinetic feedback events for the cold (blue, T $< 10^{4.5} \mathrm{K}$), warm (yellow), and hot (red, T $> 10^{5.5} \mathrm{K}$) phases. The vertical grey line marks the time of energy injection along with the injected energy. We also indicate the cold and warm gas masses before each outburst. \textbf{ Top:} mass-weighted average (solid curves), and 95th percentile (dashed) of the radial velocity of gas with outward motion (i.e., $v_\text{rad} > 0$) \textbf{Middle:} velocity dispersion computed from the 3D velocities of all gas cells within the same aperture. \textbf{Bottom:} {Gas mass fraction of the different gas phases within the aperture.} During the outburst, both the outflowing velocity and velocity dispersion rise rapidly before returning towards the pre-outburst levels within $\sim20-50$ Myr. The 95th percentile of outflowing cold gas reaches velocities $> 800~\mathrm{km\ s^{-1}}$, showing the short-lived but extreme impact of the event. {Despite this strong kinematic response, only a small fraction of the warm and cold gas is heated up.}}
    \label{fig:res_kinematics}
\end{figure*}

\subsubsection{{Instantaneous effect of AGN feedback: }a closer look at individual AGN outbursts.}
\label{sec:res_a_closer_look}

To directly link individual feedback events to the thermodynamic and kinematic response of the surrounding gas, we record each individual AGN kinetic feedback event in our simulations.

Feedback outbursts of the most massive SMBH in the central BCG occur at irregular intervals, ranging from a few Myr to several tens of Myr, with occasional quiescent periods lasting up to $> 300$ Myr without a feedback event. The injected energies span $10^{57}$–$10^{60} \mathrm{erg}$ per feedback event. Periods of elevated AGN activity (duty cycle on the order of a few Myr) are globally associated with increased gas velocity dispersion and outflowing velocities in the core (<70 ckpc) across all gas phases. However, given the relatively coarse temporal sampling of our simulation (with snapshot spacings of order $\sim 100$ Myr), multiple feedback events can occur between consecutive outputs, making it difficult to directly associate specific outbursts with corresponding gas responses. 

We therefore also output high time resolution snapshot sets, spaced by a few Myr. We do so over three time segments that encompass three major AGN feedback events with injected energies of $3\times10^{58}$ erg, occurring respectively at $z=0.24$, $z=0.33$, and two feedback events of $\sim 1\times10^{59}$ erg around $z=0.06$. This enhanced temporal cadence allows us to directly follow the response of the different gas phases to individual AGN outbursts.

Fig.~\ref{fig:res_kinematics} shows the mass-weighted average radial velocity (v$_{\rm rad,out}$) (for outflowing gas cells only) and the velocity dispersion ($\sigma_v$) versus time, and the mass fraction of the gas phases, in the vicinity of the SMBH (i.e., within 30 kpc of the SMBH). The grey vertical lines topped by red squares mark the exact time and magnitude of each feedback event. 

{Before an AGN outburst, the different gas phases exhibit similar v$_{\rm rad,out}$ values ($100$--$200$~km~s$^{-1}$). The warm and cold phases also have similar $\sigma_v$ values ($300$--$500$~km~s$^{-1}$), while the hot phase generally has somewhat lower values ($200$--$300$~km~s$^{-1}$). This similarity may indicate a degree of kinematic coupling between the different gas phases.} Following an outburst, both v$_{\rm rad,out}$ and $\sigma_v$ increase sharply across all gas phases, within less than $5$~Myr of the event, with both increasing by up to a factor of $\sim3$ and reaching several hundred km~s$^{-1}$. The response of the different phases, however, varies between outbursts, with the cold and warm phases sometimes responding more weakly than the hot phase. This differential response may be related to the spatial distribution and physical properties of the gas. The cooler phases are typically organized into dense, compact clouds that may be initially stationary, infalling, or rotating. Their response therefore depends strongly on the relative location and orientation of the clouds with respect to the outflow: clouds directly in the path of the outflow can be accelerated more efficiently, whereas those outside the outflow cone are less affected. In addition, their low volume filling factor and small effective cross-section, combined with their high densities, make them more difficult to accelerate, resulting in less efficient coupling of kinetic energy from the outflow to the cold gas \citep[e.g.,][]{2008Pope,2016McNamara}.

For example, at $z=0.238$, the feedback outburst occurs after a coherent central disk structure has formed, with $M_{\rm cold} \sim 10^{9}~\mathrm{M}_\odot$ within the central 30 kpc. In contrast, the third segment ($z=0.066$) occurs in the absence of a coherent disk, with the central region instead populated by a lower-mass collection of cool gas clouds extending up to $\sim 50$ kpc from the center. The individual clouds appear to respond more strongly to the outflow than the more massive, coherent disk, and are accelerated to higher outward velocities.

For most feedback events,  v$_{\rm rad,out}$ and $\sigma_v$ decline back to pre-outburst levels within roughly $20-50$ Myr for the cold, warm, and hot phases. This rapid decline suggests that, for the TNG kinetic-mode model, AGN-driven outflows and turbulence in the core are short-lived and require frequent duty cycles to be sustained. Similarly, our analysis of the TNG300 subbox containing a $\sim10^{15} \mathrm{M}_\odot$ cluster with 10 Myr temporal resolution shows that during periods of kinetic feedback injection by the central SMBH, the velocity dispersion of the hot gas systematically exhibits a sharp enhancement (by a factor of 2–10), following a rapid decline \citep[see also][]{2026Saha}.

{Despite the high velocities and increased turbulence associated with individual outbursts, the gas mass budget of each gas phase remains relatively stable (bottom panel, Fig.~\ref{fig:res_kinematics}). There is no immediate shredding of the cold gas, and the rapidly outflowing shocked gas is only mildly heated, with less than 1\% of the cold and warm phases being converted into the hot phase immediately after the outburst.}

\paragraph*{Study case: low-entropy gas during an outflow.}

\begin{figure*}
    \includegraphics[width=0.6\linewidth]{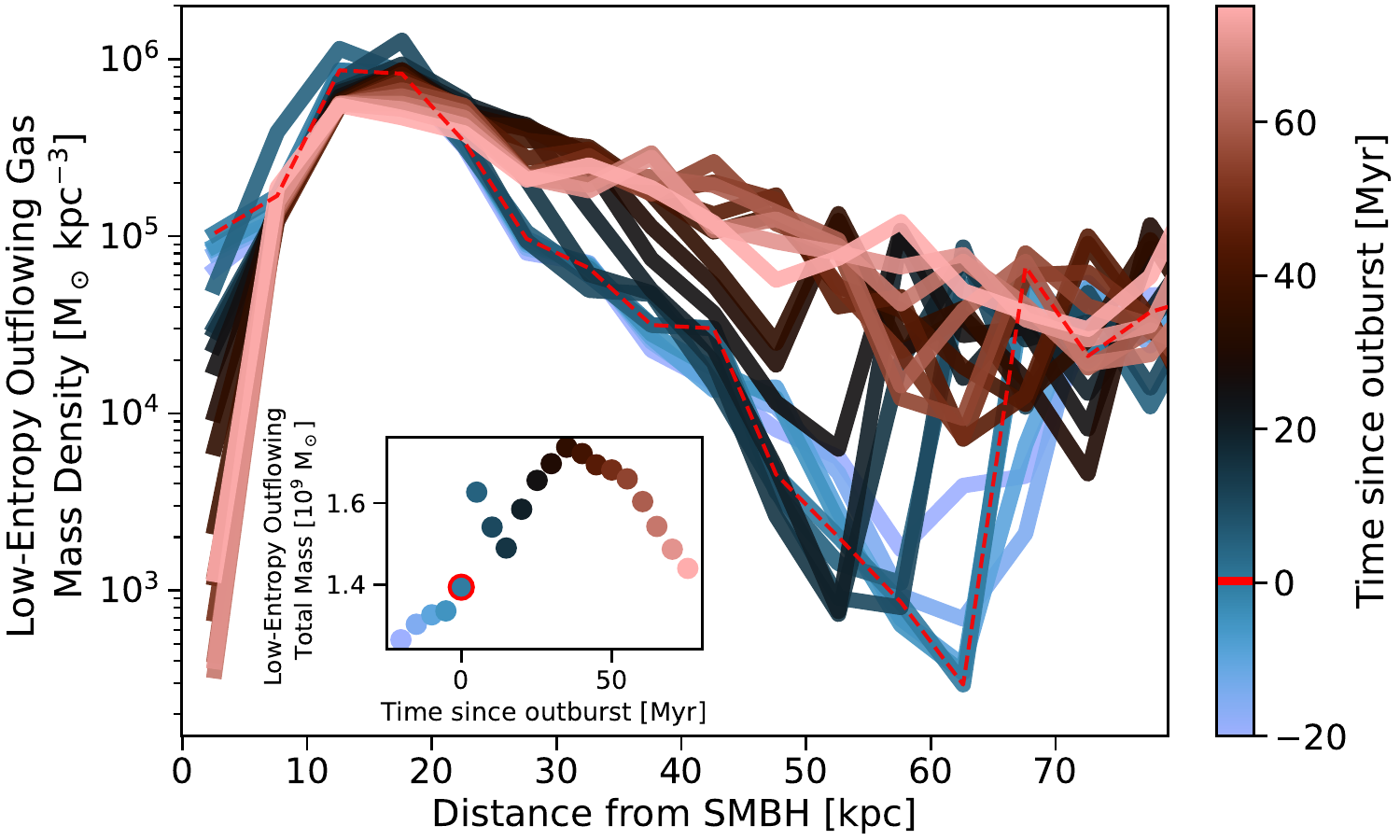}
    \includegraphics[width=1.0\textwidth]{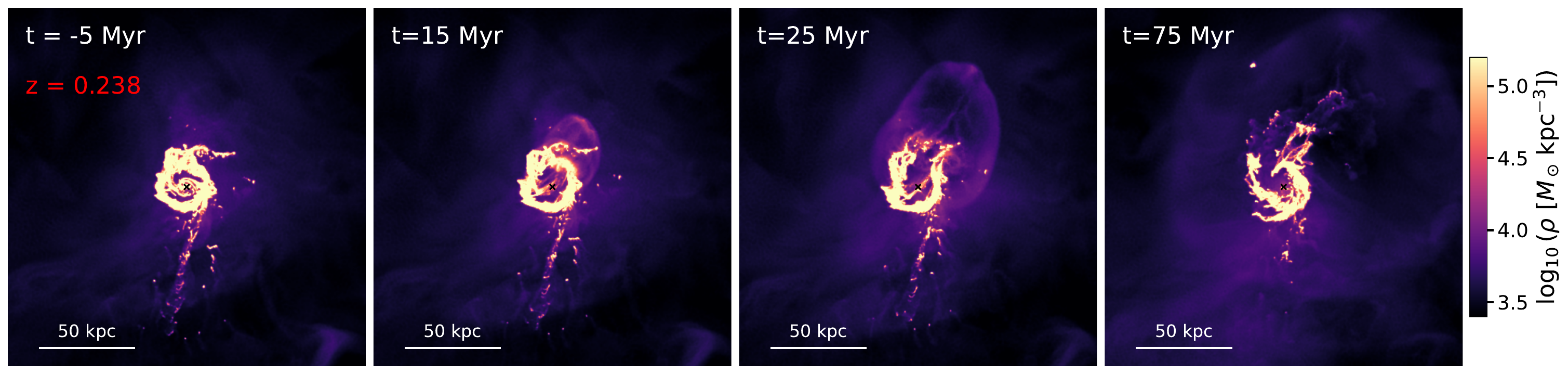}
    \includegraphics[width=1.013\textwidth]{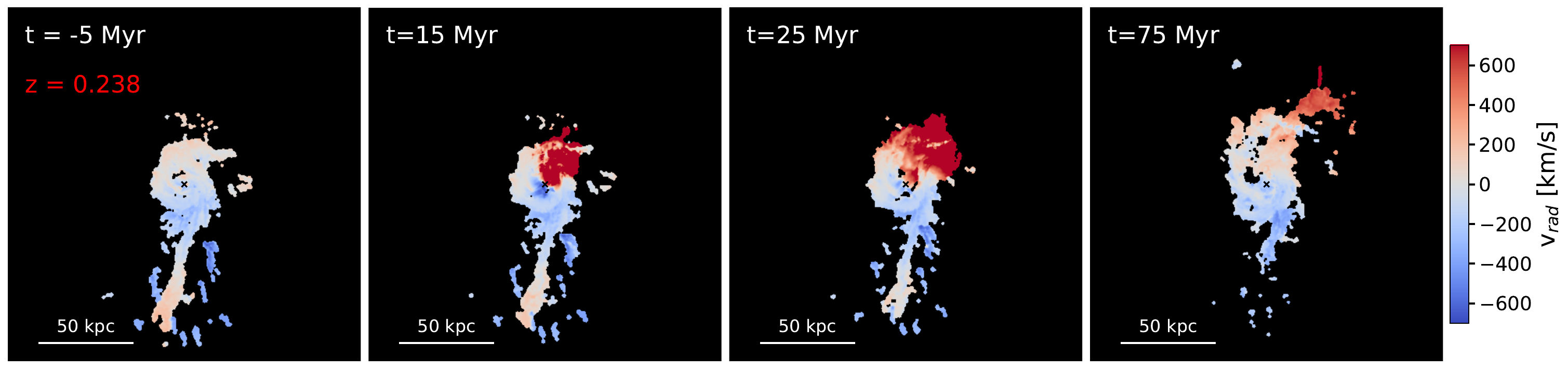}
    \caption{Uplift of low-entropy gas. \textbf{Top panel:} Mass density profiles of the outflowing low-entropy gas between $-20$ Myr and $70$ Myr relative to the feedback event (red dashed line). The inset shows the corresponding total mass of low-entropy gas within 80 kpc across time. Following the outburst, there is an increase in the mass density of outflowing low-entropy gas between $20$ and $70$ kpc, as well as a global increase in its total mass, consistent with this gas being uplifted or pushed by the outburst. \textbf{Bottom images:} Gas density and radial velocity maps (170 $\times$ 170 kpc) of the low-entropy ($<50~\mathrm{keV\ cm^2}$) gas at representative chosen times before and after an SMBH directional kinetic feedback event at $z=0.238$ (E = $3.8 \times 10^{58}$ erg). The directional kinetic energy injection produces an underdense cavity with an upward-moving shock front that partially disrupts the dense disk-like structure and entrains trailing dense gas. }
    \label{fig:res_entropy_upflift}
\end{figure*}

In Figure~\ref{fig:res_entropy_upflift}, we focus on one feedback event and show the gas density (top image row), and the radial velocity of the low-entropy gas (bottom image row), from 5 Myr before to 75 Myr after the outburst.\footnote{Although we show only a single outburst event in detail, we have examined the same quantities for two other high time resolution segments ($z = 0.33$ and $z = 0.066$) and observe similar trends. } The previous feedback event occurred $> 100$ Myr ago, and the disk was then largely undisturbed for an extended period of time prior.

Before the AGN outburst, the gas exhibits a dense, rotationally supported disk. At the time of the outburst, the SMBH kicks the gas with a directional kinetic energy injection of $3.8 \times 10^{58}$ erg. As the shock propagates through the medium, its pressurized shock front (although with a weak Mach number < 1.8 after $t = 5$ Myr) partially destroys the disky structure by $t = 15$ Myr. In the radial velocity maps, this directional outflow is visible in the low-entropy gas phase moving outward at velocities exceeding $500\ \mathrm{km\ s^{-1}}$ at $t = 15$ Myr. By $t = 25$ Myr, we see the shock front progressing outward, along with the partial destruction of the outer disk and the uplift of some dense gas. Behind the shock front, the ICM becomes visibly less dense, forming a cavity, while clumps of high-density, low-entropy gas flow outward at $300$–$500\ \mathrm{km\ s^{-1}}$, filling the cavity ($t = 25-80$ Myr). This low-entropy gas may be pushed by the AGN wind and/or uplifted and/or grow in the trailing region of the cavities/shock front, or both.

Over the following tens of Myr, the disk gradually reforms as the velocities of the outflowing low-entropy gas decrease. Simultaneously, a stream of infalling low-entropy gas develops on the side opposite to the feedback outburst direction, originating from wake of a satellite at $\gtrsim 100$ kpc during its flyby (outside of our field of view). This infalling gas, as well as the gas opposite to the feedback outburst direction, is largely unaffected by the violent outburst.

Figure~\ref{fig:res_entropy_upflift} also shows the impact of AGN feedback on the radial redistribution of low-entropy gas volume density (top). To isolate the outflowing gas, the profiles are constructed using only gas with positive radial velocity in each shell, before and during the feedback outburst depicted in the maps. In the inset, we also show the total low-entropy outflowing gas mass within this 80 kpc region over time (the integral of each profile). During the AGN outburst, we observe both a decrease in the outflowing central low-entropy gas and its gradual increase at larger radii. Initially, the low-entropy gas is concentrated in the central disk, with densities of order $10^{5-6}\ \mathrm{\mathrm{M}_\odot\ kpc^{-3}}$ within the inner $20$ kpc, and dropping to $< 10^{4}\ \mathrm{\mathrm{M}_\odot\ kpc^{-3}}$ at larger radii. Immediately after the outburst (red dotted line), the low-entropy gas mass within 10 kpc decreases sharply to $\leq10^{3}$. At larger radii, from $\sim 20$ to 70 kpc, the mass density of low-entropy, outflowing gas increases from $10^{3}$ to $10^{5}\ \mathrm{\mathrm{M}_\odot\ kpc^{-3}}$ over 70 Myr. The inset of Fig.~\ref{fig:res_entropy_upflift} further shows that the total amount of outflowing low-entropy gas peaks $\sim 45$ Myr after the feedback event and then gradually decreases as this denser gas falls back.

These results support a scenario in which AGN feedback both heats and ejects gas from the very central region ($< 10$ kpc), while uplifting and transporting a portion of low-entropy material from the central galaxy into the surrounding ICM.

\section{Discussion}\label{sec:discussion}

In this work, we simulate and analyse a low-mass galaxy cluster (h963) with a super-resolved multiphase ICM within a central 370 ckpc region around the BCG, focusing on both the cumulative and instantaneous effects of AGN kinetic feedback on the multiphase gas at $z<0.35$. To better isolate the effect of AGN feedback, we also perform a variation run in which AGN feedback is off for 3.7 Gyr (see Table~\ref{tab:tab_sim_run} for a summary). In the following discussion, we combine these results to understand the impact of feedback on the cold gas (Section \ref{sec:discussion:cooling_heating}). In particular, we seek to understand why AGN feedback has little apparent impact on the total cold gas mass, despite clear differences in its spatial distribution, star formation rate, cooling times, and thermodynamic and kinematic state. We then explore the contribution of satellites to the cold gas budget  (Section \ref{sec:discussion:satt}) and compare the simulated cool gas structures with observed optical nebulae (Section \ref{sec:discussion:comp_obs}).

\subsection{Cooling versus heating due to AGN feedback.} \label{sec:discussion:cooling_heating}

\begin{figure}
    \includegraphics[width=0.48\textwidth]{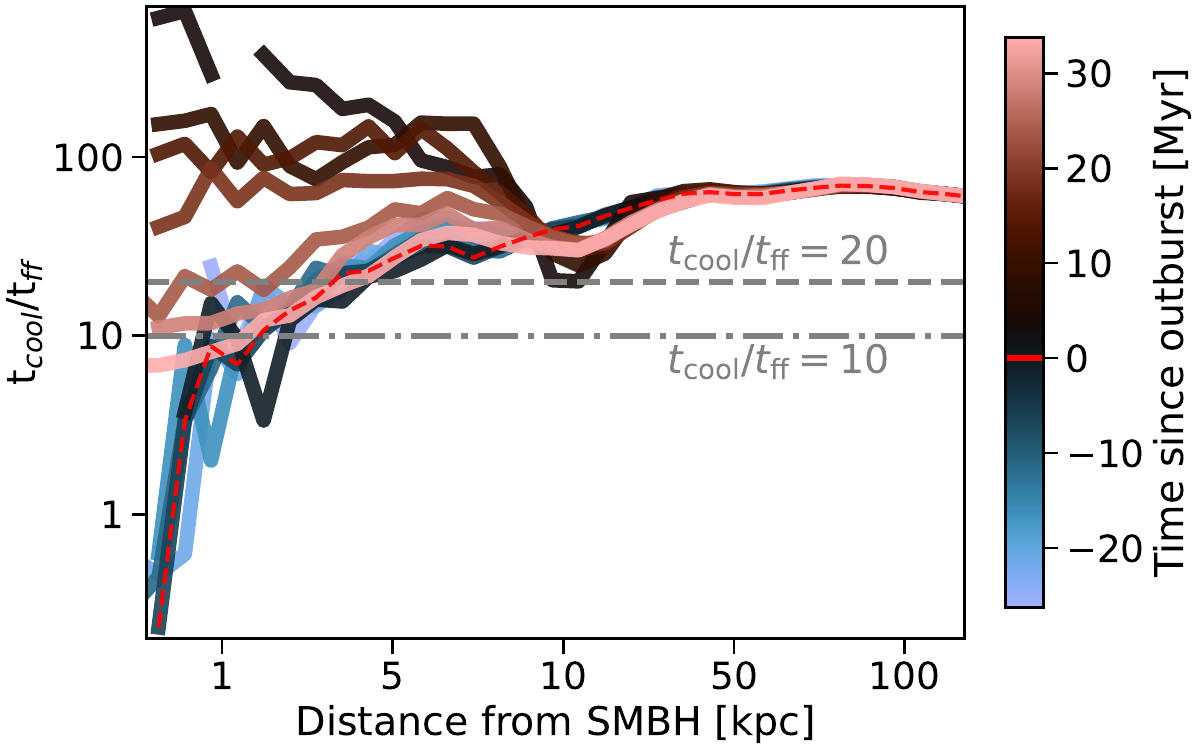}
   \caption{Radial profiles of $t_\mathrm{cool}/t_\mathrm{ff}$ measured 30 Myr before and after an AGN feedback energy injection ($\sim10^{59}$ erg). The canonical threshold for it is shown in grey. Only the hot phase is included to assess its susceptibility to thermal instability. The energy injection significantly perturbs the thermodynamic state of the inner region for $\sim$15–20 Myr, after which it relaxes back toward its pre-outburst state.}
    \label{fig:disc_tcool_tff_profile}
\end{figure}

SMBHs play a central role in quenching star formation in central galaxies both by ejecting gas and by regulating the cooling times of the gaseous halo reservoirs. In the variation run with AGN kinetic feedback turned off for several Gyr (Section~\ref{sec:results:parag:var_run}), a disky, star-forming structure forms, whereas the BCG remains quenched in the feedback-on case. The thermal and kinematic properties of the cold gas are likewise significantly altered (Fig.~\ref{fig:res_phase_space}), with AGN feedback increasing the $t_\mathrm{cool}/t_\mathrm{ff}$ ratio and turbulent Mach number of the gas. Fig.~\ref{fig:disc_tcool_tff_profile} further shows how one AGN feedback event can raise $t_{\rm cool}/t_{\rm ff}$ in the very core, suppressing thermal instabilities (TI) for a few tens of Myr following a feedback event. This increase counteracts the in-situ precipitation of hot halo gas through TI, whereby cold gas condenses out of the hot background medium when the local cooling time becomes sufficiently short compared to the dynamical time \citep[][]{2012Sharma,2012McCourt,2018Prasad,2017Voit}.

Furthermore, previous studies have shown that, over cosmic time, AGN feedback generally reduces the mass of cold gas in simulated clusters \citep{2019Nelson,2020Zinger,2022Nobels,2025Rohr_cool}. In particular, \citet{2025Rohr_cool} show that, in TNG-Cluster, the onset of kinetic feedback in BCGs {(between $z \sim 3-5$)} is followed by a reduction in the global growth of the cool ICM within $\sim1$--$2$ Gyr, measured between $0.15R_{200c}$ and $R_{200c}$ across the 352 clusters. Our analysis, however, focuses on a narrower and later evolutionary period, from $z=0.35$ to $z=0$. Over this interval, we do not find a significant decrease in the cold gas mass fraction. This does not necessarily contradict previous studies. Rather, our results likely reflect the different evolutionary stage probed here: by $z=0.35$, the initial reduction in the growth of the cool gas reservoir following the onset of kinetic feedback has already taken place, leaving a relatively stable cold gas mass fraction over the later stages of cluster evolution.

Among the 352 TNG-Cluster BCGs, the largest central cool-gas reservoirs at low redshift are found in strong cool-core systems (Fig.~\ref{fig:meth_bcgs_tngcluster}, within the central 50 kpc core region). These systems also host the most rapidly accreting SMBHs and exhibit the most vigorous kinetic feedback activity, together with high levels of turbulence and numerous X-ray cavities \citep{2024Lehle,Prunier2025a,2026Saha}. This co-occurrence \citep[also observed in multiple observational studies, e.g.,][]{2025Li_VSF,2019Olivares_mol,2022Olivares_MUSE} points to a close connection between AGN feedback and the formation and maintenance of cold gas nebulae. In h963 specifically, we do not find a significant difference in the total cold gas mass fraction between the runs with AGN kinetic feedback on or off \footnote {We have checked that the increase in star formation (SF) when the AGN feedback is off is several ($\sim$ 10) orders of magnitude less energetic than AGN feedback events; therefore, SF alone can not compensate for AGN outbursts in the energy budget.}(Figure \ref{fig:glob_quant_with_without}), although the feedback-on run hosts nearly twice as many clouds. {This suggests that AGN feedback mainly fragments and redistributes the existing cold gas. Previous studies have also suggested that AGN feedback can promote local thermal instability, leading to the condensation of new cold gas \citep[e.g.,][]{2013Gaspari,2017Voit}. The interplay between global ICM heating and local condensation could therefore contribute to regulating the cold gas reservoir, potentially explaining its similar mass in the feedback-on and feedback-off runs. Without gas tracers, however, it is difficult to establish the origin of the cold gas. In the following, we explore the physical processes through which AGN feedback could promote local gas cooling in our simulation.} 

\paragraph*{Uplift}

In our high-cadence segments of the simulation where we follow individual feedback events, {patches of low-entropy gas are uplifted to altitudes of > 10~kpc, where the ambient gas has higher entropy and $t_\mathrm{cool}/t_\mathrm{ff}$ exceeds the canonical threshold for thermal instability (Fig.~\ref{fig:disc_tcool_tff_profile}).} Specifically, in Fig.~\ref{fig:res_entropy_upflift}, following an outburst in the direction of the disk, low-entropy gas is displaced from the central region to heights of $\sim50$ kpc. The uplifted gas remains low-entropy, indicating that the cavity primarily transports the gas rather than substantially heating it. Consistent with this picture, we find only a modest decrease in the cold- and warm-gas mass fractions during the uplift phase (Fig.~\ref{fig:res_kinematics}, middle row, $z=0.238$). However, $25$--$50$~Myr later, the cold-warm gas mass fractions return to approximately their initial values, suggesting that condensation may occur following the uplift. 

We do an order-of-magnitude estimate for the example shown in Figure \ref{fig:res_entropy_upflift}, assuming that the cavity is empty and displaces gas with the density of the ambient ICM at the cavity radius. The displaced mass can be expressed as
\begin{math}
M_{\rm hot,displaced} \simeq 1.2\ n(r) m_{\rm p} V,
\end{math} where $n (r)$ is the electron density at the cavity centre, $m_{\rm p}$ is the proton mass, and $V$ is the cavity volume (assuming a simple prolate ellipsoid). 50 Myr after the outburst, this gives $M_{\rm hot,displaced} \sim 8\times10^8 M_\odot$, comparable to the mass of the outflowing cold gas $M_{\rm cold, out}\sim 2.5\times10^8 M_\odot$ (inset of Figure \ref{fig:res_entropy_upflift}, mass of outflowing gas $50$ Myr after the outburst relative to that immediately before the outburst). This is within the values quoted in \citet{tamhane2026cavitydrivengastransportcoolcore}, where molecular gas masses scale approximately linearly with the gas mass displaced by cavities across about twenty BCGs, albeit with substantial scatter.

{Overall, this supports a picture in which AGN-driven cavities can transport a substantial amount of low-entropy gas to $>$ 10 kpc radii, where it remains denser and cooler than its surroundings, making it more prone to condensation. }
 
\begin{figure}
    \hspace{-0.4cm}
    \includegraphics[width=1.12\linewidth, trim=6cm 0.2cm 7cm 0cm, clip ]{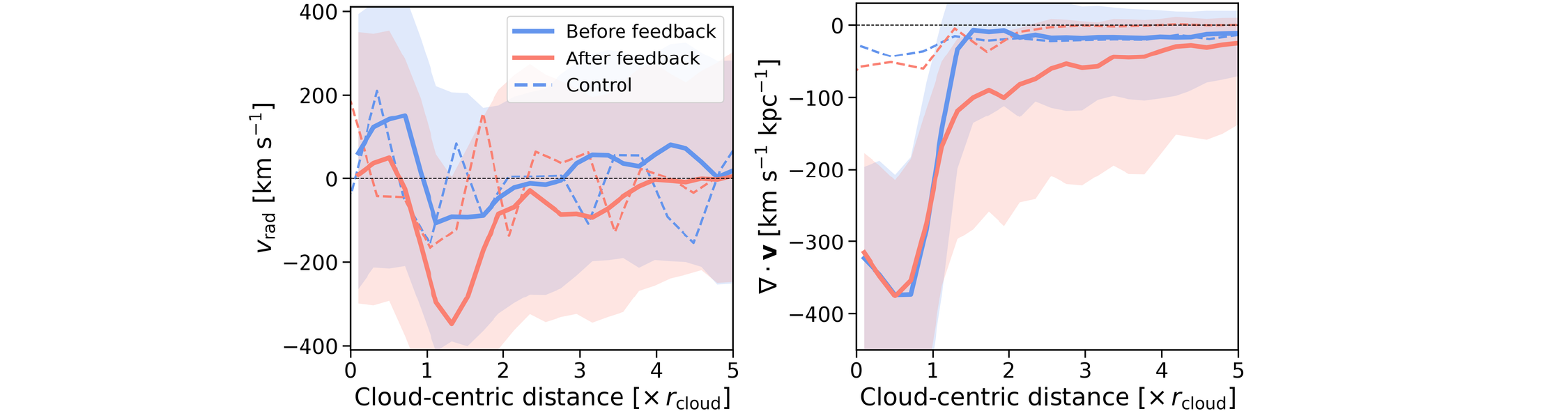} \caption{AGN-enhanced inflows onto cold clouds. Stacked cloud-centered radial profiles showing the transition from cold-cloud to ambient-gas properties within $\pm30$~Myr of a feedback event. Profiles are constructed from $>900$ clouds, with shaded regions showing the $32-68$th percentiles. Dashed lines show time-matched spherical control regions without cold clouds. \textit{Left}: Radial velocity relative to each cloud's center-of-mass motion. \textit{Right}: Velocity divergence. Negative values indicate inflows. Cold clouds exhibit local inflows within $\sim1$--$3\ r_{\rm cloud}$, which strengthen following feedback events and may enhance cloud accretion and growth.}
    \label{fig:disc_inflow_cloud_centric}
\end{figure}

\paragraph*{Mixing} {Mixing between uplifted low-entropy gas and the ambient hot ICM can produce intermediate-temperature gas with a shorter cooling time, potentially promoting further condensation \citep[e.g.,][]{2010Marinacci}. The feedback-on run hosts nearly twice as many clouds as the feedback-off run despite having a similar total cold gas mass (Fig.~\ref{fig:glob_quant_with_without} and Fig.~\ref{fig:res_nobh}). Moreover, averaged over the three events in Section~\ref{sec:res_a_closer_look}, the number of diffuse clouds (fewer than 10 connected cold cells) is $\sim$37\% higher within 30~Myr after each event than before it. This points to increased fragmentation following AGN outbursts, enhancing the cold--hot interface area and potentially creating additional seeds for cooling.}

\paragraph*{Isobaric compressions and density contrasts}

While the standard criterion for condensation is $t_\mathrm{cool}/t_\mathrm{ff} \lesssim 10\text{--}20$ \citep{2012McCourt}, this threshold can be significantly higher in the presence of large relative density fluctuations $\delta\rho/\bar{\rho} \gtrsim 1$, allowing condensation to occur even when the background $t_\mathrm{cool}/t_\mathrm{ff}$ is high \citep{2015Singh,2019Choudhury}. In particular, AGN-driven shocks and expanding cavities generate compressions in the surrounding hot ICM, locally increasing the density and reducing the cooling time. As shown in the bottom panels of Fig.~\ref{fig:res_phase_space}, the turbulent Mach number is higher in the feedback-on case (left panel), reaching values (>0.5) within the innermost 10 kpc, both above and below the canonical condensation threshold.  Consistent with this picture, \citet{2026Sotira} find that cold-gas condensation is enhanced when the turbulent Mach number of the hot phase reaches values of $\sim$0.3.

In Figure \ref{fig:disc_inflow_cloud_centric} we show signatures of enhanced mass inflows following feedback in cloud-centered radial profiles. The median profiles are obtained by stacking more than 900 clouds over a 30 Myr interval before and after a feedback event. For comparison, we also show corresponding control regions without cold clouds. \cite{2020Nelson_tng50} shows that the density and temperature contrasts between the clouds and their surroundings in the CGM drive local cooling flows toward them, promoting cloud growth through accretion. Indeed, before feedback, the left panel of Figure \ref{fig:disc_inflow_cloud_centric} (blue curve) shows local inflows at the cloud interfaces, i.e., with v$_{\rm rad}<0$ within $\sim1$--$2\ r_{\rm cloud}$. These inflows are further strengthened following feedback (light red curve), with v$_{\rm rad}<-200$ km s$^{-1}$ and negative velocity divergence, $\nabla\cdot\mathbf{v}<0$, between $\sim1$ and $3\ r_{\rm cloud}$, indicating enhanced converging motions at the cloud boundaries.

In many cloud growth models, the mass inflow rate can be expressed as \begin{math}\ \dot{M}_{\rm in} \sim A_{\rm cl} \cdot \rho_{\rm int}\cdot v_{\rm in} \end{math}, where surface area $A_{\rm cl}$, the density of the interface $\rho_{\rm int}$, and the inflow velocity $v_{\rm in}$ \citep{2024Tan}. We find that $A_{\rm cl}$ varies little following feedback. Between $1-2\ r_{\rm cloud}$, the median $v_{\rm in}$ (i.e., -v$_{\rm rad}$ onto the cloud) decreases by a factor of $\sim4$ (left panel Fig.~\ref{fig:disc_inflow_cloud_centric}) and $\rho_{\rm int}$ increases by a factor of $\sim2$ (not shown here) implying a higher inflow rate $\dot{M}_{\rm in}$. 

{These results suggest that AGN-driven turbulence may promote the growth of existing cold clouds by enhancing inflows onto their surfaces. The physical origin of these enhanced inflows, however, remains uncertain. Pressure disturbances driven by AGN outflows may contribute to the increased turbulence and vorticity observed in the central region, which persist for up to 20--30 Myr after a feedback event. Our analysis does not establish a direct causal link between these processes and cloud growth, and a more detailed investigation is left to future work.}

\paragraph*{Final thoughts and limitations}

Together, these results suggest a dual role of AGN feedback. On the one hand, {it prevents widespread in-situ condensation from the hot atmosphere by increasing} $t_\mathrm{cool}/t_\mathrm{ff}$ overall and on average across the cluster volume. On the other hand, it redistributes and uplifts low-entropy gas to higher altitudes, providing additional cool-gas seeds for mixing, and enhances large density fluctuations through shocks and sound waves in the cores. This provides a possible explanation for the coexistence of powerful AGN feedback and large central cold-gas reservoirs observed in cool-core clusters.

However, {it remains unclear whether net new cool gas forms as a result of AGN activity, or whether the changes we see in the cool gas are driven mainly by fragmentation and redistribution of pre-existing cool gas. More generally, it is unclear which processes (e.g., radiative cooling from the hot atmosphere, AGN-influenced cooling, satellites, or large-scale gas accretion) are most relevant for the origin of cool gas in cluster cores.} Quantifying the contribution of these channels would benefit from a tracer-based analysis, where we can follow individual gas parcels backward in time and reconstruct the origin of the cold gas. This is left for future work.

While our analysis focuses on a single halo and should therefore not be overgeneralized, it provides a detailed view of processes shaping the multiphase gas in cluster cores within a full cosmological context. However, cool gas nebulae are found in halos spanning a wide range of masses and assembly histories, all of which influence how cooling, AGN feedback, and environmental processes interact with the multiphase gas. Moreover, the TNG galaxy formation model does not include $< 10^4$\ K gas, nor certain physical processes such as thermal conduction or cosmic rays, and the present simulations lack the resolution to probe scales below $\sim 30$\ pc. These limitations may affect the formation, growth, and survival of cold clouds.

{Finally, the exact quantitative values, such as the number of clouds, may vary between realizations of the same halo (see also \citealt{2019Genel}). However, test runs of h963 with different refinement start times, followed until $z=0.2$ (Sect.~\ref{subsec:meth_slr}), show very similar cold gas mass fractions and cloud mass and size distributions, suggesting that such variations are limited.}

\subsection{Contribution of satellites to the cold gas budget} \label{sec:discussion:satt}
{Satellite galaxies can increase the cool ICM mass by directly depositing their ISM gas into the ICM through ram-pressure stripping \citep[e.g., in Virgo or the Norma cluster,][]{2008Kenney,2012Dasyra, 2019Jachym}. In addition, satellites can generate density perturbations in the surrounding ICM, potentially promoting the condensation of cold gas.}

{In the TNG-Cluster simulation, while satellites are an important source of cold gas in the ICM at $z>0.5$, their overall contribution declines at late times due to their decreasing abundance \citep[][]{2025Rohr_cool,2025Staffehl}\footnote{Similarly, in the CGM of massive Milky Way–like halos, \citet{2025Ramesh_origin} find with tracers that only a small fraction of cold clouds ($\sim 5\%$) originate from satellites as they spiral toward the halo center at low redshifts.}. Nevertheless, in the ICM of our particular simulated low-mass cluster, we witness several infalling dense gas structures, resembling jellyfish galaxies, with characteristic morphologies and gas tails extending radially away from the center. Gas stripped from satellites at high altitudes (outside the refinement region) may continue to fall towards the cluster core and contribute to its cold gas reservoir, provided it survives the journey. Moreover, we find that the fly-by of satellites induces density perturbations. These locally decrease $t_{\rm cool}/t_{\rm ff}$, which may in turn facilitate condensation at large radii. Some of this gas could subsequently accrete toward the center}

{Overall, satellites may contribute substantially to the cold gas reservoir through both direct gas stripping and satellite-induced condensation, but quantifying this contribution will require further analysis with tracers.}

\subsection{Comparison to observed cool gas in cluster core}\label{sec:discussion:comp_obs}

\paragraph*{Filament morphologies.} 
We compare our results with observations of BCGs, which reveal a diversity of molecular and warm ionized (H$\alpha$-emitting) filamentary structures in galaxy clusters. Observational studies generally identify a range of morphological configurations, most commonly extended, clumpy, or asymmetric filamentary structures, and more rarely compact, disk-like configurations \citep{2019Olivares_mol,2022Olivares_MUSE}. This is qualitatively consistent with our findings, where various morphological configurations are reproduced at different times (Figure~\ref{fig:res_panel}). Disk-like structures are the rarest (see also \textcolor{blue}{Lehle et al. in prep}), in agreement with observations where molecular and H$\alpha$ gas disks are detected in only a few BCGs, such as Hydra A \citep{2019Rose} and A262 \citep{2019Russel}.

In the study of \citet{2019Olivares_mol}, based on a sample of 15 cool-core clusters, filaments of molecular gas were found spanning $0.6 - 25$ kpc (ALMA, median $7$ kpc), while warm ionized (H$\alpha$, MUSE) filaments extend over larger scales of $3.8 - 45$ kpc (median $23$ kpc), tracing more extended envelopes around colder structures. These measurements are resolution-limited, e.g., ALMA resolution is ($\sim 0.5-3$ kpc) while MUSE is also affected by atmospheric seeing, and likely misses the smallest clumps. {The filament and cloud 3D extents} in h963 fall within this observed range (Figure~\ref{fig:results_cloud_stats}), although we do not explicitly model mock H$\alpha$ or molecular emission, nor apply observational detection thresholds. In our simulation, the warm phase is also typically located at the interfaces of cold gas clouds (Figure~\ref{fig:meth_refinement_img}), consistent with this picture.

Observations reveal a strong spatial association between cold molecular gas, warm ionized filaments, and AGN-driven X-ray structures. Filaments are frequently found (in projection) along the rims of X-ray cavities, beneath buoyantly rising bubbles, or interacting with radio lobes, as seen in systems such as Perseus, Phoenix, A1835, and A1795 \citep[e.g.,][]{2017Russel}. In our simulation, using thin slices that minimize projection effects, we qualitatively find that some of the low-entropy multiphase gas located along the AGN energy injection path is uplifted and remains spatially coincident with X-ray cavities (Figures \ref{fig:res_panel} and \ref{fig:res_entropy_upflift}).

\paragraph*{Kinematic structures.}

Integral field spectroscopy (e.g., with MUSE or SITELLE) enables spatially resolved measurements of the line-of-sight (LOS) kinematics of ionized nebulae using emission lines such as H$\alpha$. For example, in Perseus, \citet{2024Vigneron} report coherent velocities within individual filaments, ranging from $\sim -200$ to $300$ km s$^{-1}$. {Cool structures in h963 exhibit diverse kinematics, with LOS velocities spanning $\sim -530$ to $465$ km s$^{-1}$ and radial velocities $-645$ to $240$ km s$^{-1}$ (5-95th percentile for the low-entropy gas across redshifts)}. Although individual filaments often display coherent motions, LOS velocities alone cannot unambiguously distinguish between inflows and outflows. As a result, it remains unclear whether the observed filaments are predominantly inflowing, as found in our simulated cluster.

Velocity dispersion, inferred from emission-line widths, probes the dynamical state of the gas. Observations generally show elevated dispersions near the AGN and X-ray cavities, while filaments at larger radii are more quiescent. In Perseus, for example, outer nebular gas has dispersions below $\sim 50$ km s$^{-1}$, rising to $>250$ km s$^{-1}$ in the central region \citep{2024Vigneron}. Similarly, in PKS 0745$-$191 and A1835, velocity dispersions peak at $200$–$300$ km s$^{-1}$ near cavities and within the central $\sim 10$ kpc \citep{2025Li_VSF}.

Observations show that H$\alpha$ and molecular gas are spatially and kinematically correlated, indicating a possible dynamical coupling between warm and cold phases \citep[][]{2025Li_VSF}. Similarly, we find that all gas phases (cool, warm, and hot) simultaneously show increases in v$_{\rm rad}$ and $\sigma_v$ following AGN energy injections -- albeit at different amplitudes -- suggesting strongly coupled dynamics.

High H$\alpha$ velocity dispersions are often confined to the paths of radio jets (e.g., A262), indicating direct jet–gas interactions, while surrounding gas remains quiescent. Despite the simplicity of the TNG  model, we see a similar pattern in the simulation: gas along the direction of kinetic feedback is strongly affected, whereas gas on the opposite side remains largely undisturbed, with low radial velocities (Fig.~\ref{fig:res_entropy_upflift}).

\paragraph*{{Magnetic support and }star formation in filaments.}

Filaments in cool-core clusters generally show little or no star formation. Ultraviolet imaging of systems such as NGC 5044, M87, and the Centaurus cluster indicates extremely low SFR ($<10^{-3}$ M$_{\odot}$ yr$^{-1}$), suggesting that filaments are stable against gravitational collapse (e.g., \citealt{2026Tamhane}, but see \citealt{2011McDonald}).  Magnetic fields can provide additional support, { allowing filaments to accumulate cold gas without forming stars \citep[e.g.,][]{2008Fabian_Nature, 2026Tamhane}.}

In our simulated filaments, we find low to zero SFR at all redshifts ($< 4 \times 10^{-3}$ M$_\odot$ yr$^{-1}$), consistent with the non-star-forming systems. 
{Their typical hydrogen number densities are $n_{\rm H} = 0.013$--$0.021$ cm$^{-3}$ (16th--84th percentiles across the cloud population and redshifts), below the star-formation threshold adopted in TNG ($n_{\rm H} > 0.1$ cm$^{-3}$). Magnetic fields may contribute to maintaining these low densities: we find }field strengths of 2--30 $\mu$G (5th--95th percentiles of gas with entropy $<50$ keV cm$^2$), with $P_{\rm B}/P_{\rm th} > 1$ in the central parts of the filaments (row 3, Figure \ref{fig:res_panel}).

\section{Summary and conclusions}\label{sec:conclusion}

In this work, we present a cosmological magnetohydrodynamical zoom-in simulation of a $2.3 \times 10^{14} M_\odot$ cool-core galaxy cluster with super-Lagrangian refinement within the central $370$ ckpc, reaching tens-of-parsec resolution in the cold gas phase. The halo is drawn from TNG-Cluster \citep{2024Nelson} and evolved using the IllustrisTNG physical model, which produces realistic AGN-driven features \citep[e.g., cavities, shocks, turbulence,][]{Prunier2025b,2025cPrunier,2024Truong}. {This unprecedented combination of both cosmological context and spatial resolution allows us to follow the formation and evolution of cold ($T<10^{4.5}$ K) and warm ($10^{4.5}$ K $<T<10^{5.5}$ K) gas and their interaction with AGN feedback at low redshift ($z\leq0.35$) self-consistently. 

Our results confirm that several properties previously investigated in idealized cluster simulations \citep[e.g.,][]{2013Gaspari,2019Beckmann,2025Sotira,2024Fournier}, including the filamentary or disky morphology of the cold gas and its redistribution, fragmentation, and uplift by AGN-driven outflows, also emerge in a fully cosmological setting at high spatial resolution. Cool gaseous nebulae form naturally in our cluster core despite the cavities, shocks, and outflows driven by the AGN. By comparing the fiducial simulation with a re-simulation in which AGN kinetic feedback is switched off for several Gyr, we find that AGN feedback reshapes the spatial distribution and kinematics of the cold gas reservoir, {increasing the number of cold clouds by a factor of $\sim2$ while leaving the total cold gas mass largely unchanged.} {Uplift, mixing, and compression may favor local condensation, although we cannot establish whether AGN activity leads to net new cold gas formation or primarily fragments and redistributes pre-existing cold gas.} Assessing the relative importance of these processes is further complicated by the realism of the cosmological setting, where AGN-driven motions interact with satellite infall and large-scale motions associated with structure formation.

The present paper provides an initial characterization of the simulation and its multiphase gas properties, laying the groundwork for follow-up studies. We summarize our main findings below:

\begin{itemize}

    \item Although the intracluster medium is filled with hot X-ray-emitting gas, it also contains a substantial amount of cooler, denser, low-entropy  ($K < 50~\mathrm{keV\ cm^2}$) gas at $z \leq 0.35$. The cold gas reservoir has a total mass of $10^{8} - 10^{10}\ \mathrm{M}_\odot$ and contributes $1 - 10\%$ of the total gas mass within the central 50 ckpc (Figures~\ref{fig:meth_refinement_img} \& \ref{fig:res_covering_frac}).\\

    \item The cold gas covering fraction, number of discrete clouds, and small-scale morphology all depend on numerical resolution. Our super-resolved simulation, with gas resolution of $2.3 \times 10^{4} M_\odot$, produces a larger population of small clouds than the original halo in TNG-Cluster, which has a gas mass resolution of $1.2 \times 10^{7} M_\odot$. However, the total cold gas mass is relatively converged. {Most of the cool gas mass is contained in the most massive clouds (>10$^{6.8}$ M$_\odot$), explaining why the total cold gas mass remains relatively similar between the two simulations} (left panel, Figure~\ref{fig:res_covering_frac}).\\
   
    \item This cooler gas is organized into predominantly elongated filamentary cloud structures, made of cold gas surrounded by a layer of warmer gas (Figures~\ref{fig:meth_refinement_img} \&~\ref{fig:res_panel}). Occasionally, they form rotationally supported disks. These clouds have characteristic extents on the order of kpc to tens of kpc, and are found from the cluster center out to radii of a few hundred kpc. They have complex morphologies and kinematics, resembling the molecular and H$\alpha$ ionised filaments observed in many cool-core galaxy clusters (Figures~\ref{fig:res_panel}, \ref{fig:results_cloud_stats}, and discussion in Section \ref{sec:discussion:comp_obs}).\\

    \item We isolate the impact of AGN feedback by re-simulating the same cluster with {AGN kinetic feedback turned off} from $z \leq 0.6$ for 3.7 Gyr (variation run, see Table \ref{tab:tab_sim_run}). AGN feedback suppresses star formation and prevents the formation of a massive cold star-forming disk (Fig.~\ref{fig:glob_quant_with_without}). It also redistributes cold gas, increasing its covering fraction, and drives outflowing cold gas (Fig.~\ref{fig:res_nobh}). AGN-driven heating increases gas turbulent motion in the core, and raises the cooling time in the cluster core, both on large scales and Gyr timescales (Fig.~ \ref{fig:res_phase_space}) and immediately after outbursts in the inner 30 kpc (Fig.~ \ref{fig:disc_tcool_tff_profile}), thereby influencing the conditions for thermal instability and the formation of the cold gas phase.\\

    \item AGN-driven outflows couple to the different gas phases. Following an outburst, the outflow velocities and velocity dispersions of the cold, warm, and hot phases increase sharply by factors of up to $\sim 3$, reaching hundreds of km~s$^{-1}$. This coupling varies between feedback events, with different gas phases responding differently depending on the configuration and kinematics of the gas prior to the outburst, as well as the direction of the AGN feedback. Overall, the direct kinematic impact of AGN feedback is short-lived, typically lasting $\lesssim 50$\ Myr (Figures~\ref{fig:res_kinematics} \& \ref{fig:res_entropy_upflift}).\\

    \item Individual episodes of AGN feedback can uplift and redistribute low-entropy gas from the central 10 kpc on $\sim 10$\ Myr timescales, producing high-velocity outflows of low-entropy gas at $\geq 500~\mathrm{km\ s^{-1}}$ (Figures ~\ref{fig:results_cloud_stats} \&  \ref{fig:res_entropy_upflift}). We identify cases of low-entropy filaments trailing behind or entrained by cavities and pressure fronts (Figure \ref{fig:res_panel} \& \ref{fig:res_entropy_upflift}).\\
    
    \item {Cool filamentary and cloud-like gas likely originates through multiple channels, including condensation from the hot ICM, stripped gas, and cooling associated with infalling satellites (Section \ref{sec:discussion:satt}) and AGN-driven processes (Section \ref{sec:discussion:cooling_heating}), which fragment and redistribute pre-existing cold gas while potentially creating conditions favorable to local condensation.} 
\end{itemize}

Our combination of TNG-Cluster halos with the GIBLE super-Lagrangian refinement technique offers a fully cosmological laboratory to study gas in cluster cores across cosmic time.  In the future, we can study the small-scale properties of ICM turbulence across different gas phases. This is especially compelling with the advent of high-energy-resolution, spatially resolved X-ray spectroscopy with \textit{XRISM} and, in the future, \textit{HUBS} and \textit{NewAthena}. We also plan to investigate the impact of numerical resolution on magnetic field structure and evolution, examine the spatial correlation between H$\alpha$ and X-ray surface brightness, and further constrain the origin of the cold gas using Lagrangian tracer methods.

\section*{Acknowledgements}
MP thanks Martin Fournier, Phillipe Salome, Pierre Guillard, Bipradeep Saha, and Sunna Gottschewski for enlightening discussions on cool gas and simulation aspects. MP acknowledges funding from the Physics Department of the University of Montreal (UdeM) and the Centre for Research in Astrophysics of Quebec (CRAQ). DN acknowledges funding from the Deutsche Forschungsgemeinschaft (DFG) through an Emmy Noether Research Group (grant number NE 2441/1-1). MP and AP acknowledge funding from the European Union (ERC, COSMIC-KEY, 101087822, PI: Pillepich). JHL acknowledges funding from the Canada Research Chairs and from the Discovery grant program from the Natural Sciences and Engineering Research Council of Canada (NSERC). KL acknowledges funding from the Hector Fellow Academy through a Research Career Development Award. RR acknowledges support by the World Premier International Research Center Initiative (WPI), MEXT, Japan. The analysis presented in this work has used the Vera cluster of the Max Planck Institute for Astronomy (MPIA), operated by the Max Planck Computational Data Facility (MPCDF). 

\section*{Data Availability}

The IllustrisTNG simulations, including TNG-Cluster, are publicly available in their entirety from \url{www.tng-project.org/data/} \citep{2019NelsonPublicReleaseTNG}. Data directly related to this publication are available on request from the corresponding author.

\bibliographystyle{mnras}
\bibliography{Litt_Bibli}

@ARTICLE{2019Pillepich,
       author = {{Pillepich}, Annalisa and {Nelson}, Dylan and {Springel}, Volker and {Pakmor}, R{\"u}diger and {Torrey}, Paul and {Weinberger}, Rainer and {Vogelsberger}, Mark and {Marinacci}, Federico and {Genel}, Shy and {van der Wel}, Arjen and et al.},
        title = "{First results from the TNG50 simulation: the evolution of stellar and gaseous discs across cosmic time}",
      journal = {\mnras},
         year = 2019,
        month = dec,
       volume = {490},
       number = {3},
        pages = {3196-3233},
          doi = {10.1093/mnras/stz2338},
archivePrefix = {arXiv},
       eprint = {1902.05553},
 primaryClass = {astro-ph.GA},
       adsurl = {https://scixplorer.org/abs/2019MNRAS.490.3196P}
}

@ARTICLE{2022Dutta,
       author = {{Dutta}, Alankar and {Sharma}, Prateek and {Nelson}, Dylan},
        title = "{Cooling flows around cold clouds in the circumgalactic medium: steady-state models and comparison with TNG50}",
      journal = {\mnras},
         year = 2022,
        month = mar,
       volume = {510},
       number = {3},
        pages = {3561-3574},
          doi = {10.1093/mnras/stab3653},
archivePrefix = {arXiv},
       eprint = {2107.02722},
 primaryClass = {astro-ph.GA},
       adsurl = {https://ui.adsabs.harvard.edu/abs/2022MNRAS.510.3561D}
}

@ARTICLE{2026Lachlan,
       author = {{Lancaster}, Lachlan and {Mohapatra}, Rajsekhar and {Fielding}, Drummond B. and {Bryan}, Greg L.},
        title = "{Ceci n'est pas une Couche de M{\'e}lange: The Meaning of Resolved Turbulent Radiative Mixing}",
      journal = {\apj},
         year = 2026,
        month = jul,
       volume = {1005},
       number = {2},
          eid = {222},
        pages = {222},
          doi = {10.3847/1538-4357/ae7714},
archivePrefix = {arXiv},
       eprint = {2606.04087},
 primaryClass = {astro-ph.GA},
       adsurl = {https://ui.adsabs.harvard.edu/abs/2026ApJ..1005..222L}
}

@ARTICLE{2026Biba,
       author = {{Biba}, Jonas and {Nelson}, Dylan},
        title = "{Turbulence in the multi-phase circumgalactic medium: bridging TNG50 and idealized kpc-scale simulations}",
      journal = {arXiv e-prints},
         year = 2026,
        month = jun,
          eid = {arXiv:2606.13777},
        pages = {arXiv:2606.13777},
          doi = {10.48550/arXiv.2606.13777},
archivePrefix = {arXiv},
       eprint = {2606.13777},
 primaryClass = {astro-ph.GA},
       adsurl = {https://ui.adsabs.harvard.edu/abs/2026arXiv260613777B}
}

@ARTICLE{2019Beckmann,
       author = {{Beckmann}, R.~S. and {Dubois}, Y. and {Guillard}, P. and {Salome}, P. and {Olivares}, V. and {Polles}, F. and {Cadiou}, C. and {Combes}, F. and {Hamer}, S. and {Lehnert}, M.~D. and {Pineau des Forets}, G.},
        title = "{Dense gas formation and destruction in a simulated Perseus-like galaxy cluster with spin-driven black hole feedback}",
      journal = {\aap},
         year = 2019,
        month = nov,
       volume = {631},
          eid = {A60},
        pages = {A60},
          doi = {10.1051/0004-6361/201936188},
archivePrefix = {arXiv},
       eprint = {1909.01329},
 primaryClass = {astro-ph.GA},
       adsurl = {https://ui.adsabs.harvard.edu/abs/2019A&A...631A..60B}
}

@ARTICLE{2026Gottschewski,
       author = {{Gottschewski}, Sunna and {Braspenning}, Joey and {Pillepich}, Annalisa and {Lehle}, Katrin and {Nelson}, Dylan},
        title = "{The star-forming past and quenching of brightest cluster galaxies in TNG-Cluster}",
      journal = {arXiv e-prints},
         year = 2026,
        month = oct,
          eid = {arXiv:2610.06634},
        pages = {arXiv:2610.06634},
archivePrefix = {arXiv},
       eprint = {2610.06634},
 primaryClass = {astro-ph.GA},
       adsurl = {https://ui.adsabs.harvard.edu/abs/2026arXiv261006634G}
}

@ARTICLE{2026Saha,
       author = {{Saha}, Bipradeep and {Pillepich}, Annalisa and {Braspenning}, Joey and {Prunier}, Marine and {Chatzigiannakis}, Dimitris and {Nelson}, Dylan},
        title = "{Bulk vs. turbulent motions at the centres of galaxy clusters: AGN-driven turbulence according to TNG-Cluster}",
      journal = {arXiv e-prints},
         year = 2026,
        month = jun,
          eid = {arXiv:2606.05355},
        pages = {arXiv:2606.05355},
          doi = {10.48550/arXiv.2606.05355},
archivePrefix = {arXiv},
       eprint = {2606.05355},
 primaryClass = {astro-ph.GA},
       adsurl = {https://ui.adsabs.harvard.edu/abs/2026arXiv260605355S}
}

@ARTICLE{2010Marinacci,
       author = {{Marinacci}, Federico and {Binney}, James and {Fraternali}, Filippo and {Nipoti}, Carlo and {Ciotti}, Luca and {Londrillo}, Pasquale},
        title = "{The mode of gas accretion on to star-forming galaxies}",
      journal = {\mnras},
         year = 2010,
        month = may,
       volume = {404},
       number = {3},
        pages = {1464-1474},
          doi = {10.1111/j.1365-2966.2010.16352.x},
archivePrefix = {arXiv},
       eprint = {1001.2446},
 primaryClass = {astro-ph.GA},
       adsurl = {https://ui.adsabs.harvard.edu/abs/2010MNRAS.404.1464M}
}

@misc{tamhane2026cavitydrivengastransportcoolcore,
      title={Cavity-driven Gas Transport in Cool-Core Clusters: Implications for Multiphase Gas Cooling}, 
      author={Prathamesh Tamhane and Brian McNamara and Paul Nulsen},
      year={2026},
      eprint={2609.26967},
      archivePrefix={arXiv},
      primaryClass={astro-ph.GA},
      url={https://arxiv.org/abs/2609.26967}, 
}

@ARTICLE{2025Lehle,
       author = {{Lehle}, Katrin and {Nelson}, Dylan and {Pillepich}, Annalisa},
        title = "{What drives cluster cool-core transformations? A population-level analysis using TNG-Cluster}",
      journal = {\aap},
         year = 2025,
        month = oct,
       volume = {702},
          eid = {A25},
        pages = {A25},
          doi = {10.1051/0004-6361/202554374},
archivePrefix = {arXiv},
       eprint = {2503.01969},
 primaryClass = {astro-ph.GA},
       adsurl = {https://ui.adsabs.harvard.edu/abs/2025A&A...702A..25L}
}

@ARTICLE{2025Olivares_Nature,
       author = {{Olivares}, Valeria and {Picquenot}, Adrien and {Su}, Yuanyuan and {Gaspari}, Massimo and {Gendron-Marsolais}, Marie-Lou and {Polles}, Fiorella L. and {Nulsen}, Paul},
        title = "{An H{\ensuremath{\alpha}}-X-ray surface-brightness correlation for filaments in cooling-flow clusters}",
      journal = {Nature Astronomy},
         year = 2025,
        month = mar,
       volume = {9},
        pages = {449-457},
          doi = {10.1038/s41550-024-02473-8},
archivePrefix = {arXiv},
       eprint = {2501.01902},
 primaryClass = {astro-ph.GA},
       adsurl = {https://ui.adsabs.harvard.edu/abs/2025NatAs...9..449O}
}

@ARTICLE{2025Rhea,
       author = {{Rhea}, Carter Lee and {Hlavacek-Larrondo}, Julie and {Gendron-Marsolais}, Marie-Lou and {Vigneron}, Benjamin and {Donahue}, Megan and {Thilloy}, Auriane and {Rousseau-Nepton}, Laurie and {Mezcua}, Mar and {Werner}, Norbert and {Barrera-Ballesteros}, Jorge and {Choi}, Hyunseop and {Edge}, Alastair and {Fabian}, Andrew and {Voit}, G. Mark},
        title = "{Mapping the Filamentary Nebula of NGC 1275 with Multiwavelength SITELLE Observations}",
      journal = {\aj},
         year = 2025,
        month = apr,
       volume = {169},
       number = {4},
          eid = {203},
        pages = {203},
          doi = {10.3847/1538-3881/adb732},
archivePrefix = {arXiv},
       eprint = {2502.05406},
 primaryClass = {astro-ph.GA},
       adsurl = {https://ui.adsabs.harvard.edu/abs/2025AJ....169..203R}
}

@ARTICLE{2025Li_VSF,
       author = {{Li}, Muzi and {McNamara}, B.~R. and {Coil}, Alison L. and {Marie-Jo{\"e}lle Gingras} and {Brighenti}, Fabrizio and {Russell}, H.~R. and {Tamhaneh}, Prathamesh D. and {Oh}, S. Peng and {Perrotta}, Serena},
        title = "{Velocity Structure Correlations between the Nebular, Molecular, and Atmospheric Gases in the Cores of Four Cool Core Clusters}",
      journal = {\apj},
         year = 2025,
        month = may,
       volume = {984},
       number = {1},
          eid = {22},
        pages = {22},
          doi = {10.3847/1538-4357/adc102},
archivePrefix = {arXiv},
       eprint = {2503.14486},
 primaryClass = {astro-ph.GA},
       adsurl = {https://ui.adsabs.harvard.edu/abs/2025ApJ...984...22L}
}

@ARTICLE{2017Russel,
       author = {{Russell}, H.~R. and {McNamara}, B.~R. and {Fabian}, A.~C. and {Nulsen}, P.~E.~J. and {Combes}, F. and {Edge}, A.~C. and {Hogan}, M.~T. and {McDonald}, M. and {Salom{\'e}}, P. and {Tremblay}, G. and {Vantyghem}, A.~N.},
        title = "{Close entrainment of massive molecular gas flows by radio bubbles in the central galaxy of Abell 1795}",
      journal = {\mnras},
         year = 2017,
        month = dec,
       volume = {472},
       number = {4},
        pages = {4024-4037},
          doi = {10.1093/mnras/stx2255},
archivePrefix = {arXiv},
       eprint = {1708.08935},
 primaryClass = {astro-ph.GA},
       adsurl = {https://ui.adsabs.harvard.edu/abs/2017MNRAS.472.4024R}
}

@ARTICLE{2024Tan,
       author = {{Tan}, Brent and {Fielding}, Drummond B.},
        title = "{Cloud atlas: navigating the multiphase landscape of tempestuous galactic winds}",
      journal = {\mnras},
         year = 2024,
        month = feb,
       volume = {527},
       number = {4},
        pages = {9683-9714},
          doi = {10.1093/mnras/stad3793},
archivePrefix = {arXiv},
       eprint = {2305.14424},
 primaryClass = {astro-ph.GA},
       adsurl = {https://ui.adsabs.harvard.edu/abs/2024MNRAS.527.9683T}
}

@ARTICLE{2013Gaspari,
       author = {{Gaspari}, M. and {Ruszkowski}, M. and {Oh}, S. Peng},
        title = "{Chaotic cold accretion on to black holes}",
      journal = {\mnras},
         year = 2013,
        month = jul,
       volume = {432},
       number = {4},
        pages = {3401-3422},
          doi = {10.1093/mnras/stt692},
archivePrefix = {arXiv},
       eprint = {1301.3130},
 primaryClass = {astro-ph.CO},
       adsurl = {https://ui.adsabs.harvard.edu/abs/2013MNRAS.432.3401G}
}

@ARTICLE{2005Pizzolato,
       author = {{Pizzolato}, Fabio and {Soker}, Noam},
        title = "{On the Nature of Feedback Heating in Cooling Flow Clusters}",
      journal = {\apj},
         year = 2005,
        month = oct,
       volume = {632},
       number = {2},
        pages = {821-830},
          doi = {10.1086/444344},
archivePrefix = {arXiv},
       eprint = {astro-ph/0407042},
 primaryClass = {astro-ph},
       adsurl = {https://ui.adsabs.harvard.edu/abs/2005ApJ...632..821P}
}

@ARTICLE{2015Singh,
       author = {{Singh}, Ashmeet and {Sharma}, Prateek},
        title = "{The cold mode: a phenomenological model for the evolution of density perturbations in the intracluster medium}",
      journal = {\mnras},
         year = 2015,
        month = jan,
       volume = {446},
       number = {2},
        pages = {1895-1906},
          doi = {10.1093/mnras/stu2264},
archivePrefix = {arXiv},
       eprint = {1409.1220},
 primaryClass = {astro-ph.GA},
       adsurl = {https://ui.adsabs.harvard.edu/abs/2015MNRAS.446.1895S}
}

@ARTICLE{2016McNamara,
       author = {{McNamara}, B.~R. and {Russell}, H.~R. and {Nulsen}, P.~E.~J. and {Hogan}, M.~T. and {Fabian}, A.~C. and {Pulido}, F. and {Edge}, A.~C.},
        title = "{A Mechanism for Stimulating AGN Feedback by Lifting Gas in Massive Galaxies}",
      journal = {\apj},
         year = 2016,
        month = oct,
       volume = {830},
       number = {2},
          eid = {79},
        pages = {79},
          doi = {10.3847/0004-637X/830/2/79},
archivePrefix = {arXiv},
       eprint = {1604.04629},
 primaryClass = {astro-ph.GA},
       adsurl = {https://ui.adsabs.harvard.edu/abs/2016ApJ...830...79M}
}

@ARTICLE{Ramesh_Nelson_23_TNG50_II,
       author = {{Ramesh}, Rahul and {Nelson}, Dylan and {Pillepich}, Annalisa},
        title = "{The circumgalactic medium of Milky Way-like galaxies in the TNG50 simulation - II. Cold, dense gas clouds and high-velocity cloud analogs}",
      journal = {\mnras},
         year = 2023,
        month = jun,
       volume = {522},
       number = {1},
        pages = {1535-1555},
          doi = {10.1093/mnras/stad951},
archivePrefix = {arXiv},
       eprint = {2303.16215},
 primaryClass = {astro-ph.GA},
       adsurl = {https://ui.adsabs.harvard.edu/abs/2023MNRAS.522.1535R}
}

@ARTICLE{2012McCourt,
       author = {{McCourt}, Michael and {Sharma}, Prateek and {Quataert}, Eliot and {Parrish}, Ian J.},
        title = "{Thermal instability in gravitationally stratified plasmas: implications for multiphase structure in clusters and galaxy haloes}",
      journal = {\mnras},
         year = 2012,
        month = feb,
       volume = {419},
       number = {4},
        pages = {3319-3337},
          doi = {10.1111/j.1365-2966.2011.19972.x},
archivePrefix = {arXiv},
       eprint = {1105.2563},
 primaryClass = {astro-ph.CO},
       adsurl = {https://ui.adsabs.harvard.edu/abs/2012MNRAS.419.3319M}
}

@ARTICLE{2025Augustin,
       author = {{Augustin}, Ramona and {Tumlinson}, Jason and {Peeples}, Molly S. and {O'Shea}, Brian W. and {Smith}, Britton D. and {Lochhaas}, Cassandra and {Wright}, Anna C. and {Acharyya}, Ayan and {Werk}, Jessica K. and {Lehner}, Nicolas and {Corlies}, Lauren and {Simons}, Raymond C. and {Howk}, J. Christopher and {O'Meara}, John M.},
        title = "{FOGGIE. X. Characterizing the Small-scale Structure of the Circumgalactic Medium and Its Imprint on Observables}",
      journal = {\apj},
         year = 2025,
        month = nov,
       volume = {993},
       number = {1},
          eid = {52},
        pages = {52},
          doi = {10.3847/1538-4357/ae0462},
archivePrefix = {arXiv},
       eprint = {2501.06551},
 primaryClass = {astro-ph.GA},
       adsurl = {https://ui.adsabs.harvard.edu/abs/2025ApJ...993...52A}
}

@ARTICLE{2019Suresh,
       author = {{Suresh}, Joshua and {Nelson}, Dylan and {Genel}, Shy and {Rubin}, Kate H.~R. and {Hernquist}, Lars},
        title = "{Zooming in on accretion - II. Cold circumgalactic gas simulated with a super-Lagrangian refinement scheme}",
      journal = {\mnras},
         year = 2019,
        month = mar,
       volume = {483},
       number = {3},
        pages = {4040-4059},
          doi = {10.1093/mnras/sty3402},
archivePrefix = {arXiv},
       eprint = {1811.01949},
 primaryClass = {astro-ph.GA},
       adsurl = {https://ui.adsabs.harvard.edu/abs/2019MNRAS.483.4040S}
}

@ARTICLE{2014LiBryan,
       author = {{Li}, Yuan and {Bryan}, Greg L.},
        title = "{Modeling Active Galactic Nucleus Feedback in Cool-core Clusters: The Formation of Cold Clumps}",
      journal = {\apj},
         year = 2014,
        month = jul,
       volume = {789},
       number = {2},
          eid = {153},
        pages = {153},
          doi = {10.1088/0004-637X/789/2/153},
archivePrefix = {arXiv},
       eprint = {1310.7580},
 primaryClass = {astro-ph.CO},
       adsurl = {https://ui.adsabs.harvard.edu/abs/2014ApJ...789..153L}
}

@ARTICLE{2017Voit,
       author = {{Voit}, G. Mark and {Meece}, Greg and {Li}, Yuan and {O'Shea}, Brian W. and {Bryan}, Greg L. and {Donahue}, Megan},
        title = "{A Global Model for Circumgalactic and Cluster-core Precipitation}",
      journal = {\apj},
         year = 2017,
        month = aug,
       volume = {845},
       number = {1},
          eid = {80},
        pages = {80},
          doi = {10.3847/1538-4357/aa7d04},
archivePrefix = {arXiv},
       eprint = {1607.02212},
 primaryClass = {astro-ph.GA},
       adsurl = {https://ui.adsabs.harvard.edu/abs/2017ApJ...845...80V}
}

@ARTICLE{2024Fournier,
       author = {{Fournier}, M. and {Grete}, P. and {Br{\"u}ggen}, M. and {Glines}, F.~W. and {O'Shea}, B.~W.},
        title = "{The properties of magnetised cold filaments in a cool-core galaxy cluster}",
      journal = {\aap},
         year = 2024,
        month = nov,
       volume = {691},
          eid = {A239},
        pages = {A239},
          doi = {10.1051/0004-6361/202451031},
archivePrefix = {arXiv},
       eprint = {2406.05044},
 primaryClass = {astro-ph.GA},
       adsurl = {https://ui.adsabs.harvard.edu/abs/2024A&A...691A.239F}
}

@ARTICLE{2022Zhang,
       author = {{Zhang}, Congyao and {Zhuravleva}, Irina and {Gendron-Marsolais}, Marie-Lou and {Churazov}, Eugene and {Schekochihin}, Alexander A. and {Forman}, William R.},
        title = "{Bubble-driven gas uplift in galaxy clusters and its velocity features}",
      journal = {\mnras},
         year = 2022,
        month = nov,
       volume = {517},
       number = {1},
        pages = {616-631},
          doi = {10.1093/mnras/stac2282},
archivePrefix = {arXiv},
       eprint = {2203.04259},
 primaryClass = {astro-ph.HE},
       adsurl = {https://ui.adsabs.harvard.edu/abs/2022MNRAS.517..616Z}
}

@ARTICLE{2019Nelson,
       author = {{Nelson}, Dylan and {Pillepich}, Annalisa and {Springel}, Volker and {Pakmor}, R{\"u}diger and {Weinberger}, Rainer and {Genel}, Shy and {Torrey}, Paul and {Vogelsberger}, Mark and {Marinacci}, Federico and {Hernquist}, Lars},
        title = "{First results from the TNG50 simulation: galactic outflows driven by supernovae and black hole feedback}",
      journal = {\mnras},
         year = 2019,
        month = dec,
       volume = {490},
       number = {3},
        pages = {3234-3261},
          doi = {10.1093/mnras/stz2306},
archivePrefix = {arXiv},
       eprint = {1902.05554},
 primaryClass = {astro-ph.GA},
       adsurl = {https://ui.adsabs.harvard.edu/abs/2019MNRAS.490.3234N}
}

@ARTICLE{2024Ramesh_GibleI,
       author = {{Ramesh}, Rahul and {Nelson}, Dylan},
        title = "{Zooming in on the circumgalactic medium with GIBLE: Resolving small-scale gas structure in cosmological simulations}",
      journal = {\mnras},
         year = 2024,
        month = feb,
       volume = {528},
       number = {2},
        pages = {3320-3339},
          doi = {10.1093/mnras/stae237},
archivePrefix = {arXiv},
       eprint = {2307.11143},
 primaryClass = {astro-ph.GA},
       adsurl = {https://ui.adsabs.harvard.edu/abs/2024MNRAS.528.3320R}
}

@ARTICLE{2024Ramesh_draping,
       author = {{Ramesh}, Rahul and {Nelson}, Dylan and {Fielding}, Drummond and {Br{\"u}ggen}, Marcus},
        title = "{Zooming in on the circumgalactic medium with GIBLE. The topology and draping of magnetic fields around cold clouds}",
      journal = {\aap},
         year = 2024,
        month = apr,
       volume = {684},
          eid = {L16},
        pages = {L16},
          doi = {10.1051/0004-6361/202348786},
archivePrefix = {arXiv},
       eprint = {2404.01370},
 primaryClass = {astro-ph.GA},
       adsurl = {https://ui.adsabs.harvard.edu/abs/2024A&A...684L..16R}
}

@ARTICLE{2026Bennett,
       author = {{Bennett}, Jake S. and {Smith}, Aaron and {Arrigoni-Battaia}, Fabrizio and {Sijacki}, Debora and {Lochhaas}, Cassandra and {Hernquist}, Lars},
        title = "{The birth of the intracluster medium: the evolution of multiphase gas and Lyman-$α$ haloes in a simulated $z\sim3$ protocluster}",
      journal = {arXiv e-prints},
         year = 2026,
        month = mar,
          eid = {arXiv:2603.16991},
        pages = {arXiv:2603.16991},
          doi = {10.48550/arXiv.2603.16991},
archivePrefix = {arXiv},
       eprint = {2603.16991},
 primaryClass = {astro-ph.GA},
       adsurl = {https://ui.adsabs.harvard.edu/abs/2026arXiv260316991B}
}

@ARTICLE{2017Weinberger,
       author = {{Weinberger}, Rainer and {Springel}, Volker and {Hernquist}, Lars and {Pillepich}, Annalisa and {Marinacci}, Federico and {Pakmor}, R{\"u}diger and {Nelson}, Dylan and {Genel}, Shy and {Vogelsberger}, Mark and {Naiman}, Jill and {Torrey}, Paul},
        title = "{Simulating galaxy formation with black hole driven thermal and kinetic feedback}",
      journal = {\mnras},
         year = 2017,
        month = mar,
       volume = {465},
       number = {3},
        pages = {3291-3308},
          doi = {10.1093/mnras/stw2944},
archivePrefix = {arXiv},
       eprint = {1607.03486},
 primaryClass = {astro-ph.GA},
       adsurl = {https://ui.adsabs.harvard.edu/abs/2017MNRAS.465.3291W}
}

@ARTICLE{2019Hummels,
       author = {{Hummels}, Cameron B. and {Smith}, Britton D. and {Hopkins}, Philip F. and {O'Shea}, Brian W. and {Silvia}, Devin W. and {Werk}, Jessica K. and {Lehner}, Nicolas and {Wise}, John H. and {Collins}, David C. and {Butsky}, Iryna S.},
        title = "{The Impact of Enhanced Halo Resolution on the Simulated Circumgalactic Medium}",
      journal = {\apj},
         year = 2019,
        month = sep,
       volume = {882},
       number = {2},
          eid = {156},
        pages = {156},
          doi = {10.3847/1538-4357/ab378f},
archivePrefix = {arXiv},
       eprint = {1811.12410},
 primaryClass = {astro-ph.GA},
       adsurl = {https://ui.adsabs.harvard.edu/abs/2019ApJ...882..156H}
}

@ARTICLE{2018Gaspari_Shaken,
       author = {{Gaspari}, M. and {McDonald}, M. and {Hamer}, S.~L. and {Brighenti}, F. and {Temi}, P. and {Gendron-Marsolais}, M. and {Hlavacek-Larrondo}, J. and {Edge}, A.~C. and {Werner}, N. and {Tozzi}, P. and {Sun}, M. and {Stone}, J.~M. and {Tremblay}, G.~R. and {Hogan}, M.~T. and {Eckert}, D. and {Ettori}, S. and {Yu}, H. and {Biffi}, V. and {Planelles}, S.},
        title = "{Shaken Snow Globes: Kinematic Tracers of the Multiphase Condensation Cascade in Massive Galaxies, Groups, and Clusters}",
      journal = {\apj},
         year = 2018,
        month = feb,
       volume = {854},
       number = {2},
          eid = {167},
        pages = {167},
          doi = {10.3847/1538-4357/aaaa1b},
archivePrefix = {arXiv},
       eprint = {1709.06564},
 primaryClass = {astro-ph.GA},
       adsurl = {https://ui.adsabs.harvard.edu/abs/2018ApJ...854..167G}
}

@ARTICLE{2018Prasad,
       author = {{Prasad}, Deovrat and {Sharma}, Prateek and {Babul}, Arif},
        title = "{Cool-core Clusters: The Role of BCG, Star Formation, and AGN-driven Turbulence}",
      journal = {\apj},
         year = 2018,
        month = aug,
       volume = {863},
       number = {1},
          eid = {62},
        pages = {62},
          doi = {10.3847/1538-4357/aacce8},
archivePrefix = {arXiv},
       eprint = {1801.04282},
 primaryClass = {astro-ph.GA},
       adsurl = {https://ui.adsabs.harvard.edu/abs/2018ApJ...863...62P}
}

@ARTICLE{Prunier2025b,
       author = {{Prunier}, Marine and {Pillepich}, Annalisa and {Hlavacek-Larrondo}, Julie and {Nelson}, Dylan},
        title = "{X-ray cavities in TNG-Cluster: a direct comparison to observations}",
      journal = {\mnras},
         year = 2025,
        month = may,
       volume = {539},
       number = {2},
        pages = {1040-1064},
          doi = {10.1093/mnras/staf525},
archivePrefix = {arXiv},
       eprint = {2503.01965},
 primaryClass = {astro-ph.GA},
       adsurl = {https://ui.adsabs.harvard.edu/abs/2025MNRAS.539.1040P}
}

@article{Prunier2025a,
    author = {Prunier, Marine and Hlavacek-Larrondo, Julie and Pillepich, Annalisa and Lehle, Katrin and Nelson, Dylan},
    title = {X-ray cavities in TNG-Cluster: AGN phenomena in the full cosmological context},
    journal = {Monthly Notices of the Royal Astronomical Society},
    volume = {536},
    number = {4},
    pages = {3200-3219},
    year = {2025},
    month = {12},
    issn = {0035-8711},
    doi = {10.1093/mnras/stae2743},
    url = {https://doi.org/10.1093/mnras/stae2743},
    eprint = {https://academic.oup.com/mnras/article-pdf/536/4/3200/61108006/stae2743.pdf},
}

@ARTICLE{2008Sanders_soundwaves,
       author = {{Sanders}, J.~S. and {Fabian}, A.~C.},
        title = "{Sound waves in the intracluster medium of the Centaurus cluster}",
      journal = {\mnras},
         year = 2008,
        month = oct,
       volume = {390},
       number = {1},
        pages = {L93-L97},
          doi = {10.1111/j.1745-3933.2008.00549.x},
archivePrefix = {arXiv},
       eprint = {0808.2384},
 primaryClass = {astro-ph},
       adsurl = {https://ui.adsabs.harvard.edu/abs/2008MNRAS.390L..93S}
}

@ARTICLE{Fabian2006_Perseus_Shock,
       author = {{Fabian}, A.~C. and {Sanders}, J.~S. and {Taylor}, G.~B. and {Allen}, S.~W. and {Crawford}, C.~S. and {Johnstone}, R.~M. and {Iwasawa}, K.},
        title = "{A very deep Chandra observation of the Perseus cluster: shocks, ripples and conduction}",
      journal = {\mnras},
         year = 2006,
        month = feb,
       volume = {366},
       number = {2},
        pages = {417-428},
          doi = {10.1111/j.1365-2966.2005.09896.x},
archivePrefix = {arXiv},
       eprint = {astro-ph/0510476},
 primaryClass = {astro-ph},
       adsurl = {https://ui.adsabs.harvard.edu/abs/2006MNRAS.366..417F}
}

@ARTICLE{2009Cavagnolo,
       author = {{Cavagnolo}, Kenneth W. and {Donahue}, Megan and {Voit}, G. Mark and {Sun}, Ming},
        title = "{Intracluster Medium Entropy Profiles for a Chandra Archival Sample of Galaxy Clusters}",
      journal = {\apjs},
         year = 2009,
        month = may,
       volume = {182},
       number = {1},
        pages = {12-32},
          doi = {10.1088/0067-0049/182/1/12},
archivePrefix = {arXiv},
       eprint = {0902.1802},
 primaryClass = {astro-ph.CO},
       adsurl = {https://ui.adsabs.harvard.edu/abs/2009ApJS..182...12C}
}

@ARTICLE{2021Wang,
       author = {{Wang}, C. and {Ruszkowski}, M. and {Pfrommer}, C. and {Oh}, S. Peng and {Yang}, H.-Y.~K.},
        title = "{Non-Kolmogorov turbulence in multiphase intracluster medium driven by cold gas precipitation and AGN jets}",
      journal = {\mnras},
         year = 2021,
        month = jun,
       volume = {504},
       number = {1},
        pages = {898-909},
          doi = {10.1093/mnras/stab966},
archivePrefix = {arXiv},
       eprint = {2012.11085},
 primaryClass = {astro-ph.GA},
       adsurl = {https://ui.adsabs.harvard.edu/abs/2021MNRAS.504..898W}
}

@ARTICLE{2019Qiu,
       author = {{Qiu}, Yu and {Bogdanovi{\'c}}, Tamara and {Li}, Yuan and {Park}, KwangHo and {Wise}, John H.},
        title = "{The Interplay of Kinetic and Radiative Feedback in Galaxy Clusters}",
      journal = {\apj},
         year = 2019,
        month = may,
       volume = {877},
       number = {1},
          eid = {47},
        pages = {47},
          doi = {10.3847/1538-4357/ab18fd},
archivePrefix = {arXiv},
       eprint = {1810.01857},
 primaryClass = {astro-ph.GA},
       adsurl = {https://ui.adsabs.harvard.edu/abs/2019ApJ...877...47Q}
}

@ARTICLE{2024Nelson,
       author = {{Nelson}, Dylan and {Pillepich}, Annalisa and {Ayromlou}, Mohammadreza and {Lee}, Wonki and {Lehle}, Katrin and {Rohr}, Eric and {Truong}, Nhut},
        title = "{Introducing the TNG-Cluster simulation: Overview and the physical properties of the gaseous intracluster medium}",
      journal = {\aap},
         year = 2024,
        month = jun,
       volume = {686},
          eid = {A157},
        pages = {A157},
          doi = {10.1051/0004-6361/202348608},
archivePrefix = {arXiv},
       eprint = {2311.06338},
 primaryClass = {astro-ph.GA},
       adsurl = {https://ui.adsabs.harvard.edu/abs/2024A&A...686A.157N}
}

@ARTICLE{2024Truong,
       author = {{Truong}, Nhut and {Pillepich}, Annalisa and {Nelson}, Dylan and {Zhuravleva}, Irina and {Lee}, Wonki and {Ayromlou}, Mohammadreza and {Lehle}, Katrin},
        title = "{X-ray-inferred kinematics of the core intracluster medium in Perseus-like clusters: Insights from the TNG-Cluster simulation}",
      journal = {\aap},
         year = 2024,
        month = jun,
       volume = {686},
          eid = {A200},
        pages = {A200},
          doi = {10.1051/0004-6361/202348562},
archivePrefix = {arXiv},
       eprint = {2311.06334},
 primaryClass = {astro-ph.GA},
       adsurl = {https://ui.adsabs.harvard.edu/abs/2024A&A...686A.200T}
}

@ARTICLE{2022DonahueVoit,
       author = {{Donahue}, Megan and {Voit}, G. Mark},
        title = "{Baryon cycles in the biggest galaxies}",
      journal = {\physrep},
         year = 2022,
        month = aug,
       volume = {973},
        pages = {1-109},
          doi = {10.1016/j.physrep.2022.04.005},
archivePrefix = {arXiv},
       eprint = {2204.08099},
 primaryClass = {astro-ph.GA},
       adsurl = {https://ui.adsabs.harvard.edu/abs/2022PhR...973....1D}
}

@ARTICLE{2021Bourne_weak_sim,
       author = {{Bourne}, Martin A. and {Sijacki}, Debora},
        title = "{AGN jet feedback on a moving mesh: gentle cluster heating by weak shocks and lobe disruption}",
      journal = {\mnras},
         year = 2021,
        month = sep,
       volume = {506},
       number = {1},
        pages = {488-513},
          doi = {10.1093/mnras/stab1662},
archivePrefix = {arXiv},
       eprint = {2008.12784},
 primaryClass = {astro-ph.HE},
       adsurl = {https://ui.adsabs.harvard.edu/abs/2021MNRAS.506..488B}
}

@ARTICLE{2011Gaspari_Melioli,
       author = {{Gaspari}, M. and {Melioli}, C. and {Brighenti}, F. and {D'Ercole}, A.},
        title = "{The dance of heating and cooling in galaxy clusters: three-dimensional simulations of self-regulated active galactic nuclei outflows}",
      journal = {\mnras},
         year = 2011,
        month = feb,
       volume = {411},
       number = {1},
        pages = {349-372},
          doi = {10.1111/j.1365-2966.2010.17688.x},
archivePrefix = {arXiv},
       eprint = {1007.0674},
 primaryClass = {astro-ph.CO},
       adsurl = {https://ui.adsabs.harvard.edu/abs/2011MNRAS.411..349G}
}

@ARTICLE{2019Olivares_mol,
       author = {{Olivares}, V. and {Salome}, P. and {Combes}, F. and {Hamer}, S. and {Guillard}, P. and {Lehnert}, M.~D. and {Polles}, F.~L. and {Beckmann}, R.~S. and {Dubois}, Y. and {Donahue}, M. and {Edge}, A. and {Fabian}, A.~C. and {McNamara}, B. and {Rose}, T. and {Russell}, H.~R. and {Tremblay}, G. and {Vantyghem}, A. and {Canning}, R.~E.~A. and {Ferland}, G. and {Godard}, B. and {Peirani}, S. and {Pineau des Forets}, G.},
        title = "{Ubiquitous cold and massive filaments in cool core clusters}",
      journal = {\aap},
         year = 2019,
        month = nov,
       volume = {631},
          eid = {A22},
        pages = {A22},
          doi = {10.1051/0004-6361/201935350},
archivePrefix = {arXiv},
       eprint = {1902.09164},
 primaryClass = {astro-ph.GA},
       adsurl = {https://ui.adsabs.harvard.edu/abs/2019A&A...631A..22O}
}

@ARTICLE{2014Guo,
       author = {{Guo}, Fulai and {Mathews}, William G.},
        title = "{Hot versus Cold: the Dichotomy in Spherical Accretion of Cooling Flows onto Supermassive Black Holes in Elliptical Galaxies, Galaxy Groups, and Clusters}",
      journal = {\apj},
         year = 2014,
        month = jan,
       volume = {780},
       number = {2},
          eid = {126},
        pages = {126},
          doi = {10.1088/0004-637X/780/2/126},
archivePrefix = {arXiv},
       eprint = {1305.2958},
 primaryClass = {astro-ph.CO},
       adsurl = {https://ui.adsabs.harvard.edu/abs/2014ApJ...780..126G}
}

@ARTICLE{2021Chadayammuri,
       author = {{Chadayammuri}, Urmila and {Tremmel}, Michael and {Nagai}, Daisuke and {Babul}, Arif and {Quinn}, Thomas},
        title = "{Fountains and storms: the effects of AGN feedback and mergers on the evolution of the intracluster medium in the ROMULUSC simulation}",
      journal = {\mnras},
         year = 2021,
        month = jun,
       volume = {504},
       number = {3},
        pages = {3922-3937},
          doi = {10.1093/mnras/stab1010},
archivePrefix = {arXiv},
       eprint = {2001.06532},
 primaryClass = {astro-ph.GA},
       adsurl = {https://ui.adsabs.harvard.edu/abs/2021MNRAS.504.3922C}
}

@ARTICLE{2006Mayer,
       author = {{Mayer}, Lucio and {Mastropietro}, Chiara and {Wadsley}, James and {Stadel}, Joachim and {Moore}, Ben},
        title = "{Simultaneous ram pressure and tidal stripping; how dwarf spheroidals lost their gas}",
      journal = {\mnras},
         year = 2006,
        month = jul,
       volume = {369},
       number = {3},
        pages = {1021-1038},
          doi = {10.1111/j.1365-2966.2006.10403.x},
archivePrefix = {arXiv},
       eprint = {astro-ph/0504277},
 primaryClass = {astro-ph},
       adsurl = {https://ui.adsabs.harvard.edu/abs/2006MNRAS.369.1021M}
}

@ARTICLE{2014Sparks,
       author = {{Sparks}, W.~B. and {Pringle}, J.~E. and {Carswell}, R.~F. and {Voit}, G.~M. and {Donahue}, M. and {Cracraft}, M. and {Meyer}, E.~T. and {Hough}, J.~H. and {Manset}, N.},
        title = "{Polarization Diagnostics for Cool Core Cluster Emission Lines}",
      journal = {\apj},
         year = 2014,
        month = jan,
       volume = {780},
       number = {1},
          eid = {66},
        pages = {66},
          doi = {10.1088/0004-637X/780/1/66},
archivePrefix = {arXiv},
       eprint = {1311.2831},
 primaryClass = {astro-ph.CO},
       adsurl = {https://ui.adsabs.harvard.edu/abs/2014ApJ...780...66S}
}

@ARTICLE{2025Ramesh_origin,
       author = {{Ramesh}, Rahul and {Nelson}, Dylan and {Fielding}, Drummond and {Br{\"u}ggen}, Marcus},
        title = "{Zooming in on the circumgalactic medium with GIBLE: Tracing the origin and evolution of cold clouds}",
      journal = {\aap},
         year = 2025,
        month = jul,
       volume = {699},
          eid = {A10},
        pages = {A10},
          doi = {10.1051/0004-6361/202451303},
archivePrefix = {arXiv},
       eprint = {2407.00172},
 primaryClass = {astro-ph.GA},
       adsurl = {https://ui.adsabs.harvard.edu/abs/2025A&A...699A..10R}
}

@ARTICLE{2008DuboisTessier,
       author = {{Dubois}, Y. and {Teyssier}, R.},
        title = "{Cosmological MHD simulation of a cooling flow cluster}",
      journal = {\aap},
         year = 2008,
        month = may,
       volume = {482},
       number = {2},
        pages = {L13-L16},
          doi = {10.1051/0004-6361:200809513},
archivePrefix = {arXiv},
       eprint = {0802.0490},
 primaryClass = {astro-ph},
       adsurl = {https://ui.adsabs.harvard.edu/abs/2008A&A...482L..13D}
}

@ARTICLE{2008Pope,
       author = {{Pope}, Edward C.~D. and {Pittard}, Julian M. and {Hartquist}, Thomas W. and {Falle}, Sam A.~E.~G.},
        title = "{The generation of optical emission-line filaments in galaxy clusters}",
      journal = {\mnras},
         year = 2008,
        month = apr,
       volume = {385},
       number = {4},
        pages = {1779-1791},
          doi = {10.1111/j.1365-2966.2008.12963.x},
archivePrefix = {arXiv},
       eprint = {0801.2164},
 primaryClass = {astro-ph},
       adsurl = {https://ui.adsabs.harvard.edu/abs/2008MNRAS.385.1779P}
}

@ARTICLE{2008Cavagnolo,
       author = {{Cavagnolo}, Kenneth W. and {Donahue}, Megan and {Voit}, G. Mark and {Sun}, Ming},
        title = "{An Entropy Threshold for Strong H{\ensuremath{\alpha}} and Radio Emission in the Cores of Galaxy Clusters}",
      journal = {\apjl},
         year = 2008,
        month = aug,
       volume = {683},
       number = {2},
        pages = {L107},
          doi = {10.1086/591665},
archivePrefix = {arXiv},
       eprint = {0806.0382},
 primaryClass = {astro-ph},
       adsurl = {https://ui.adsabs.harvard.edu/abs/2008ApJ...683L.107C}
}

@ARTICLE{2008Revaz,
       author = {{Revaz}, Y. and {Combes}, F. and {Salom{\'e}}, P.},
        title = "{Formation of cold filaments in cooling flow clusters}",
      journal = {\aap},
         year = 2008,
        month = jan,
       volume = {477},
       number = {3},
        pages = {L33-L36},
          doi = {10.1051/0004-6361:20078915},
archivePrefix = {arXiv},
       eprint = {0711.4051},
 primaryClass = {astro-ph},
       adsurl = {https://ui.adsabs.harvard.edu/abs/2008A&A...477L..33R}
}

@ARTICLE{2025Reefe,
       author = {{Reefe}, Michael and {McDonald}, Michael and {Chatzikos}, Marios and {Seebeck}, Jerome and {Mushotzky}, Richard and {Veilleux}, Sylvain and {Allen}, Steven W. and {Bayliss}, Matthew and {Calzadilla}, Michael and {Canning}, Rebecca and {Donahue}, Megan and {Floyd}, Benjamin and {Gaspari}, Massimo and {Hlavacek-Larrondo}, Julie and {McNamara}, Brian and {Russell}, Helen and {Sarkar}, Arnab and {Sharon}, Keren and {Somboonpanyakul}, Taweewat},
        title = "{Cold Gas and Star Formation in the Phoenix Cluster with JWST}",
      journal = {\apj},
         year = 2025,
        month = aug,
       volume = {989},
       number = {2},
          eid = {156},
        pages = {156},
          doi = {10.3847/1538-4357/aded90},
archivePrefix = {arXiv},
       eprint = {2501.08527},
 primaryClass = {astro-ph.GA},
       adsurl = {https://ui.adsabs.harvard.edu/abs/2025ApJ...989..156R}
}

@ARTICLE{2015Brighenti,
       author = {{Brighenti}, Fabrizio and {Mathews}, William G. and {Temi}, Pasquale},
        title = "{Hot Gaseous Atmospheres in Galaxy Groups and Clusters Are Both Heated and Cooled by X-Ray Cavities}",
      journal = {\apj},
         year = 2015,
        month = apr,
       volume = {802},
       number = {2},
          eid = {118},
        pages = {118},
          doi = {10.1088/0004-637X/802/2/118},
archivePrefix = {arXiv},
       eprint = {1501.07647},
 primaryClass = {astro-ph.GA},
       adsurl = {https://ui.adsabs.harvard.edu/abs/2015ApJ...802..118B}
}

@ARTICLE{2014Hamer,
       author = {{Hamer}, S.~L. and {Edge}, A.~C. and {Swinbank}, A.~M. and {Oonk}, J.~B.~R. and {Mittal}, R. and {McNamara}, B.~R. and {Russell}, H.~R. and {Bremer}, M.~N. and {Combes}, F. and {Fabian}, A.~C. and {Nesvadba}, N.~P.~H. and {O'Dea}, C.~P. and {Baum}, S.~A. and {Salom{\'e}}, P. and {Tremblay}, G. and {Donahue}, M. and {Ferland}, G.~J. and {Sarazin}, C.~L.},
        title = "{Cold gas dynamics in Hydra-A: evidence for a rotating disc}",
      journal = {\mnras},
         year = 2014,
        month = jan,
       volume = {437},
       number = {1},
        pages = {862-878},
          doi = {10.1093/mnras/stt1949},
archivePrefix = {arXiv},
       eprint = {1310.4501},
 primaryClass = {astro-ph.CO},
       adsurl = {https://ui.adsabs.harvard.edu/abs/2014MNRAS.437..862H}
}

@ARTICLE{2012Dasyra,
       author = {{Dasyra}, K.~M. and {Combes}, F. and {Salom{\'e}}, P. and {Braine}, J.},
        title = "{Survival of molecular gas in Virgo's hot intracluster medium: CO near M 86}",
      journal = {\aap},
         year = 2012,
        month = apr,
       volume = {540},
          eid = {A112},
        pages = {A112},
          doi = {10.1051/0004-6361/201218775},
archivePrefix = {arXiv},
       eprint = {1202.1763},
 primaryClass = {astro-ph.CO},
       adsurl = {https://ui.adsabs.harvard.edu/abs/2012A&A...540A.112D}
}

@ARTICLE{2019Jachym,
       author = {{J{\'a}chym}, Pavel and {Kenney}, Jeffrey D.~P. and {Sun}, Ming and {Combes}, Fran{\c{c}}oise and {Cortese}, Luca and {Scott}, Tom C. and {Sivanandam}, Suresh and {Brinks}, Elias and {Roediger}, Elke and {Palou{\v{s}}}, Jan and {Fumagalli}, Michele},
        title = "{ALMA Unveils Widespread Molecular Gas Clumps in the Ram Pressure Stripped Tail of the Norma Jellyfish Galaxy}",
      journal = {\apj},
         year = 2019,
        month = oct,
       volume = {883},
       number = {2},
          eid = {145},
        pages = {145},
          doi = {10.3847/1538-4357/ab3e6c},
archivePrefix = {arXiv},
       eprint = {1905.13249},
 primaryClass = {astro-ph.GA},
       adsurl = {https://ui.adsabs.harvard.edu/abs/2019ApJ...883..145J}
}

@ARTICLE{2008Kenney,
       author = {{Kenney}, Jeffrey D.~P. and {Tal}, Tomer and {Crowl}, Hugh H. and {Feldmeier}, John and {Jacoby}, George H.},
        title = "{A Spectacular H{\ensuremath{\alpha}} Complex in Virgo: Evidence for a Collision between M86 and NGC 4438 and Implications for the Collisional ISM Heating of Ellipticals}",
      journal = {\apjl},
         year = 2008,
        month = nov,
       volume = {687},
       number = {2},
        pages = {L69},
          doi = {10.1086/593300},
archivePrefix = {arXiv},
       eprint = {0810.0711},
 primaryClass = {astro-ph},
       adsurl = {https://ui.adsabs.harvard.edu/abs/2008ApJ...687L..69K}
}

@ARTICLE{2024Duan_uplift,
       author = {{Duan}, Xiaodong and {Guo}, Fulai},
        title = "{Cold Filaments Formed in Hot Wake Flows Uplifted by Active Galactic Nucleus Bubbles in Galaxy Clusters}",
      journal = {\apj},
         year = 2024,
        month = sep,
       volume = {972},
       number = {1},
          eid = {41},
        pages = {41},
          doi = {10.3847/1538-4357/ad5bdf},
archivePrefix = {arXiv},
       eprint = {2403.02807},
 primaryClass = {astro-ph.GA},
       adsurl = {https://ui.adsabs.harvard.edu/abs/2024ApJ...972...41D}
}

@ARTICLE{2026Poitras,
       author = {{Poitras}, Camille and {Gendron-Marsolais}, Marie-Lou and {Olivares}, Valeria and {Li}, Yuan and {Picquenot}, Adrien and {Simionescu}, Aurora and {Fossati}, Matteo and {Boselli}, Alessandro and {Mu{\~n}oz}, Laura Hermosa and {Cazzoli}, Sara and {Hlavacek-Larrondo}, Julie and {Richard-Laferri{\`e}re}, Annabelle},
        title = "{Observations of AGN-driven feedback: dynamics and ionization of the filaments in M87}",
      journal = {\mnras},
         year = 2026,
        month = jan,
       volume = {545},
       number = {2},
          eid = {staf1981},
        pages = {staf1981},
          doi = {10.1093/mnras/staf1981},
archivePrefix = {arXiv},
       eprint = {2512.17563},
 primaryClass = {astro-ph.GA},
       adsurl = {https://ui.adsabs.harvard.edu/abs/2026MNRAS.545f1981P}
}

@ARTICLE{2019Choudhury,
       author = {{Choudhury}, Prakriti Pal and {Sharma}, Prateek and {Quataert}, Eliot},
        title = "{Multiphase gas in the circumgalactic medium: relative role of t$_{cool}$/t$_{ff}$ and density fluctuations}",
      journal = {\mnras},
         year = 2019,
        month = sep,
       volume = {488},
       number = {3},
        pages = {3195-3210},
          doi = {10.1093/mnras/stz1857},
archivePrefix = {arXiv},
       eprint = {1901.02903},
 primaryClass = {astro-ph.GA},
       adsurl = {https://ui.adsabs.harvard.edu/abs/2019MNRAS.488.3195C}
}

@ARTICLE{2019Rose,
       author = {{Rose}, Tom and {Edge}, A.~C. and {Combes}, F. and {Gaspari}, M. and {Hamer}, S. and {Nesvadba}, N. and {Russell}, H. and {Tremblay}, G.~R. and {Baum}, S.~A. and {O'Dea}, C. and {Peck}, A.~B. and {Sarazin}, C. and {Vantyghem}, A. and {Bremer}, M. and {Donahue}, M. and {Fabian}, A.~C. and {Ferland}, G. and {McNamara}, B.~R. and {Mittal}, R. and {Oonk}, J.~B.~R. and {Salom{\'e}}, P. and {Swinbank}, A.~M. and {Voit}, M.},
        title = "{Deep and narrow CO absorption revealing molecular clouds in the Hydra-A brightest cluster galaxy}",
      journal = {\mnras},
         year = 2019,
        month = may,
       volume = {485},
       number = {1},
        pages = {229-238},
          doi = {10.1093/mnras/stz406},
archivePrefix = {arXiv},
       eprint = {1902.01863},
 primaryClass = {astro-ph.GA},
       adsurl = {https://ui.adsabs.harvard.edu/abs/2019MNRAS.485..229R}
}

@ARTICLE{2019Russel,
       author = {{Russell}, H.~R. and {McNamara}, B.~R. and {Fabian}, A.~C. and {Nulsen}, P.~E.~J. and {Combes}, F. and {Edge}, A.~C. and {Madar}, M. and {Olivares}, V. and {Salom{\'e}}, P. and {Vantyghem}, A.~N.},
        title = "{Driving massive molecular gas flows in central cluster galaxies with AGN feedback}",
      journal = {\mnras},
         year = 2019,
        month = dec,
       volume = {490},
       number = {3},
        pages = {3025-3045},
          doi = {10.1093/mnras/stz2719},
archivePrefix = {arXiv},
       eprint = {1902.09227},
 primaryClass = {astro-ph.GA},
       adsurl = {https://ui.adsabs.harvard.edu/abs/2019MNRAS.490.3025R}
}

@ARTICLE{2008Fabian_Nature,
       author = {{Fabian}, A.~C. and {Johnstone}, R.~M. and {Sanders}, J.~S. and {Conselice}, C.~J. and {Crawford}, C.~S. and {Gallagher}, III, J.~S. and {Zweibel}, E.},
        title = "{Magnetic support of the optical emission line filaments in NGC 1275}",
      journal = {\nat},
         year = 2008,
        month = aug,
       volume = {454},
       number = {7207},
        pages = {968-970},
          doi = {10.1038/nature07169},
archivePrefix = {arXiv},
       eprint = {0808.2712},
 primaryClass = {astro-ph},
       adsurl = {https://ui.adsabs.harvard.edu/abs/2008Natur.454..968F}
}

@ARTICLE{2026Lehle,
       author = {{Lehle}, Katrin and {Ruszkowski}, Mateusz and {Nelson}, Dylan and {Prunier}, Marine and {Pillepich}, Annalisa},
        title = "{Magnetic fields in the intracluster medium with TNG-Cluster: Properties, morphology, and tangential anisotropy}",
      journal = {\aap},
         year = 2026,
        month = jan,
       volume = {705},
          eid = {A41},
        pages = {A41},
          doi = {10.1051/0004-6361/202556486},
archivePrefix = {arXiv},
       eprint = {2507.12517},
 primaryClass = {astro-ph.GA},
       adsurl = {https://ui.adsabs.harvard.edu/abs/2026A&A...705A..41L}
}

@ARTICLE{2026Tamhane,
       author = {{Tamhane}, Prathamesh and {Sun}, Ming and {Waldron}, William and {Hosogi}, Kokoro and {da Silva}, Patricia and {Le}, Huan and {Gaspari}, Massimo and {Combes}, Francoise and {Werner}, Norbert and {Schellenberger}, Gerrit and {Fabian}, Andrew and {Canning}, Rebecca and {David}, Laurence and {Donahue}, Megan and {Voit}, Mark},
        title = "{HST view of NGC 5044: Constraints on filament widths, magnetic support, multiphase structure, and comparison with cluster environments}",
      journal = {\pasa},
         year = 2026,
        month = mar,
       volume = {43},
          eid = {e036},
        pages = {e036},
          doi = {10.1017/pasa.2026.10172},
archivePrefix = {arXiv},
       eprint = {2603.00249},
 primaryClass = {astro-ph.GA},
       adsurl = {https://ui.adsabs.harvard.edu/abs/2026PASA...43...36T}
}

@ARTICLE{2011McDonald,
       author = {{McDonald}, Michael and {Veilleux}, Sylvain and {Rupke}, David S.~N. and {Mushotzky}, Richard and {Reynolds}, Christopher},
        title = "{Star Formation Efficiency in the Cool Cores of Galaxy Clusters}",
      journal = {\apj},
         year = 2011,
        month = jun,
       volume = {734},
       number = {2},
          eid = {95},
        pages = {95},
          doi = {10.1088/0004-637X/734/2/95},
archivePrefix = {arXiv},
       eprint = {1104.0665},
 primaryClass = {astro-ph.CO},
       adsurl = {https://ui.adsabs.harvard.edu/abs/2011ApJ...734...95M}
}

@ARTICLE{2026Chatzigiannakis,
       author = {{Chatzigiannakis}, Dimitris and {Pillepich}, Annalisa and {Simionescu}, Aurora and {Truong}, Nhut and {Nelson}, Dylan},
        title = "{The temperature and metallicity distributions of the ICM: insights with TNG-Cluster for XRISM-like observations}",
      journal = {\mnras},
         year = 2026,
        month = apr,
       volume = {547},
       number = {2},
          eid = {stag061},
        pages = {stag061},
          doi = {10.1093/mnras/stag061},
archivePrefix = {arXiv},
       eprint = {2503.01983},
 primaryClass = {astro-ph.GA},
       adsurl = {https://ui.adsabs.harvard.edu/abs/2026MNRAS.547ag061C}
}

@ARTICLE{2025Rintoul,
       author = {{Rintoul}, Thomas A. and {van de Voort}, Freeke and {Hannington}, Andrew T. and {Pakmor}, R{\"u}diger and {Bieri}, Rebekka and {Werhahn}, Maria and {Talbot}, Rosie Y.},
        title = "{The role of magnetic fields in ram pressure stripping of satellite galaxies in the circumgalactic medium around massive galaxies}",
      journal = {\mnras},
         year = 2025,
        month = nov,
       volume = {543},
       number = {4},
        pages = {4321-4334},
          doi = {10.1093/mnras/staf1718},
archivePrefix = {arXiv},
       eprint = {2506.18983},
 primaryClass = {astro-ph.GA},
       adsurl = {https://ui.adsabs.harvard.edu/abs/2025MNRAS.543.4321R}
}

@ARTICLE{2020Zinger,
       author = {{Zinger}, Elad and {Pillepich}, Annalisa and {Nelson}, Dylan and {Weinberger}, Rainer and {Pakmor}, R{\"u}diger and {Springel}, Volker and {Hernquist}, Lars and {Marinacci}, Federico and {Vogelsberger}, Mark},
        title = "{Ejective and preventative: the IllustrisTNG black hole feedback and its effects on the thermodynamics of the gas within and around galaxies}",
      journal = {\mnras},
         year = 2020,
        month = nov,
       volume = {499},
       number = {1},
        pages = {768-792},
          doi = {10.1093/mnras/staa2607},
archivePrefix = {arXiv},
       eprint = {2004.06132},
 primaryClass = {astro-ph.GA},
       adsurl = {https://ui.adsabs.harvard.edu/abs/2020MNRAS.499..768Z}
}

@ARTICLE{2022Nobels,
       author = {{Nobels}, Folkert S.~J. and {Schaye}, Joop and {Schaller}, Matthieu and {Bah{\'e}}, Yannick M. and {Chaikin}, Evgenii},
        title = "{The interplay between AGN feedback and precipitation of the intracluster medium in simulations of galaxy groups and clusters}",
      journal = {\mnras},
         year = 2022,
        month = oct,
       volume = {515},
       number = {4},
        pages = {4838-4859},
          doi = {10.1093/mnras/stac2061},
archivePrefix = {arXiv},
       eprint = {2204.02205},
 primaryClass = {astro-ph.GA},
       adsurl = {https://ui.adsabs.harvard.edu/abs/2022MNRAS.515.4838N}
}

@ARTICLE{2019Lakhchaura,
       author = {{Lakhchaura}, K. and {Truong}, N. and {Werner}, N.},
        title = "{Correlations between supermassive black holes, hot atmospheres, and the total masses of early-type galaxies}",
      journal = {\mnras},
         year = 2019,
        month = sep,
       volume = {488},
       number = {1},
        pages = {L134-L142},
          doi = {10.1093/mnrasl/slz114},
archivePrefix = {arXiv},
       eprint = {1904.10513},
 primaryClass = {astro-ph.GA},
       adsurl = {https://scixplorer.org/abs/2019MNRAS.488L.134L}
}

@ARTICLE{2018Pillepich,
       author = {{Pillepich}, Annalisa and {Springel}, Volker and {Nelson}, Dylan and {Genel}, Shy and {Naiman}, Jill and {Pakmor}, R{\"u}diger and {Hernquist}, Lars and {Torrey}, Paul and {Vogelsberger}, Mark and {Weinberger}, Rainer and {Marinacci}, Federico},
        title = "{Simulating galaxy formation with the IllustrisTNG model}",
      journal = {\mnras},
         year = 2018,
        month = jan,
       volume = {473},
       number = {3},
        pages = {4077-4106},
          doi = {10.1093/mnras/stx2656},
archivePrefix = {arXiv},
       eprint = {1703.02970},
 primaryClass = {astro-ph.GA},
       adsurl = {https://ui.adsabs.harvard.edu/abs/2018MNRAS.473.4077P}
}

@ARTICLE{2021PillepichErosita,
       author = {{Pillepich}, Annalisa and {Nelson}, Dylan and {Truong}, Nhut and {Weinberger}, Rainer and {Martin-Navarro}, Ignacio and {Springel}, Volker and {Faber}, Sandy M. and {Hernquist}, Lars},
        title = "{X-ray bubbles in the circumgalactic medium of TNG50 Milky Way- and M31-like galaxies: signposts of supermassive black hole activity}",
      journal = {\mnras},
         year = 2021,
        month = dec,
       volume = {508},
       number = {4},
        pages = {4667-4695},
          doi = {10.1093/mnras/stab2779},
archivePrefix = {arXiv},
       eprint = {2105.08062},
 primaryClass = {astro-ph.GA},
       adsurl = {https://ui.adsabs.harvard.edu/abs/2021MNRAS.508.4667P}
}

@ARTICLE{2025cPrunier,
       author = {{Prunier}, Marine and {Ubertosi}, Francesco and {Hlavacek-Larrondo}, Julie and {Pillepich}, Annalisa},
        title = "{X-ray shocks in the cool cores of galaxy clusters: insights from TNG-Cluster}",
      journal = {\mnras},
         year = 2025,
        month = dec,
       volume = {544},
       number = {4},
        pages = {4188-4207},
          doi = {10.1093/mnras/staf1912},
archivePrefix = {arXiv},
       eprint = {2509.25314},
 primaryClass = {astro-ph.GA},
       adsurl = {https://ui.adsabs.harvard.edu/abs/2025MNRAS.544.4188P}
}

@ARTICLE{2024Omoruyi,
       author = {{Omoruyi}, Osase and {Tremblay}, Grant R. and {Combes}, Francoise and {Davis}, Timothy A. and {Gladders}, Michael D. and {Vikhlinin}, Alexey and {Nulsen}, Paul and {Kharb}, Preeti and {Baum}, Stefi A. and {O'Dea}, Christopher P. and {Sharon}, Keren and {Terrazas}, Bryan A. and {Nevin}, Rebecca and {Schechter}, Aimee L. and {Zuhone}, John A. and {McDonald}, Michael and {Dahle}, Hakon and {Bayliss}, Matthew B. and {Connor}, Thomas and {Florian}, Michael and {Rigby}, Jane R. and {Vaddi}, Sravani},
        title = "{``Beads-on-a-string'' Star Formation Tied to One of the Most Powerful Active Galactic Nucleus Outbursts Observed in a Cool-core Galaxy Cluster}",
      journal = {\apj},
         year = 2024,
        month = mar,
       volume = {963},
       number = {1},
          eid = {1},
        pages = {1},
          doi = {10.3847/1538-4357/ad1101},
archivePrefix = {arXiv},
       eprint = {2312.06762},
 primaryClass = {astro-ph.GA},
       adsurl = {https://ui.adsabs.harvard.edu/abs/2024ApJ...963....1O}
}

@ARTICLE{2020Nelson_tng50,
       author = {{Nelson}, Dylan and {Sharma}, Prateek and {Pillepich}, Annalisa and {Springel}, Volker and {Pakmor}, R{\"u}diger and {Weinberger}, Rainer and {Vogelsberger}, Mark and {Marinacci}, Federico and {Hernquist}, Lars},
        title = "{Resolving small-scale cold circumgalactic gas in TNG50}",
      journal = {\mnras},
         year = 2020,
        month = oct,
       volume = {498},
       number = {2},
        pages = {2391-2414},
          doi = {10.1093/mnras/staa2419},
archivePrefix = {arXiv},
       eprint = {2005.09654},
 primaryClass = {astro-ph.GA},
       adsurl = {https://ui.adsabs.harvard.edu/abs/2020MNRAS.498.2391N}
}

@ARTICLE{2014Zhuravleva,
       author = {{Zhuravleva}, I. and {Churazov}, E. and {Schekochihin}, A.~A. and {Allen}, S.~W. and {Ar{\'e}valo}, P. and {Fabian}, A.~C. and {Forman}, W.~R. and {Sanders}, J.~S. and {Simionescu}, A. and {Sunyaev}, R. and {Vikhlinin}, A. and {Werner}, N.},
        title = "{Turbulent heating in galaxy clusters brightest in X-rays}",
      journal = {\nat},
         year = 2014,
        month = nov,
       volume = {515},
       number = {7525},
        pages = {85-87},
          doi = {10.1038/nature13830},
archivePrefix = {arXiv},
       eprint = {1410.6485},
 primaryClass = {astro-ph.HE},
       adsurl = {https://ui.adsabs.harvard.edu/abs/2014Natur.515...85Z}
}

@ARTICLE{Fabian_hiddend1,
       author = {{Fabian}, A.~C. and {Ferland}, G.~J. and {Sanders}, J.~S. and {McNamara}, B.~R. and {Pinto}, C. and {Walker}, S.~A.},
        title = "{Hidden cooling flows in clusters of galaxies}",
      journal = {\mnras},
         year = 2022,
        month = sep,
       volume = {515},
       number = {3},
        pages = {3336-3345},
          doi = {10.1093/mnras/stac2003},
archivePrefix = {arXiv},
       eprint = {2207.04951},
 primaryClass = {astro-ph.GA},
       adsurl = {https://ui.adsabs.harvard.edu/abs/2022MNRAS.515.3336F}
}

@ARTICLE{2026Sotira,
       author = {{Sotira}, Stefano and {Bourne}, Martin A. and {Sijacki}, Debora and {Vazza}, Franco and {Brighenti}, Fabrizio},
        title = "{Cold gas formation triggered by active galactic nuclei jet feedback in galaxy cluster cores}",
      journal = {arXiv e-prints},
         year = 2026,
        month = jan,
          eid = {arXiv:2601.14391},
        pages = {arXiv:2601.14391},
          doi = {10.48550/arXiv.2601.14391},
archivePrefix = {arXiv},
       eprint = {2601.14391},
 primaryClass = {astro-ph.GA},
       adsurl = {https://ui.adsabs.harvard.edu/abs/2026arXiv260114391S}
}

@ARTICLE{2025Sotira,
       author = {{Sotira}, S. and {Vazza}, F. and {Brighenti}, F.},
        title = "{Impact of active galactic nucleus feedback modes on the turbulent properties of the multiphase intracluster medium}",
      journal = {\aap},
         year = 2025,
        month = may,
       volume = {697},
          eid = {A138},
        pages = {A138},
          doi = {10.1051/0004-6361/202452549},
archivePrefix = {arXiv},
       eprint = {2410.07314},
 primaryClass = {astro-ph.GA},
       adsurl = {https://ui.adsabs.harvard.edu/abs/2025A&A...697A.138S}
}

@ARTICLE{2018Tremblay,
       author = {{Tremblay}, G.~R. and {Combes}, F. and {Oonk}, J.~B.~R. and {Russell}, H.~R. and {McDonald}, M.~A. and {Gaspari}, M. and {Husemann}, B. and {Nulsen}, P.~E.~J. and {McNamara}, B.~R. and {Hamer}, S.~L. and {O'Dea}, C.~P. and {Baum}, S.~A. and {Davis}, T.~A. and {Donahue}, M. and {Voit}, G.~M. and {Edge}, A.~C. and {Blanton}, E.~L. and {Bremer}, M.~N. and {Bulbul}, E. and {Clarke}, T.~E. and {David}, L.~P. and {Edwards}, L.~O.~V. and {Eggerman}, D. and {Fabian}, A.~C. and {Forman}, W. and {Jones}, C. and {Kerman}, N. and {Kraft}, R.~P. and {Li}, Y. and {Powell}, M. and {Randall}, S.~W. and {Salom{\'e}}, P. and {Simionescu}, A. and {Su}, Y. and {Sun}, M. and {Urry}, C.~M. and {Vantyghem}, A.~N. and {Wilkes}, B.~J. and {ZuHone}, J.~A.},
        title = "{A Galaxy-scale Fountain of Cold Molecular Gas Pumped by a Black Hole}",
      journal = {\apj},
         year = 2018,
        month = sep,
       volume = {865},
       number = {1},
          eid = {13},
        pages = {13},
          doi = {10.3847/1538-4357/aad6dd},
archivePrefix = {arXiv},
       eprint = {1808.00473},
 primaryClass = {astro-ph.GA},
       adsurl = {https://ui.adsabs.harvard.edu/abs/2018ApJ...865...13T}
}

@ARTICLE{2015LiBryan,
       author = {{Li}, Yuan and {Bryan}, Greg L. and {Ruszkowski}, Mateusz and {Voit}, G. Mark and {O'Shea}, Brian W. and {Donahue}, Megan},
        title = "{Cooling, AGN Feedback, and Star Formation in Simulated Cool-core Galaxy Clusters}",
      journal = {\apj},
         year = 2015,
        month = oct,
       volume = {811},
       number = {2},
          eid = {73},
        pages = {73},
          doi = {10.1088/0004-637X/811/2/73},
archivePrefix = {arXiv},
       eprint = {1503.02660},
 primaryClass = {astro-ph.GA},
       adsurl = {https://ui.adsabs.harvard.edu/abs/2015ApJ...811...73L}
}

@ARTICLE{Rohr2024_cool,
       author = {{Rohr}, Eric and {Pillepich}, Annalisa and {Nelson}, Dylan and {Ayromlou}, Mohammadreza and {P{\'e}roux}, C{\'e}line and {Zinger}, Elad},
        title = "{The cooler past of the intracluster medium in TNG-Cluster}",
      journal = {mnras},
         year = 2024,
        month = nov,
          doi = {10.1093/mnras/stae2536},
archivePrefix = {arXiv},
       eprint = {2410.19900},
 primaryClass = {astro-ph.GA},
       adsurl = {https://ui.adsabs.harvard.edu/abs/2024arXiv241019900R}
}

@ARTICLE{2016Shin,
       author = {{Shin}, Jaejin and {Woo}, Jong-Hak and {Mulchaey}, John S.},
        title = "{A Systematic Search for X-Ray Cavities in Galaxy Clusters, Groups, and Elliptical Galaxies}",
      journal = {\apjs},
         year = 2016,
        month = dec,
       volume = {227},
       number = {2},
          eid = {31},
        pages = {31},
          doi = {10.3847/1538-4365/227/2/31},
archivePrefix = {arXiv},
       eprint = {1610.03487},
 primaryClass = {astro-ph.CO},
       adsurl = {https://ui.adsabs.harvard.edu/abs/2016ApJS..227...31S}
}

@ARTICLE{2008Salome,
       author = {{Salom{\'e}}, P. and {Combes}, F. and {Revaz}, Y. and {Edge}, A.~C. and {Hatch}, N.~A. and {Fabian}, A.~C. and {Johnstone}, R.~M.},
        title = "{Cold gas in the Perseus cluster core: excitation of molecular gas in filaments}",
      journal = {\aap},
         year = 2008,
        month = jun,
       volume = {484},
       number = {2},
        pages = {317-325},
          doi = {10.1051/0004-6361:200809493},
archivePrefix = {arXiv},
       eprint = {0804.2113},
 primaryClass = {astro-ph},
       adsurl = {https://ui.adsabs.harvard.edu/abs/2008A&A...484..317S}
}

@ARTICLE{2018Ehlert,
       author = {{Ehlert}, K. and {Weinberger}, R. and {Pfrommer}, C. and {Pakmor}, R. and {Springel}, V.},
        title = "{Simulations of the dynamics of magnetized jets and cosmic rays in galaxy clusters}",
      journal = {\mnras},
         year = 2018,
        month = dec,
       volume = {481},
       number = {3},
        pages = {2878-2900},
          doi = {10.1093/mnras/sty2397},
archivePrefix = {arXiv},
       eprint = {1806.05679},
 primaryClass = {astro-ph.CO},
       adsurl = {https://ui.adsabs.harvard.edu/abs/2018MNRAS.481.2878E}
}

@ARTICLE{2004Birzan,
       author = {{B{\^\i}rzan}, L. and {Rafferty}, D.~A. and {McNamara}, B.~R. and {Wise}, M.~W. and {Nulsen}, P.~E.~J.},
        title = "{A Systematic Study of Radio-induced X-Ray Cavities in Clusters, Groups, and Galaxies}",
      journal = {\apj},
         year = 2004,
        month = jun,
       volume = {607},
       number = {2},
        pages = {800-809},
          doi = {10.1086/383519},
archivePrefix = {arXiv},
       eprint = {astro-ph/0402348},
 primaryClass = {astro-ph},
       adsurl = {https://ui.adsabs.harvard.edu/abs/2004ApJ...607..800B}
}

@ARTICLE{2007Pfrommer,
       author = {{Pfrommer}, C. and {En{\ss}lin}, T.~A. and {Springel}, V. and {Jubelgas}, M. and {Dolag}, K.},
        title = "{Simulating cosmic rays in clusters of galaxies - I. Effects on the Sunyaev-Zel'dovich effect and the X-ray emission}",
      journal = {\mnras},
         year = 2007,
        month = jun,
       volume = {378},
       number = {2},
        pages = {385-408},
          doi = {10.1111/j.1365-2966.2007.11732.x},
archivePrefix = {arXiv},
       eprint = {astro-ph/0611037},
 primaryClass = {astro-ph},
       adsurl = {https://ui.adsabs.harvard.edu/abs/2007MNRAS.378..385P}
}

@ARTICLE{Sijacki2008,
       author = {{Sijacki}, Debora and {Pfrommer}, Christoph and {Springel}, Volker and {En{\ss}lin}, Torsten A.},
        title = "{Simulations of cosmic-ray feedback by active galactic nuclei in galaxy clusters}",
      journal = {\mnras},
         year = 2008,
        month = jul,
       volume = {387},
       number = {4},
        pages = {1403-1415},
          doi = {10.1111/j.1365-2966.2008.13310.x},
archivePrefix = {arXiv},
       eprint = {0801.3285},
 primaryClass = {astro-ph},
       adsurl = {https://ui.adsabs.harvard.edu/abs/2008MNRAS.387.1403S}
}

@ARTICLE{2024Lehle,
       author = {{Lehle}, Katrin and {Nelson}, Dylan and {Pillepich}, Annalisa and {Truong}, Nhut and {Rohr}, Eric},
        title = "{The heart of galaxy clusters: Demographics and physical properties of cool-core and non-cool-core halos in the TNG-Cluster simulation}",
      journal = {\aap},
         year = 2024,
        month = jul,
       volume = {687},
          eid = {A129},
        pages = {A129},
          doi = {10.1051/0004-6361/202348609},
archivePrefix = {arXiv},
       eprint = {2311.06333},
 primaryClass = {astro-ph.GA},
       adsurl = {https://ui.adsabs.harvard.edu/abs/2024A&A...687A.129L}
}

@ARTICLE{2019NelsonPublicReleaseTNG,
       author = {{Nelson}, Dylan and {Springel}, Volker and {Pillepich}, Annalisa and {Rodriguez-Gomez}, Vicente and {Torrey}, Paul and {Genel}, Shy and {Vogelsberger}, Mark and {Pakmor}, Ruediger and {Marinacci}, Federico and {Weinberger}, Rainer and {Kelley}, Luke and {Lovell}, Mark and {Diemer}, Benedikt and {Hernquist}, Lars},
        title = "{The IllustrisTNG simulations: public data release}",
      journal = {Computational Astrophysics and Cosmology},
         year = 2019,
        month = may,
       volume = {6},
       number = {1},
          eid = {2},
        pages = {2},
          doi = {10.1186/s40668-019-0028-x},
archivePrefix = {arXiv},
       eprint = {1812.05609},
 primaryClass = {astro-ph.GA},
       adsurl = {https://ui.adsabs.harvard.edu/abs/2019ComAC...6....2N}
}

@ARTICLE{2022Olivares_MUSE,
       author = {{Olivares}, V. and {Salom{\'e}}, P. and {Hamer}, S.~L. and {Combes}, F. and {Gaspari}, M. and {Kolokythas}, K. and {O'Sullivan}, E. and {Beckmann}, R.~S. and {Babul}, A. and {Polles}, F.~L. and {Lehnert}, M. and {Loubser}, S.~I. and {Donahue}, M. and {Gendron-Marsolais}, M.-L. and {Lagos}, P. and {Pineau des Forets}, G. and {Godard}, B. and {Rose}, T. and {Tremblay}, G. and {Ferland}, G. and {Guillard}, P.},
        title = "{Gas condensation in brightest group galaxies unveiled with MUSE. Morphology and kinematics of the ionized gas}",
      journal = {\aap},
         year = 2022,
        month = oct,
       volume = {666},
          eid = {A94},
        pages = {A94},
          doi = {10.1051/0004-6361/202142475},
archivePrefix = {arXiv},
       eprint = {2201.07838},
 primaryClass = {astro-ph.GA},
       adsurl = {https://ui.adsabs.harvard.edu/abs/2022A&A...666A..94O}
}

@ARTICLE{2023Ramesh_tng50,
       author = {{Ramesh}, Rahul and {Nelson}, Dylan and {Pillepich}, Annalisa},
        title = "{The circumgalactic medium of Milky Way-like galaxies in the TNG50 simulation - I: halo gas properties and the role of SMBH feedback}",
      journal = {\mnras},
         year = 2023,
        month = jan,
       volume = {518},
       number = {4},
        pages = {5754-5777},
          doi = {10.1093/mnras/stac3524},
archivePrefix = {arXiv},
       eprint = {2211.00020},
 primaryClass = {astro-ph.GA},
       adsurl = {https://ui.adsabs.harvard.edu/abs/2023MNRAS.518.5754R}
}

@ARTICLE{2003SpringelHernquist,
       author = {{Springel}, Volker and {Hernquist}, Lars},
        title = "{Cosmological smoothed particle hydrodynamics simulations: a hybrid multiphase model for star formation}",
      journal = {\mnras},
         year = 2003,
        month = feb,
       volume = {339},
       number = {2},
        pages = {289-311},
          doi = {10.1046/j.1365-8711.2003.06206.x},
archivePrefix = {arXiv},
       eprint = {astro-ph/0206393},
 primaryClass = {astro-ph},
       adsurl = {https://ui.adsabs.harvard.edu/abs/2003MNRAS.339..289S}
}

@ARTICLE{2012Sharma,
       author = {{Sharma}, Prateek and {McCourt}, Michael and {Quataert}, Eliot and {Parrish}, Ian J.},
        title = "{Thermal instability and the feedback regulation of hot haloes in clusters, groups and galaxies}",
      journal = {\mnras},
         year = 2012,
        month = mar,
       volume = {420},
       number = {4},
        pages = {3174-3194},
          doi = {10.1111/j.1365-2966.2011.20246.x},
archivePrefix = {arXiv},
       eprint = {1106.4816},
 primaryClass = {astro-ph.CO},
       adsurl = {https://ui.adsabs.harvard.edu/abs/2012MNRAS.420.3174S}
}

@ARTICLE{2025Staffehl,
       author = {{Staffehl}, Milan and {Nelson}, Dylan and {Ayromlou}, Mohammadreza and {Rohr}, Eric and {Pillepich}, Annalisa},
        title = "{The abundance and origin of cool gas in galaxy clusters in the TNG-Cluster simulation}",
      journal = {\aap},
         year = 2025,
        month = nov,
       volume = {703},
          eid = {A33},
        pages = {A33},
          doi = {10.1051/0004-6361/202554373},
archivePrefix = {arXiv},
       eprint = {2503.01960},
 primaryClass = {astro-ph.GA},
       adsurl = {https://ui.adsabs.harvard.edu/abs/2025A&A...703A..33S}
}

@ARTICLE{2025Rohr_cool,
       author = {{Rohr}, Eric and {Pillepich}, Annalisa and {Nelson}, Dylan and {Ayromlou}, Mohammadreza and {P{\'e}roux}, C{\'e}line and {Zinger}, Elad},
        title = "{The cooler past of the intracluster medium in TNG-cluster}",
      journal = {\mnras},
         year = 2025,
        month = jan,
       volume = {536},
       number = {2},
        pages = {1226-1250},
          doi = {10.1093/mnras/stae2536},
archivePrefix = {arXiv},
       eprint = {2410.19900},
 primaryClass = {astro-ph.GA},
       adsurl = {https://ui.adsabs.harvard.edu/abs/2025MNRAS.536.1226R}
}

@ARTICLE{2025Fournier,
       author = {{Fournier}, M. and {Grete}, P. and {Br{\"u}ggen}, M. and {O'Shea}, B.~W. and {Prasad}, D. and {Wibking}, B.~D. and {Glines}, F.~W. and {Mohapatra}, R.},
        title = "{XMAGNET: Velocity structure functions of active galactic nucleus-driven turbulence in the multiphase intracluster medium}",
      journal = {\aap},
         year = 2025,
        month = jun,
       volume = {698},
          eid = {A121},
        pages = {A121},
          doi = {10.1051/0004-6361/202554278},
archivePrefix = {arXiv},
       eprint = {2502.19486},
 primaryClass = {astro-ph.GA},
       adsurl = {https://ui.adsabs.harvard.edu/abs/2025A&A...698A.121F}
}

@ARTICLE{2014Panagoulia_vol,
       author = {{Panagoulia}, E.~K. and {Fabian}, A.~C. and {Sanders}, J.~S. and {Hlavacek-Larrondo}, J.},
        title = "{A volume-limited sample of X-ray galaxy groups and clusters - II. X-ray cavity dynamics}",
      journal = {\mnras},
         year = 2014,
        month = oct,
       volume = {444},
       number = {2},
        pages = {1236-1259},
          doi = {10.1093/mnras/stu1499},
archivePrefix = {arXiv},
       eprint = {1407.6614},
 primaryClass = {astro-ph.CO},
       adsurl = {https://ui.adsabs.harvard.edu/abs/2014MNRAS.444.1236P}
}

@ARTICLE{2024Vigneron,
       author = {{Vigneron}, Benjamin and {Hlavacek-Larrondo}, Julie and {Rhea}, Carter Lee and {Gendron-Marsolais}, Marie-Lou and {Lim}, Jeremy and {Reinheimer}, Jake and {Li}, Yuan and {Drissen}, Laurent and {Bryan}, Greg L. and {Donahue}, Megan and {Edge}, Alastair and {Fabian}, Andrew and {Hamer}, Stephen and {Martin}, Thomas and {McDonald}, Michael and {McNamara}, Brian and {Richard-Lafferri{\`e}re}, Annabelle and {Rousseau-Nepton}, Laurie and {Voit}, G. Mark and {Webb}, Tracy and {Werner}, Norbert},
        title = "{High-spectral-resolution Observations of the Optical Filamentary Nebula Surrounding NGC 1275}",
      journal = {\apj},
         year = 2024,
        month = feb,
       volume = {962},
       number = {1},
          eid = {96},
        pages = {96},
          doi = {10.3847/1538-4357/ad0fd8},
archivePrefix = {arXiv},
       eprint = {2311.16247},
 primaryClass = {astro-ph.GA},
       adsurl = {https://ui.adsabs.harvard.edu/abs/2024ApJ...962...96V}
}

@ARTICLE{1999Crawford,
       author = {{Crawford}, C.~S. and {Allen}, S.~W. and {Ebeling}, H. and {Edge}, A.~C. and {Fabian}, A.~C.},
        title = "{The ROSAT Brightest Cluster Sample - III. Optical spectra of the central cluster galaxies}",
      journal = {\mnras},
         year = 1999,
        month = jul,
       volume = {306},
       number = {4},
        pages = {857-896},
          doi = {10.1046/j.1365-8711.1999.02583.x},
archivePrefix = {arXiv},
       eprint = {astro-ph/9903057},
 primaryClass = {astro-ph},
       adsurl = {https://ui.adsabs.harvard.edu/abs/1999MNRAS.306..857C}
}

@article{mohr_properties_1999,
	title = {Properties of the {Intracluster} {Medium} in an ensemble of {Nearby} {Galaxy} {Clusters}},
	volume = {517},
	journal = {Astrophysical Journal},
	author = {Mohr, J.J. and Mathiesen, B. and Evrard, A.E.},
	year = {1999},
	pages = {627--649},
}

@ARTICLE{2024Fresco,
       author = {{Fresco}, A. and {P{\'e}roux}, C. and {Merloni}, A. and {Comparat}, J. and {Szakacs}, R. and {Weng}, S.},
        title = "{Searching for cold gas traced by MgII quasar absorbers in massive X-ray-selected galaxy clusters}",
      journal = {\aap},
         year = 2024,
        month = apr,
       volume = {684},
          eid = {A136},
        pages = {A136},
          doi = {10.1051/0004-6361/202346866},
archivePrefix = {arXiv},
       eprint = {2404.10773},
 primaryClass = {astro-ph.GA},
       adsurl = {https://ui.adsabs.harvard.edu/abs/2024A&A...684A.136F}
}

@ARTICLE{2019Genel,
       author = {{Genel}, Shy and {Bryan}, Greg L. and {Springel}, Volker and {Hernquist}, Lars and {Nelson}, Dylan and {Pillepich}, Annalisa and {Weinberger}, Rainer and {Pakmor}, R{\"u}diger and {Marinacci}, Federico and {Vogelsberger}, Mark},
        title = "{A Quantification of the Butterfly Effect in Cosmological Simulations and Implications for Galaxy Scaling Relations}",
      journal = {\apj},
         year = 2019,
        month = jan,
       volume = {871},
       number = {1},
          eid = {21},
        pages = {21},
          doi = {10.3847/1538-4357/aaf4bb},
archivePrefix = {arXiv},
       eprint = {1807.07084},
 primaryClass = {astro-ph.GA},
       adsurl = {https://ui.adsabs.harvard.edu/abs/2019ApJ...871...21G}
}


\end{document}